\documentclass[article]{JHEP3}

\usepackage{amsmath,amssymb}
\usepackage[dvips]{graphics}
\usepackage{epsfig}
\usepackage{psfrag}

\newcommand{\re}{{\mathbb{R}}{\mathrm{e}}}
\newcommand{\im}{{\mathbb{I}}{\mathrm{m}}}

\newcommand{\be}{\begin{equation}}
\newcommand{\ee}{\end{equation}}
\newcommand{\bea}{\begin{eqnarray}}
\newcommand{\eea}{\end{eqnarray}}
\newcommand{\bean}{\begin{eqnarray*}}
\newcommand{\eean}{\end{eqnarray*}}

\def\beq{\begin{equation}}

\def\eeq{\end{equation}}

\relax

\newcommand{\AI}{A_{\rm I}}

\newcommand{\AII}{A_{\rm II}}
\newcommand{\BII}{B_{\rm II}}

\newcommand{\CO}{\mathcal{O}}

\newcommand{\vecto}[2]{\left( \begin{array}{c} #1 \\ #2 \end{array}
\right) }
\newcommand{\matrto}[4]{\left( \begin{array}{cc} #1 & #2 \\
#3 & #4 \end{array} \right) }

\newcommand{\hypgeo}{{}_2 \mbox{F}_1}

\preprint{
{\small{\textsf{CERN-PH-TH/2007-091}}}
}

\title{Greybody Factors for $d$--Dimensional Black Holes}

\author{Troels Harmark$^{a}$, Jos\'e Nat\'ario$^{b}$ and Ricardo Schiappa$^{c}$
\\
$^{a}$The Niels Bohr Institute, \\
Blegdamsvej 17, 2100 Copenhagen \O, Denmark\\
\\
$^{b}$CAMGSD, Departamento de Matem\'atica, Instituto Superior T\'ecnico,\\
Av. Rovisco Pais 1, 1049--001 Lisboa, Portugal\\
\\
$^{c}$Theory Division, Department of Physics, CERN,\\
CH--1211 Gen\`eve 23, Switzerland\\
\\
\email{harmark@nbi.dk}, \quad
\email{jnatar@math.ist.utl.pt}, \quad
\email{ricardos@mail.cern.ch}
}

\abstract{Gravitational greybody factors are analytically computed for static, spherically symmetric black holes in $d$--dimensions, including black holes with charge and in the presence of a cosmological constant (where a proper definition of greybody factors for both asymptotically de Sitter and Anti--de Sitter spacetimes is provided). This calculation includes both the low--energy case---where the frequency of the scattered wave is small and real---and the asymptotic case---where the frequency of the scattered wave is very large along the imaginary axis---addressing gravitational perturbations as described by the Ishibashi--Kodama master equations, and yielding full transmission and reflection scattering coefficients for all considered spacetime geometries. At low frequencies a general method is developed, which can be employed for all three types of spacetime asymptotics, and which is independent of the details of the black hole. For asymptotically de Sitter black holes the greybody factor is different for even or odd spacetime dimension, and proportional to the ratio of the areas of the event and cosmological horizons. For asymptotically Anti--de Sitter black holes the greybody factor has a rich structure in which there are several critical frequencies where it equals either one (pure transmission) or zero (pure reflection, with these frequencies corresponding to the normal modes of pure Anti--de Sitter spacetime). At asymptotic frequencies the computation of the greybody factor uses a technique inspired by monodromy matching, and some universality is hidden in the transmission and reflection coefficients. For either charged or asymptotically de Sitter black holes the greybody factors are given by non--trivial functions, while for asymptotically Anti--de Sitter black holes the greybody factor precisely equals one (corresponding to pure blackbody emission).}

\keywords{Black Holes, $d$--Dimensional General Relativity and Greybody Factors}

\begin{document}



\vfill

\eject



\section{Introduction and Discussion}


Hawking radiation lies at the frontier between classical general relativity and quantum field theory, and may be a key towards unlocking the mysteries of a theory of quantum gravity. Classical macroscopic black holes in general relativity obey laws that are parallel to the laws of thermodynamics \cite{bch73}. As one sets out to describe quantum fields in black hole backgrounds, or, more generally, in the vicinity of any horizon, this similarity to thermodynamics becomes an exact connection as one unveils that black holes have a temperature and an entropy associated to them \cite{h75, h76}. Thermal radiation, sourced at the black hole event horizon, is emitted into the surrounding space with the consequence that the semi--classical black hole slowly looses its mass and eventually evaporates. At the precise location of the event horizon the Hawking radiation is blackbody radiation. However, this radiation still has to traverse a non--trivial, curved spacetime geometry before it eventually reaches an observer and is detected (\textit{e.g.}, an observer located at asymptotic infinity in an asymptotically flat spacetime). The surrounding spacetime thus works as a potential barrier for the radiation, giving a deviation from the blackbody radiation spectrum as seen by an asymptotic observer. The relative factor between the asymptotic radiation spectrum and the spectrum of blackbody radiation is dubbed the greybody factor.

The famous calculation of Hawking radiation \cite{h75} uses a semi--classical approximation to show that black holes have an exact thermal spectrum, where the expectation value $\langle n(\omega) \rangle$ for the number of particles of a given species, emitted in a mode with frequency $\omega$, is given by
\begin{equation}\label{n(omega)}
\langle n(\omega) \rangle =
\frac{\gamma(\omega)}{e^{\frac{\omega}{T_H}} \pm 1},
\end{equation}
\noindent
where $T_H$ is the Hawking temperature, the plus (minus) sign describes fermions (bosons), and where $\gamma(\omega)$ is the greybody factor, \textit{i.e.}, the probability for an outgoing wave, in the $\omega$--mode, to reach infinity. This coincides, as we shall see, with the absorption probability, \textit{i.e.}, the probability for an incoming wave, in the $\omega$--mode, to be absorbed by the black hole. If one integrates this expression over all spectra it leads to the total black hole emission rate. Observe that, should $\gamma(\omega)$ be a constant, the black hole emission spectrum would be exactly that of a blackbody. It is the non--triviality of $\gamma(\omega)$, the greybody factor, which leads to deviations of blackbody emission and the consequent greybody radiation.

Early calculations of these greybody factors were done in \cite{p76, u76}. The set--up of the computation is very simple to understand, although extracting exact results may be of some difficulty. Scattering of particles off black holes is described via linearized wave equations, which describe the ``particle perturbation'' to the black hole geometry (see, \textit{e.g.}, \cite{aj00} for a review). In this work we shall consider gravitons. Such linear perturbation theory to black hole geometries was first studied, in a four--dimensional context, in \cite{rw57, z70, z74} and such framework has recently been extended to $d$--dimensions\footnote{In this work it is always the case that $d>3$.}, and for any spherically symmetric black hole (with or without charge and with or without a cosmological constant), in \cite{ik03a, ik03b}. The resulting equation describing gravitational perturbations to black hole geometries, in this spherically symmetric context, can always be written as a one--dimensional Schr\"odinger--like equation, where the one--dimensional coordinate is the so--called tortoise coordinate (describing the spacetime geometry outside the black hole) and where the potential describes both the black hole geometry and which type of perturbation one is addressing (which type of particle, or, in the present case of considering gravitons, which type of gravitational perturbation: tensor, vector, or scalar, as described by their tensor properties on the ${\mathbb{S}}^{d-2}$ sphere \cite{ik03a, ik03b}).

In general the potentials are extremely complicated and an exact solution to the Schr\"odinger problem (with the appropriate boundary conditions for the scattering problem) is out of reach: one always needs to rely on some numerical work, or some approximation scheme, as the ones first set out in \cite{p76, u76}. Some notable exceptions appear in string theory, when studying the scalar--wave absorption cross--section of the non--dilatonic extremal D3--brane, or the extremal D1D5--system, where one can obtain exact solutions in terms of Mathieu functions \cite{gh98, clpt99}.

In terms of the associated Schr\"odinger problem the absorption probability $\gamma(\omega)$ may be written as $\gamma(\omega) = | T(\omega) |^2$, where $T(\omega)$ is the transmission coefficient in the considered spacetime geometry, \textit{i.e.}, the transmission coefficient for the considered potential (we shall be more specific on this precise relation in the following). In this parlance, the greybody factor is the tunneling probability for the barrier described by the given potential. It is the goal of this paper to compute $\gamma(\omega)$ for a wide variety of situations.

In an asymptotically flat spacetime, a quantity which is closely related to the greybody factor is the absorption cross--section, $\sigma(\omega)$, which follows from application of the optical theorem as (see, \textit{e.g.}, \cite{g97})
\begin{equation}
\sigma (\omega) = \gamma(\omega) |\Psi(\omega)|^2.
\end{equation}
\noindent
In here, $\Psi(\omega)$ is the projection of the incoming spherical wave--function into the asymptotic plane--wave. It is important to point out that such a result can only hold in asymptotically flat spacetimes: indeed only in this case can one define incoming and outgoing asymptotic particle states, and thus an S--matrix. When dealing with either asymptotically de Sitter (dS) or asymptotically Anti--de Sitter (AdS) spacetimes, there is no good notion of an S--matrix and thus one cannot define an absorption cross--section. This is also related to the question of what are the good perturbative quantum gravity observables in these spaces: while in asymptotically dS spacetimes this is a subtle question \cite{w01}, in asymptotically AdS spacetimes it is well known that the observables are not associated to an S--matrix but to boundary correlation functions, as shown via the AdS/CFT correspondence \cite{agmoo99}. As such, our focus in this paper precisely lies with greybody factors, whose notion we shall extended to both asymptotically dS and AdS spacetimes.

In this work we shall address the calculation of greybody factors (or absorption probabilities) for static, spherically symmetric black holes in $d$--dimensions. This naturally includes charged black holes and a possible cosmological constant. We consider scalar and gravitational perturbations\footnote{The interest on gravitational perturbations is clear, as gravitational--wave astronomy becomes a reality in the near--future.} (of tensor, vector, and scalar type) as described by the Ishibashi--Kodama (IK) master equations \cite{ik03a, ik03b}, and will compute the full transmission and reflection coefficients of the associated Schr\"odinger problem for all considered spacetime geometries, in certain specific regimes of the frequency, $\omega$. As we have said before, exact solutions are virtually impossible to obtain. Here, we choose to focus on two approximations that have provided for interesting results in the past: scalar field perturbations in the low--energy case (which originated in the work of \cite{u76}) and gravitational perturbations in the asymptotic case (which originated in \cite{n03}).

In the low frequency approximation $\omega \ll T_H$ and $\omega R_H \ll 1$, with $R_H$ the radius of the event horizon, there is a universal result for asymptotically flat black holes \cite{dgm96}: the absorption cross--section for the s--wave of a minimally coupled massless scalar field is given by the area of the event horizon, $\sigma = A_H$. When considering the sub--leading contributions to the scattering (\textit{i.e.}, the higher partial--waves), where the wave angular--momentum component has $\ell > 0$ (and which is the case for gravitational perturbations), the result for the cross--section changes although universality is maintained. In this paper we first give a simple but general derivation of the leading contribution to the absorption cross--section for scalar waves of asymptotically flat black holes which are static and spherically symmetric. This goes beyond the results of \cite{dgm96}, which is only concerned with Schwarzschild black holes (for a general number of spacetime dimensions). As in \cite{dgm96}, our derivation gives the universal result that the cross--section is $\sigma = A_H$. 

We subsequently use our analysis of asymptotically flat black holes in order to study the leading s--wave contribution to the greybody factor for black holes in both asymptotically dS and asymptotically AdS spacetimes. As we shall discuss later, the literature concerning greybody factors for black holes in non--asymptotically flat spacetimes, such as dS or AdS, is rather sparse, and this problem has not been fully considered in the past literature. We shall fill such a gap in the present paper, by devising a general computational method which can be applied for all three types of spacetime asymptotics.

For dS black holes we find the greybody factor for low frequencies $\omega \ll T_H$ and $\omega R_H \ll 1$. This is done in the case of small dS black holes, \textit{i.e.}, with the horizon radius $R_H$ being much smaller than the distance--scale set by the cosmological horizon. This approximation is necessary in order to separate the region near the event horizon from the asymptotic region where we are approximately in dS spacetime. We obtain the non--trivial result
\begin{equation} \label{uni1}
\gamma ( \omega ) = 4 h(\hat{\omega}) \frac{A_H}{A_C} .
\end{equation}
\noindent
Here $A_H$ and $A_C$ are the areas of the event and cosmological horizons, respectively, while $h(\hat{\omega})$ is a non--linear function of $\hat{\omega}$ (which is the frequency measured in units of the scale set by the cosmological constant). The function $h(\hat{\omega})$ has a different expression for even or odd spacetime dimension, but in both cases is a monotonically increasing function of $\hat{\omega}$, within its region of validity, with $h(0)=1$. As we shall see later, the function $h(\hat{\omega})$ generalizes a result for $d$--dimensional Schwarzschild dS black holes previously obtained in \cite{kgb05}. The result for the greybody factor \eqref{uni1} is obtained for scalar waves and is valid for all static and spherically symmetric asymptotically dS black holes.

For AdS black holes, we find the greybody factor in two different regimes. One is the regime of low frequencies and small black holes: $\omega \ll T_H$ and $\omega R_H \ll 1$, and with $R_H$ much smaller than the distance--scale set by the cosmological constant. The other regime is with $\omega \ll T_H$, and with $\omega$ much smaller than the energy--scale set by the cosmological constant. To better explain our results let us introduce here $\hat{\omega}$ as the frequency  measured in units of the scale set by the cosmological constant, and $\hat{A}_H$ as the area of the black hole measured in units of the scale set by the cosmological constant. We find, in both regimes, that for $\hat{\omega}^{d-2} \ll \hat{A}_H$ we have
\begin{equation}
\gamma ( \hat{\omega} ) \propto \frac{\hat{\omega}^{d-2}}{\hat{A}_H}.
\end{equation}
\noindent
This holds in particular for large AdS black holes (with $\hat{A}_H \gg 1$) when $\hat{\omega} \ll 1$. For small AdS black holes (with $\hat{A}_H \ll 1$) we find a rather rich structure for the greybody factor $\gamma(\hat{\omega})$. For $\hat{\omega}^{d-2} \sim \hat{A}_H$ there is a critical frequency such that $\gamma(\hat{\omega}) = 1$, corresponding to pure transmission of  radiation. For $\hat{\omega} = 2n+d-1$, with $n \in \{ 0, 1, 2, \cdots \}$, we find $\gamma(\hat{\omega}) = 0$ corresponding to pure reflection of the radiation. Interestingly enough, these frequencies precisely correspond to the normal frequencies of scalar wave perturbations in the \textit{pure} AdS spacetime \cite{ns04}. Finally, for $(2n+d-1-\hat{\omega})^2 \sim \hat{A}_H$, we find other critical points where $\gamma(\hat{\omega}) = 1$. These results for the greybody factor are obtained for scalar waves and are valid for all static and spherically symmetric asymptotically AdS black holes. The rich structure displayed by the greybody factor for AdS black holes is particularly interesting in view of the AdS/CFT correspondence, which relates AdS black hole phenomena---such as the above---to thermal gauge theory (see, \textit{e.g.}, \cite{agmoo99}). We expect that this structure will also appear in the dual gauge--theory thermal correlation--functions, and it would be rather interesting to further pursue this question in the future.

In the asymptotic limit, where the frequency is very large along the imaginary axis $\omega \to + i \infty$ (more precisely $|R_H \omega | \gg 1$), there are no simple universal results for the greybody factors (but we will return to this point in the following). However, it is interesting to observe that \cite{n03}, for the Schwarzschild black hole, $\gamma(\omega)$ is such that the expectation value for the number of emitted gravitons becomes
\begin{equation}
\langle n(\omega) \rangle = \frac{1}{e^{\frac{\omega}{T_H}} + 3}.
\end{equation}
\noindent
This led \cite{n03} to an interesting conjecture. To understand the conjecture of \cite{n03}, let us first go back to \cite{ms96} (see \cite{k97} as well, for some related work), where it was shown that---for certain five--dimensional black holes---the greybody factors act in such a way that the black hole spectroscopy at asymptotic infinity mimics the excitation spectrum of the microscopic string, \textit{i.e.}, for the asymptotic observer the \textit{greybody} black hole radiation looks like a microscopic \textit{string}, at least at small energies. This means that greybody factors actually carry some information on what concerns the quantum structure of black holes! One may now return to \cite{n03} which, based on the low--frequency results of \cite{ms96}, speculates whether one may likewise be able to infer on the microscopical description of the black hole, at asymptotic frequencies, by analyzing the above greybody factors. If this would be the case, the black hole microscopics at asymptotic frequencies would have to involve new degrees of freedom with rather exotic statistics, at least for the Schwarzschild black hole (further studies along similar lines were later pursued in \cite{ks04}).

What our asymptotic results show is that, for the Reissner--Nordstr\"om (RN) or the asymptotically dS cases (with or without charge), these new microscopic degrees of freedom would have to involve even more exotic statistics than in the Schwarzschild case (see the relevant formulae in the main body of the text). However, for the asymptotically AdS case, and for both neutral and charged AdS black holes, $\gamma(\omega)$ is such that the expectation value for the number of emitted gravitons at asymptotic frequencies becomes
\begin{equation}
\langle n(\omega) \rangle = \frac{1}{e^{\frac{\omega}{T_H}} - 1}.
\end{equation}
\noindent
This is pure blackbody radiation. It would be rather interesting to further study the microscopic dual of AdS black holes, at large imaginary frequencies. It may just be that these are very simple degrees of freedom, as suggested by the result above.

In order to discuss universality of greybody factors in the asymptotic limit one first needs to review some basics of black hole scattering theory. As it turns out, universality in the asymptotic limit is hidden in some transmission and reflection scattering coefficients. As we have said, gravitational perturbations are described by a one--dimensional Schr\"odinger--like equation, with some potential $V(x)$ associated to the background spacetime geometry and to each type of perturbation. Here, $x$ is the tortoise coordinate $dx = \frac{dr}{f(r)}$ with $f(r) = - g_{00}$ the radial function in the metric. Explicit formulae for black hole metrics and for the potentials associated to tensor, vector and scalar type perturbations may be found in the appendices of \cite{ns04}. Now, let $\Phi_\omega$ be the solution of this Schr\"odinger--like equation,
\begin{equation}
- \frac{d^2 \Phi_\omega}{dx^2} + V(x) \Phi_\omega = \omega^2 \Phi_\omega,
\end{equation}
\noindent
with complex frequency $\omega \in \mathbb{C}$, which describes the scattering of an incoming wave originating at $x=+\infty$ (\textit{i.e.}, spatial infinity for asymptotically flat black holes or the cosmological horizon for asymptotically dS black holes). Therefore one has
\begin{eqnarray}
\Phi_\omega &\sim& e^{i \omega x} + R e^{-i\omega x}, \qquad x \to + \infty, \nonumber \\
\Phi_\omega &\sim& T e^{i \omega x}, \qquad x \to - \infty,
\end{eqnarray}
\noindent
where $R(\omega)$ and $T(\omega)$ are the reflection and transmission coefficients, respectively. Notice that $\Phi_{-\omega}$ solves the exact same equation as $\Phi_\omega$, but now satisfies
\begin{eqnarray}
\Phi_{-\omega} &\sim& e^{-i \omega x} + \widetilde{R} e^{i\omega x}, \qquad x \to + \infty, \nonumber \\
\Phi_{-\omega} &\sim& \widetilde{T} e^{-i \omega x}, \qquad x \to - \infty,
\end{eqnarray}
\noindent
for some other reflection and transmission coefficients,
$\widetilde{R}(\omega)$ and $\widetilde{T}(\omega)$. In these
conditions, it is easy to check that the flux
\begin{equation}
J = \frac{1}{2i} \left( \Phi_{-\omega}\frac{d \Phi_\omega}{dx} - \Phi_{\omega}\frac{d
\Phi_{-\omega}}{dx} \right)
\end{equation}
\noindent does not depend on $x$. Evaluating it at both $x \to \pm
\infty$ then yields
\begin{equation}
R \widetilde{R} + T \widetilde{T} = 1.
\end{equation}
\noindent If $\omega \in \mathbb{R}$ then clearly $\Phi_{-\omega}=\Phi_{\omega}^*$, and hence $\widetilde{R}=R^*$ and $\widetilde{T}=T^*$. Consequently we obtain the familiar formula $|R|^2 + |T|^2 = 1$. So far this is all elementary quantum mechanics.

Now let $\Phi'_\omega$ be the solution of the Schr\"odinger--like equation with complex frequency $\omega \in \mathbb{C}$ which describes the scattering of an outgoing wave originating at $x=-\infty$ (\textit{i.e.}, the outer black hole horizon). Then we must have
\begin{eqnarray}
\Phi'_\omega &\sim& T' e^{-i\omega x}, \qquad x \to + \infty, \nonumber \\
\Phi'_\omega &\sim& e^{-i \omega x} + R' e^{i \omega x}, \qquad x \to - \infty,
\end{eqnarray}
\noindent
where $T'(\omega)$ and $R'(\omega)$ are the transmission and reflection coefficients. But since the space of solutions of the Schr\"odinger--like equation has dimension $2$, $\Phi'_\omega$ must be a linear combination of $\Phi_{\omega}$ and $\Phi_{-\omega}$. In fact,
\begin{equation}
\Phi'_\omega = - \frac{\widetilde{R}}{\widetilde{T}} \Phi_\omega + \frac1{\widetilde{T}} \Phi_{-\omega},
\end{equation}
\noindent
and consequently
\begin{eqnarray}
R' &=& - \frac{T}{\widetilde{T}}\widetilde{R}, \nonumber \\
T' &=& T.
\end{eqnarray}
\noindent
Notice that if $\omega$ is real then $|R'|=|R|$, but this does not have to hold for complex $\omega$. However, it is always true that $T'=T$. Finally, notice that $\Phi'_{-\omega}$ still solves the Schr\"odinger--like equation, satisfying
\begin{eqnarray}
\Phi'_{-\omega} &\sim& \widetilde{T}' e^{i \omega x}, \qquad x \to + \infty, \nonumber \\
\Phi'_{-\omega} &\sim& e^{i \omega x} + \widetilde{R}' e^{-i\omega x}, \qquad x \to - \infty,
\end{eqnarray}
\noindent
for yet some other reflection and transmission coefficients $\widetilde{R}'(\omega)$ and $\widetilde{T}'(\omega)$. Again, since the space of solutions of the Schr\"odinger--like equation has dimension $2$, $\Phi'_{-\omega}$ must be a linear combination of $\Phi_{\omega}$ and $\Phi_{-\omega}$, and one may easily check that
\begin{equation}
\Phi'_{-\omega} = \frac1T \Phi_\omega - \frac{R}{T} \Phi_{-\omega}.
\end{equation}
\noindent
Consequently,
\begin{eqnarray}
\widetilde{R}' &=& - \frac{\widetilde{T}}{T} R, \nonumber \\
\widetilde{T}' &=& \widetilde{T}.
\end{eqnarray}
\noindent
Therefore we have both $T\widetilde{T} = T'\widetilde{T}'$ and $R\widetilde{R} = R'\widetilde{R}'$. The greybody factors, for generic complex frequency $\omega \in {\mathbb{C}}$, are naturally defined as $\gamma(\omega) = T(\omega) \widetilde{T}(\omega)$, generalizing the real frequency formula $\gamma(\omega) = | T(\omega) |^2$. In particular, it turns out that the greybody factors for the two scattering problems are precisely the same.

The results above hold for asymptotically flat or asymptotically dS spacetimes. The case of asymptotically AdS spacetimes is a bit more involved. We present here some brief comments, and will return to this point in the main body of the paper. For asymptotically AdS spacetimes, the tortoise coordinate varies from $- \infty$ at the horizon to a fixed constant at spatial infinity, which we shall choose to be zero. For $x \sim 0$ one finds the asymptotic expansion \cite{ns04}
\begin{equation}
\Phi (x) \sim C_+ \sqrt{2\pi\omega x}\ J_{\frac{j_\infty}{2}} \left( \omega x \right) + C_- \sqrt{2\pi\omega x}\ J_{-\frac{j_\infty}{2}} \left( \omega x \right),
\end{equation}
\noindent
where $j_\infty = d-1, d-3, d-5$ for tensor type, vector type and scalar type perturbations, $J_\nu$ is a Bessel function of the first kind, and $C_\pm$ are (complex) integration constants (for most of the conventions in this paper, we refer the reader to \cite{ns04}). This means that if $\re(\omega)>0$ then for $x \ll -1$ we have the asymptotic expansion
\begin{equation}\label{116}
\Phi (x) \sim \left( C_+ e^{i\beta_+} + C_- e^{i\beta_-}\right) e^{i \omega x} + \left( C_+ e^{-i\beta_+} + C_- e^{-i\beta_-}\right) e^{-i \omega x},
\end{equation}
\noindent
where $\beta_\pm = \frac{\pi}4 (1 \pm j_\infty)$ (again, see \cite{ns04} for further details). We can then use this expansion in order to define the transmission and reflection coefficients at infinity, in terms of the coefficients of the Bessel functions. For example, for incoming waves one has
\begin{equation}
\left(
\begin{matrix}
e^{i\beta_+} & e^{i\beta_-} \\
e^{-i\beta_+} & e^{-i\beta_-}
\end{matrix}
\right) \left(
\begin{matrix}
C_+ \\ C_-
\end{matrix}
\right) = \left(
\begin{matrix}
R \\ 1
\end{matrix}
\right).
\end{equation}
\noindent
We will actually only need to use the fact that the solution at infinity has the plane--wave expansion (\ref{116}) for $x \ll -1$.

One may now return to the question concerning universality of greybody factors in the asymptotic limit. As we have advertised earlier, universality in the asymptotic limit is hidden in some of the transmission and reflection scattering coefficients which were defined above. Indeed, what our results show is that, for \textit{all} considered spacetime geometries, it is the case that $\widetilde{T}=1$ and thus $\gamma(\omega) = T(\omega)$. Moreover, for all the asymptotically flat spacetime geometries
\begin{equation}
\widetilde{R} = -2i \cos \left( \frac{\pi j}{2} \right),
\end{equation}
\noindent
where $j$ is a parameter such that $j=0$ ($j=\frac{d-3}{2d-5}$) corresponds to tensor and scalar type gravitational perturbations of uncharged (charged) black holes and $j=2$ ($j=\frac{3d-7}{2d-5}$) corresponds to vector type gravitational perturbations of uncharged (charged) black holes. Still, for all asymptotically flat cases, we find that it is also the case that
\begin{equation}
T = 1 + 2i \cos \left( \frac{\pi j}{2} \right) R,
\end{equation}
\noindent
resulting in the universal relation for the asymptotic greybody factor
\begin{equation}
\gamma(\omega) = T(\omega) \widetilde{T}(\omega) = 1 + 2i \cos \left( \frac{\pi j}{2} \right) R(\omega).
\end{equation}
\noindent
While asymptotically dS spacetimes do not show a great deal of universality, besides the $\widetilde{T}=1$ coefficient, the same does not happen in asymptotically AdS geometries, which show universality in both scattering coefficients and the greybody factor. Indeed, for all the AdS geometries, we find that $\widetilde{T}=1$ and $T=1$, resulting in the universal greybody factor of $\gamma = 1$. Moreover, it is also the case that $\widetilde{R}=0$ and
\begin{equation}
R = 2i \cos \left( \frac{\pi j}{2} \right),
\end{equation}
\noindent
with $j$ as above. It is interesting to observe the similarity with the asymptotically flat case. We leave a deeper understanding of these observed traces of universality for future research on these matters.

There is a vast literature on greybody factors and absorption cross--sections, natural consequences of the study of scattering by black holes (see \cite{aj00} for a review). Let us here briefly review some recent research, with direct relevance for the present work. Black holes in asymptotically flat $d$--dimensional spacetimes, such as the Schwarzschild and the RN solutions, have been rather well understood, at least on what concerns scalar field emission. Starting with the Schwarzschild solution, some recent interest has arisen via so--called brane--world scenarios. In this context, some studies have focused on brane black holes, and greybody factors have been computed: \cite{km02a, km02b} uses the matching--solutions technique of \cite{u76} at low frequencies $R_H \omega \ll 1$ in order to study scalar, spinor and vector particle emission, both brane localized or into the bulk; a topic which was further explored in \cite{hk03, jp06a}. An open issue in the previous papers concerns graviton emission. One would want a full analysis of this situation as well, specially as gravitational--wave astronomy becomes a reality in the near--future. The study of graviton emission from $d$--dimensional Schwarzschild black holes has recently been addressed in a couple of papers \cite{cns05, ccg05a, ccg05b, cekt06, p06, dkss06, jp06b}, again in the $R_H \omega \ll 1$ regime. We should point out that the analysis in \cite{cekt06}, for the low frequency greybody emission of the Schwarzschild black hole, is very close to the one we do in the present paper. On what concerns the RN solution, much less work has been done in the literature. An exception is \cite{jp05}, which studies scalar emission from $d$--dimensional RN black holes, in the usual $R_H \omega \ll 1$ regime. Part of the goal of the present work is to hopefully fill in some of the gaps in the literature, regarding gravitational greybody factors for $d$--dimensional spherically symmetric black holes.

On what concerns black holes in non--asymptotically flat spacetimes, such as dS or AdS black holes, the literature is much sparser. Black holes in asymptotically dS spacetime were first studied in \cite{gh77}, focusing on the Schwarzschild dS black hole. However, it was not until \cite{kgb05} that greybody factors for these black holes were studied, in a fully $d$--dimensional context. This work focused on emission of scalar fields, computing both greybody factors and differential energy--emission rates on a brane and on the bulk. The regime of initial interest was the low--frequency regime of $R_H \omega \ll 1$, but the authors of \cite{kgb05} did not use the standard matching--solutions technique of \cite{u76}. Instead, they chose to focus on the strict $\omega \to 0$ limit, finding that, unlike the simpler Schwarzschild case, the absorption probability of the Schwarzschild dS black hole goes to a constant---and not to zero---as the frequency vanishes. This is also what we find in this paper, as we  extend the calculation beyond the strict $\omega \rightarrow 0$ limit by employing the matching techniques of \cite{u76}, which we have generalized to non--asymptotically flat spacetimes. Furthermore, the authors of \cite{kgb05} claimed that this results in an divergent cross--section as $\omega \to 0$. As we have mentioned before, an S--matrix exists only in flat space, as it requires asymptotic particle states to be defined, and thus one can not talk about cross--section in an asymptotically dS spacetime. Indeed, in \cite{kgb05} the authors make use of the \textit{flat} space optical theorem to define the cross--section in terms of the greybody factor, an expression which clearly can not be applied to non--asymptotically flat spacetimes.

On what concerns black holes in asymptotically AdS spacetimes, these were first studied in \cite{hp83}, focusing on the Schwarzschild AdS black hole, but to date no reasonable greybody calculations for these black holes have been performed in a $d$--dimensional context. The only exception is \cite{hk00}, which however focused on the geometrical optics approximation (\textit{i.e.}, in the very high energy regime, at real frequencies). The present work thus hopes to fill in the gaps in the literature, on what concerns greybody factors for $d$--dimensional black holes in non--asymptotically flat spacetimes.

On the technical side, our computation at low frequencies is very much based on the matching--solutions technique which was first introduced in \cite{u76}. We do present some important improvements on this method, as we extend it to non--asymptotically flat spacetimes, and we believe the description we present---still based on the basic idea of matching solutions far and near the black hole horizon---is one of the simplest approaches in the literature. On what concerns asymptotic frequencies, the technique for computing greybody factors in asymptotically flat spacetimes was first developed in \cite{n03}, very much based on the monodromy methods introduced in \cite{mn03}. Here we generalize those methods also for non--asymptotically flat spacetimes, again very much based on monodromy methods, this time around monodromy techniques which were first introduced in \cite{cns04, ns04}. In both frequency regimes we shall briefly review the simple and well--known Schwarzschild case, for both completeness and pedagogical purposes. A reader who seeks further details on the techniques we use may also consult the aforementioned references.

At very high and real frequencies, the greybody factor must approach the geometrical--optics limit, a result which is independent of the emitted particle's spin. It would certainly be of interest if future work could provide for a full classification of greybody factors---as we do in this paper---also in this geometrical optics regime. This would greatly enhance our knowledge of generic black hole greybody factors in arbitrary dimension\footnote{See \cite{s76} for such a calculation in asymptotically flat spacetimes.}. Another point of interest would be to really test the conjectures in \cite{n03} (which were further refined in \cite{ks04}). Focusing on the more stringy cases, from the full list we provide in this paper, one could envisage actually obtaining new results for the microscopics of specific black holes in the asymptotic frequency regime. Two particularly promising cases are the extremal RN black hole, which has been well studied in the string theoretic framework, and the AdS black hole, which seems to point towards pure blackbody radiation in the large imaginary frequency regime. Still on what concerns the asymptotic frequency regime, we should note that recent work \cite{ss06} has shown that standard quantum mechanical perturbation theory methods allow for a perturbative study of imaginary frequency regimes, coming back from infinity. This is a very interesting calculation and it would certainly be of great interest in future research also to extend it to the greybody calculation we perform in this paper. Another natural extension of all the calculations in this work deals with the consideration of scalar, spinor or vector fields, both massless and massive. Such extensions would provide for a very complete and detailed knowledge of Hawking emission from $d$--dimensional spherically symmetric black holes. One other generalization of our results deals with the extension of the present results to higher--derivative corrected black holes in string theory. Some preliminary steps along these directions have been recently taken in \cite{gbk05, ms06}, and it would certainly be of interest to further proceed along these lines. Finally, one last but still very promising venue of future research, deals with applications of our results for AdS black holes in the
context of the AdS/CFT correspondence \cite{agmoo99}. Indeed, our results shed new light, at both small and large frequencies, on the behavior of thermal correlation functions in the dual gauge theory, and it would be of great interest to provide an explicit calculation of the correlation function dual to the greybody factor we have computed.


\section{Asymptotically Flat Spacetimes} \label{sec:flat}


We begin with the study of asymptotically flat spacetimes, considering both the Schwarzschild and the RN solutions for $d$--dimensional black holes (we refer the reader to the appendices of \cite{ns04} for a complete description of these geometries). The boundary conditions for the scattering process which computes greybody factors in asymptotically flat spacetimes are very simple to understand and are schematically depicted in Figure~\ref{Flat}. Blackbody radiation is produced at the black hole horizon, with part of this radiation traveling all the way to infinity, and the rest being reflected back to the black hole due to the interaction with the non--trivial spacetime geometry outside of the black hole. This non--trivial geometry translates to the potential in the Schr\"odinger--like equation, and these potentials have been described in \cite{ik03b} (again, we refer the reader to the appendices of \cite{ns04} for a complete listing of all these potentials). We plot the potential for scalar field and tensor type gravitational perturbations in the Schwarzschild geometry in Figure~\ref{S6Pot}.

An important point to have in mind concerns the stability of black holes in asymptotically flat spacetimes to tensor, vector and scalar perturbations, as discussed in \cite{ik03b}. For black holes without charge, all types of perturbations are stable in any dimension. Working in generic dimension $d$ we are guaranteed to always have a stable solution. For charged black holes, tensor and vector perturbations are stable in any dimension. Scalar perturbations are stable in four and five dimensions but there is no proof of stability in dimension $d \ge 6$. As we work in generic dimension $d$ we are thus not guaranteed to always have a stable solution. Our results will apply if and only if the spacetime under consideration is stable.

\FIGURE[ht]{\label{Flat}
    \centering
    \psfrag{H+}{$\mathcal{H}^+$}
    \psfrag{H-}{$\mathcal{H}^-$}
    \psfrag{I+}{$\mathcal{I}^+$}
    \psfrag{I-}{$\mathcal{I}^-$}
    \epsfxsize=0.6\textwidth
    \leavevmode
    \epsfbox{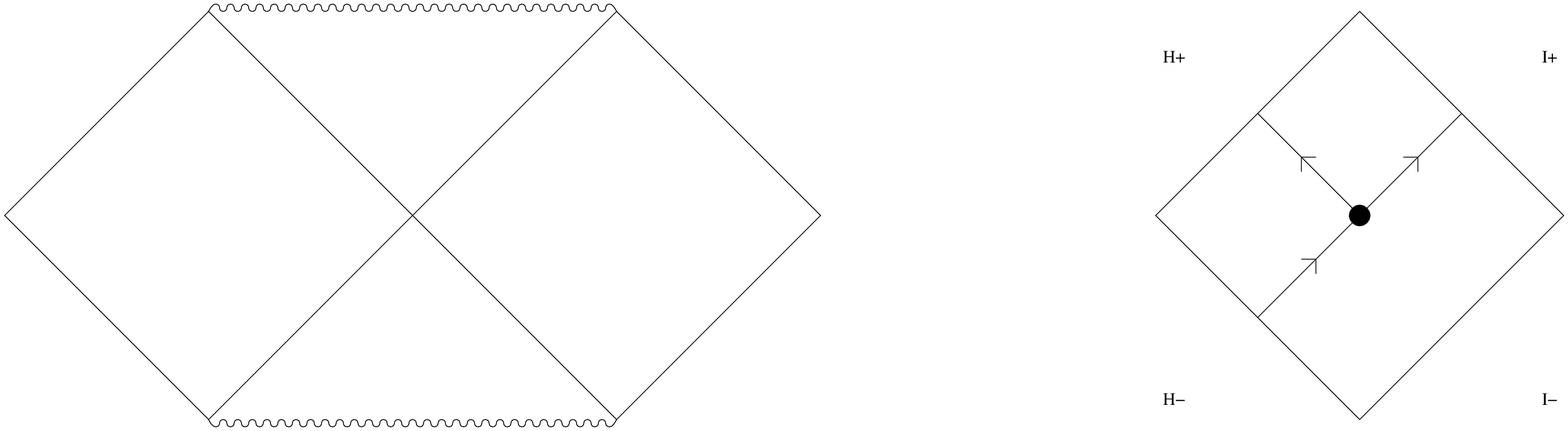}
\caption{Penrose diagram for the Schwarzschild spacetime, along with the schematics of the emission problem in the region covered by the tortoise coordinate. The solid line represents emission from the black hole event horizon. The dot represents the scattering of the emitted wave in the spacetime geometry. The symbol $\mathcal{H}^-$ represents the past black hole event horizon while $\mathcal{H}^+$ represents the future black hole event horizon.}
}

\FIGURE[ht]{\label{S6Pot}
    \centering
    \epsfxsize=0.3\textwidth
    \leavevmode
    \epsfbox{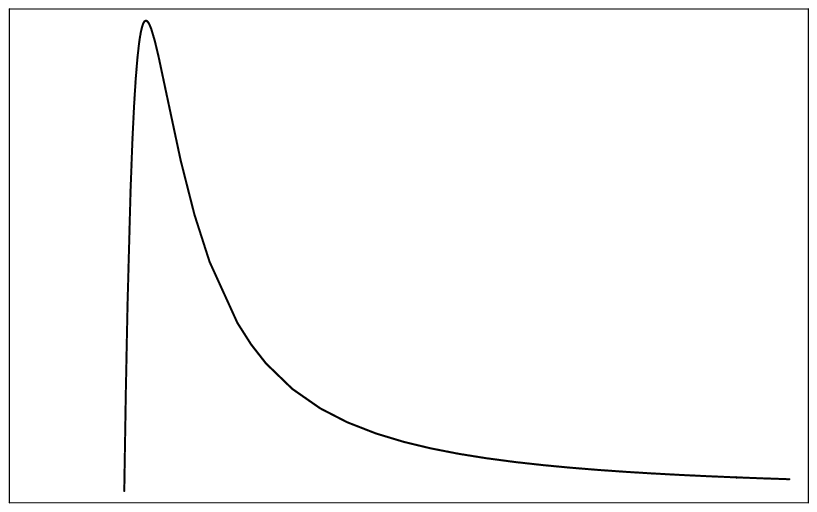}
    $\quad$
    \epsfxsize=0.3\textwidth
    \leavevmode
    \epsfbox{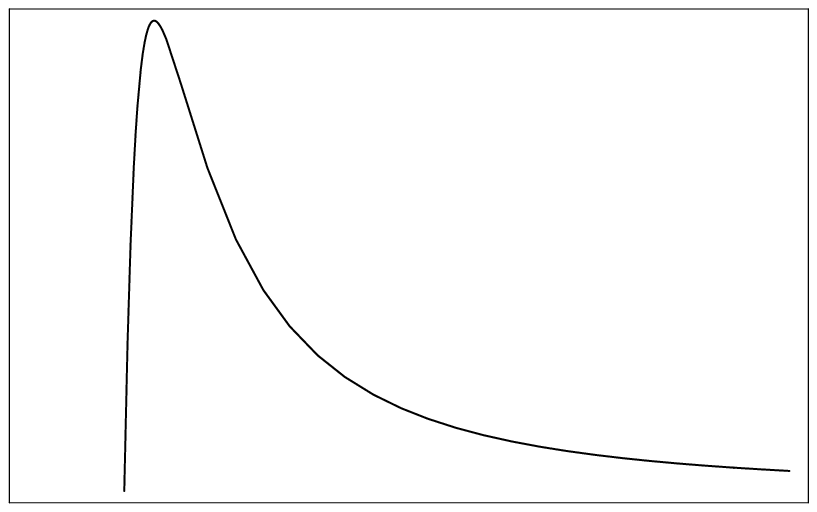}
    $\quad$
    \epsfxsize=0.3\textwidth
    \leavevmode
    \epsfbox{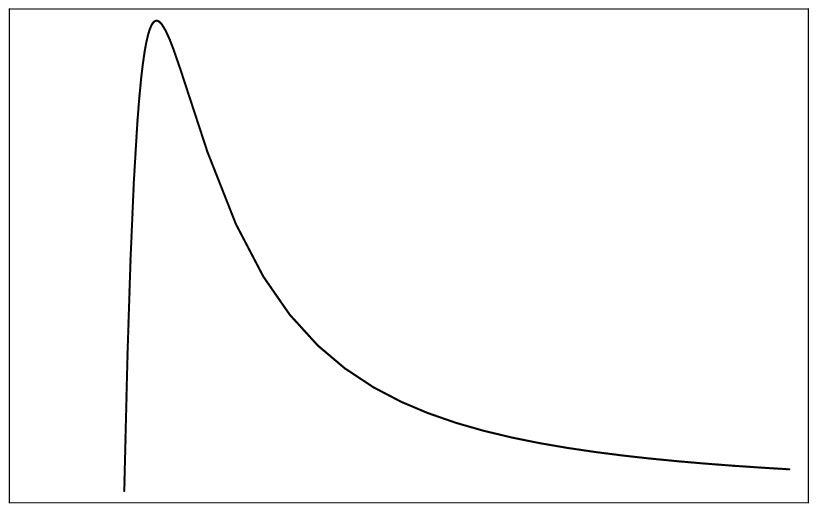}
\caption{Potential for Schwarzschild scalar field and tensor--type perturbations in dimension $d=6$. Plot is in the radial coordinate from the black hole event horizon to asymptotic infinity, with $\ell=0,2,4$, respectively.}
}


\subsection{Greybody Factors at Low Frequency} \label{sec:greysch}


We begin our calculations by considering the greybody factor and the absorption cross--section for the $d$--dimensional Schwarzschild black hole. This is done in a suitably general fashion that can be readily generalized not only to other asymptotically flat black holes, such as the RN solution, but also to asymptotically dS and AdS black holes, as we shall consider below in sections \ref{sec:ds} and \ref{sec:ads}. In the following, we shall consider the derivation of the greybody factor in complete generality, and only when considering the asymptotic region we shall reduce our considerations to the specific asymptotically flat case.

For black holes in asymptotically flat spacetime, the greybody factor or absorption cross--section is well--known for most cases \cite{u76, dgm96, km02a, km02b}. Nonetheless, we shall go through the derivation in the following, in part to set up the subsequent analysis for the non--asymptotically flat cases, but also because our derivation is rather simple and general, treating \textit{all} static and spherically symmetric black holes at the same time, and thus making universality of the greybody factor manifest from scratch. The analysis we present  is mainly based in the methods first presented in \cite{u76} for the four--dimensional Schwarzschild black hole and later discussed in \cite{dgm96}, where higher--dimensional Schwarzschild black holes were considered.

The process we shall study in the following is that of the absorption of a scalar wave by a black hole. Therefore, we study scalar waves in a given black hole spacetime. The scalar wave propagates from infinity throughout spacetime, is partly reflected by the potential barrier of the black hole, and near the horizon the transmitted radiation appears as purely incoming radiation into the black hole. As described in the introduction, the greybody factor for low frequency scattering is identical to the absorption probability of a given black hole, since the scattering and absorption processes are reverse to each other and since, at low energy, we consider real frequencies. The specific physical process which is measured with the greybody factor is the emission of radiation from the black hole, which gets partly reflected by the potential barrier just outside of the black hole horizon, and where the transmitted radiation will appear as purely outgoing radiation in the asymptotic region of spacetime. Thus, finding the absorption cross--section is equivalent to finding the greybody factor, in the case of low frequencies.

We begin by considering a general, static and spherically symmetric, $d$--dimensional black hole metric of the form
\begin{equation} \label{bhmet}
g = - f(r)\, dt \otimes dt + f(r)^{-1}\, dr \otimes dr + r^2 d\Omega_{d-2}^2.
\end{equation}
\noindent
All the black holes considered in this paper have a metric of this form, and due to spherical symmetry $f(r)$ is a function of the radial coordinate $r$ only. We now write
\begin{equation} \label{fdiv}
f(r) = f_a(r) + f_h(r).
\end{equation}
\noindent
Here, $f_a(r)$ is the asymptotic part of $f(r)$. In this paper we shall consider three choices of $f_a(r)$, corresponding to asymptotically flat spacetime, asymptotically dS spacetime and asymptotically AdS spacetime. The function $f_h(r)$ instead contains the physics which is specific to the black hole. As we shall see in the following, the precise form of $f_h(r)$ turns out \textit{not} to be important---meaning, for instance, that charge does not play a role in the low energy greybody factor. One naturally defines the asymptotic region to be the region where $f_a(r) \gg f_h(r)$. The horizon region is instead the region near the black hole where one has $f_a(r) \sim f_h(r)$, \textit{i.e.}, where the two functions are of the same order. We moreover define the horizon radius $R_H$ to be the largest value of $r$, in the horizon region, for which $f(r)=0$.

Near the horizon $r \simeq R_H$ one can write
\begin{equation} \label{genfexp}
f(r) \simeq 2 k_H (r-R_H), \qquad k_H \equiv \frac{1}{2} f'(R_H).
\end{equation}
\noindent
Here $k_H$ is related to the Hawking temperature $T_H$ of the black hole by
\begin{equation}
T_H = \frac{k_H}{2\pi}.
\end{equation}
\noindent
Another physical quantity of the horizon that will be relevant in the following is the area $A_H$ of the event horizon, given by
\begin{equation} \label{defarea}
A_H = \Omega_{d-2} R_H^{d-2} = \frac{2\pi^{\frac{d-1}{2}}}{\Gamma \left(\frac{d-1}{2}\right)}\, R_H^{d-2}.
\end{equation}
\noindent
This formula is quite general as it follows from the general form of the metric \eqref{bhmet}.

In the following we consider the propagation of a scalar wave in the background of the black hole spacetime \eqref{bhmet}. This scalar wave just corresponds to a scalar field of the form
\begin{equation} \label{scwave}
\Phi(t,r,\Omega) = e^{i\omega t} \Phi_{\omega,\ell} (r) Y_{\ell m}(\Omega),
\end{equation}
\noindent
where $\omega$ is the frequency of the wave and the $Y_{\ell m}(\Omega)$ are the spherical harmonics on the $(d-2)$--dimensional sphere. We can write the scalar wave equation in the black hole background \eqref{bhmet} as
\begin{equation} \label{sceq1}
\partial_r ( r^{d-2} f(r)\, \partial_r \Phi_{\omega,\ell} ) + \omega^2\, \frac{r^{d-2}}{f(r)}\, \Phi_{\omega,\ell} - \ell(\ell+d-3)\, r^{d-4} \Phi_{\omega,l} = 0.
\end{equation}
\noindent
If one now defines the tortoise coordinate $x=x(r)$ by
\begin{equation} \label{deftort}
dx = \frac{dr}{f(r)},
\end{equation}
\noindent
one may write the scalar wave equation \eqref{sceq1} in the form of a standard Schr\"odinger--like wave equation
\begin{equation} \label{sceq2}
\left[  \frac{d^2}{dx^2} + \omega^2 - V(r) \right] \left( r^{\frac{d-2}{2}} \Phi_{\omega,\ell} \right) = 0,
\end{equation}
\noindent
with $V(r)$ being the potential, given in terms of $f(r)$ by
\begin{equation} \label{potV}
V(r) = \frac{(d-2)(d-4)}{4} \frac{f(r)^2}{r^2} + \frac{d-2}{2}\, \frac{f(r) \partial_r f(r)}{r} + \ell(\ell+d-3)\, \frac{f(r)}{r^2}.
\end{equation}

Let us now consider the low frequency limit for the scalar wave \eqref{scwave}
\begin{equation} \label{lowfreq}
\omega \ll T_H, \qquad \omega R_H \ll 1.
\end{equation}
\noindent
Notice that the first inequality can also be as written as $\omega \ll k_H$. The fact that we are in the low frequency limit, as defined by \eqref{lowfreq}, enables us to match the behavior of the Schr\"odinger wave--function across broad regions of spacetime with a very high degree of accuracy, since the low frequency limit precisely means that the wave--length of the scalar wave is much larger than any of the characteristic scales associated with the black hole. In order to find the greybody factor, we find it convenient to split up the spacetime in three regions:
\begin{itemize}
\item Region I: The region near the event horizon, defined by $r \simeq R_H$ and $V(r) \ll \omega^2$.
\item Region II: The intermediate region, between regions I and III, \textit{i.e.}, between the horizon region and the asymptotic region. This region is defined by $V(r) \gg \omega^2$.
\item Region III: The asymptotic region. This region is defined by $r \gg R_H$.
\end{itemize}
\noindent
We shall then match the behavior of the wave--function \eqref{scwave} between these regions.

The leading contribution to the greybody factor, in the low frequency limit \eqref{lowfreq}, comes from the $\ell = 0$ mode. Therefore, we shall set $\ell = 0$ in the following computation of the greybody factor. This makes it significantly easier to solve the wave equation in region II. We will write, in the following,
\begin{equation}
\Phi_\omega (r) = \Phi_{\omega, \ell=0} (r),
\end{equation}
\noindent
as a short--hand notation for the radial part of the wave--function \eqref{scwave}.

At this stage there is one important thing to point out: observe that the scalar wave fulfills equivalent equations to those for tensor type perturbations of gravitational waves. However, one cannot directly employ our results below, and those in sections \ref{sec:dslow} and \ref{sec:adslow}, for the scalar--wave absorption probability to the absorption probability for tensor type gravitational perturbations, since s--wave perturbations are not available in this case. One would instead have to consider higher wave--modes with non--trivial angular dependence. This caveat will not exist in the asymptotic case.

\subsubsection*{Region I: The Horizon Region}

We define region I as the region in which $r \simeq R_H$ and $V(r) \ll \omega^2$. In this case, the potential \eqref{potV} near the horizon may be written as
\begin{equation}
V(r) = 2(d-2)\, \frac{k_H^2}{R_H } \left( r-R_H \right).
\end{equation}
\noindent
We can thus re--write the horizon region condition $V(r) \ll \omega^2$ as
\begin{equation} \label{genregI}
\frac{r-R_H}{R_H} \ll \frac{\omega^2}{k_H^2},
\end{equation}
\noindent
thus defining region I. Notice that \eqref{genregI} together with $\omega\ll k_H$ implies that $r-R_H \ll R_H$.

Since we can neglect the potential $V(r)$ as compared to the frequency squared, the scalar equation \eqref{sceq2} in region I reduces to
\begin{equation}
\left[ \frac{d^2}{dx^2} + \omega^2 \right] \left( r^{\frac{d-2}{2}} \Phi_{\omega} \right) = 0.
\end{equation}
\noindent
Clearly, the general solution for a purely incoming wave can thus be written
\begin{equation} \label{genscIa}
\left( \frac{r}{R_H} \right)^{\frac{d-2}{2}} \Phi_{\omega} = \AI\, e^{i \omega x }.
\end{equation}
\noindent
Furthermore, since we have that $r-R_H \ll R_H$, it is evident that one may also write
\begin{equation} \label{genscIb}
\Phi_{\omega} = \AI\, e^{i \omega x}.
\end{equation}
\noindent
To measure the flux near the horizon, associated to the solution above, one only needs to notice that, in terms of the tortoise coordinate $x$, one is effectively considering an one--dimensional Schr\"odinger--like equation with zero potential. Therefore, the flux per unit area is simply
\begin{equation} \label{Jhor}
j_{\rm hor} = \frac{1}{2i} \left( \Phi_{\omega}^*\frac{d\Phi_{\omega}}{dx} - \Phi_{\omega}\frac{d\Phi_{\omega}^*}{dx} \right) = \omega |\AI|^2.
\end{equation}
\noindent
The total flux near the event horizon is therefore
\begin{equation} \label{totalJhor}
J_{\rm hor} = A_H \omega |\AI|^2
\end{equation}
\noindent
where $A_H$ is the area of the horizon given in equation \eqref{defarea}.

From \eqref{genfexp} we now see that in region I we have
\begin{equation} \label{genxlim}
x \simeq \frac{1}{2 k_H} \log \left( \frac{r-R_H}{R_H} \right).
\end{equation}
\noindent
Let us now consider the case of being slightly away from the horizon, in such a manner that
\begin{equation} \label{genregIb}
\frac{r-R_H}{R_H} \gg e^{-\frac{2 k_H}{\omega}}.
\end{equation}
\noindent
This is consistent with being in region I as defined by \eqref{genregI} since combining these two conditions gives
\begin{equation}
e^{-\frac{2k_H}{\omega}} \ll \frac{\omega^2}{k_H^2},
\end{equation}
\noindent
which follows from having $\omega \ll k_H$ and is thereby a consequence of the low frequency approximation \eqref{lowfreq}. Using \eqref{genscIb} we therefore obtain that the scalar wave--function in the region defined by \eqref{genregI} and \eqref{genregIb} becomes
\begin{equation} \label{genscIc}
\Phi_{\omega} = \AI \left[ 1 + i \frac{\omega}{2 k_H} \log \left( \frac{r-R_H}{R_H} \right) \right].
\end{equation}
\noindent
where we have used \eqref{genxlim}. Below, and in the process of computing the greybody factor, we shall match \eqref{genscIc} to a general solution of the scalar wave equation in region II.

\subsubsection*{Region II: The Intermediate Region}

Region II is defined as the region where $V(r) \gg \omega^2$. The scalar wave equation \eqref{sceq1} then reduces to
\begin{equation}
\partial_r ( r^{d-2} f(r)\, \partial_r \Phi_{\omega} ) = 0.
\end{equation}
\noindent
The most general solution to this equation is
\begin{equation} \label{genwfctII}
\Phi_\omega (r) = \AII + \BII\, G(r),
\end{equation}
\noindent
with
\begin{equation}
G(r) = \int^r_{\infty} \frac{dr'}{g(r')}, \qquad g(r) \equiv r^{d-2} f(r).
\end{equation}
\noindent
For $r \simeq R_H$ we get from \eqref{genfexp} that
\begin{equation}
G(r) \simeq \frac{1}{2 R_H^{d-2} k_H} \log \left( \frac{r-R_H}{R_H} \right).
\end{equation}
\noindent
Since this is the part of region II which is closest to region I, we should now match this solution to the wave--function \eqref{genscIc} of region I. Doing this yields the matching
\begin{equation}
\AII = \AI, \qquad \BII = i \omega R_H^{d-2} \AI.
\end{equation}
\noindent
Now, for $r \gg R_H$, we have instead
\begin{equation}
G(r) \simeq \int^r_\infty \frac{dr'}{(r')^{d-2} f_a(r')},
\end{equation}
\noindent
so that in the end one obtains the final expression of
\begin{equation} \label{genphiII}
\Phi_\omega (r) = \AI \left( 1 + i \omega R_H^{d-2} \int^r_\infty \frac{dr'}{(r')^{d-2} f_a(r')} \right),
\end{equation}
\noindent
for $r \gg R_H$, in the region with $\omega^2 \ll V(r)$. This expression for the wave--function, \eqref{genphiII}, is what we shall use in the following in order match to a general solution for the scalar wave equation in the final asymptotic region (\textit{i.e.}, region III). It is important to point out that, up to this stage, we have been completely generic on which type of black hole we are considering (allowing for both charge and a cosmological constant). Furthermore, the analysis in regions I and II is \textit{completely} insensitive to the inclusion of charge. Next, we shall do the matching in the asymptotic region for the case of an asymptotically flat spacetime, while in sections \ref{sec:dslow} and \ref{sec:adslow} we shall do it for the cases of asymptotically dS and AdS spacetimes.

\subsubsection*{Region III: The Asymptotic Region in the Flat Spacetime Case}

The asymptotic region is defined by $r \gg R_H$. Here $f(r) \simeq f_a(r)$ and, for the case of an asymptotically flat black hole spacetime, one simply has
\begin{equation}
f_a (r) =1.
\end{equation}
\noindent
The general solution of the flat space wave equation, for $\ell = 0$, is given by
\begin{equation}\label{genflatsol}
\Phi_{\omega} = \rho^{\frac{3-d}{2}} \left[ C_1 H^{(1)}_{(d-3)/2} (\rho) + C_2 H^{(2)}_{(d-3)/2} (\rho) \right],
\end{equation}
\noindent
with $\rho = r \omega$ and where $H^{(1)}_\nu (\rho) = J_\nu (\rho) + i N_\nu (\rho)$ and $H^{(1)}_\nu (\rho) = J_\nu (\rho) - i N_\nu (\rho)$ are the Hankel functions, here given in terms of the Bessel functions $J_\nu (\rho)$ and $N_\nu (\rho)$. We can match this solution to the wave--function of region II, when $\rho \ll 1$. In this limit we get for the expression above
\begin{equation} \label{match1}
\Phi_{\omega} = \frac{C_1+C_2}{\Gamma \left( \frac{d-1}{2} \right) 2^{\frac{d-3}{2}}} \left[ 1+ \CO(\rho) \right] - i(C_1-C_2) \frac{\Gamma\left( \frac{d-3}{2} \right)2^{\frac{d-3}{2}}}{\pi \rho^{d-3}} \left[ 1+ \CO(\rho) \right].
\end{equation}
\noindent
On the other hand, from the wave--function \eqref{genphiII} in region II, one has
\begin{equation}\label{match2}
\Phi_\omega (r) = \AI \left( 1 - i \omega R_H^{d-2} \frac{1}{(d-3)r^{d-3}} \right).
\end{equation}
\noindent
Matching the region III wave--function \eqref{genflatsol} to the region II wave--function \eqref{match2} we finally get
\begin{equation}\label{flatABIII}
C_1 + C_2 = \Gamma \left( \frac{d-1}{2} \right) 2^{\frac{d-3}{2}} \AI, \qquad C_1-C_2 = \frac{\pi \omega^{d-2} R_H^{d-2} }{(d-3) \Gamma\left( \frac{d-3}{2} \right)2^{\frac{d-3}{2}}} \AI.
\end{equation}
\noindent
In matching \eqref{match1} and \eqref{match2} we are using the fact that $\omega R_H \ll 1$, from \eqref{lowfreq}, since we are just considering matching in the region $R_H \ll r \ll 1/\omega$. This is, furthermore, also why any terms of higher order in $\rho = r\omega$ can be ignored in $N_{(d-3)/2}(\rho)$, when matching \eqref{match1} and \eqref{match2}, since it is simple to observe that it follows from $\omega R_H \ll 1$ that $|C_1-C_2| \ll |C_1+C_2|$ in \eqref{flatABIII}.

The total incoming flux for the general wave--function solution \eqref{genflatsol} is
\begin{equation}\label{flattotalflux}
J_{\rm asy} = r^{d-2} \Omega_{d-2} \frac{1}{2i} \left( \Phi_\omega^* \frac{d\Phi_\omega}{dr} - \Phi_\omega \frac{d\Phi_\omega^*}{dr} \right) = J_{\rm in} - J_{\rm out} ,
\end{equation}
\noindent
where
\begin{equation}\label{flatfluxes}
J_{\rm in} = \frac{2}{\pi} \Omega_{d-2} \omega^{3-d} | C_1 |^2, \qquad J_{\rm out} = \frac{2}{\pi} \Omega_{d-2} \omega^{3-d} | C_2 |^2 ,
\end{equation}
\noindent
are the incoming and outgoing fluxes, respectively. Using now \eqref{flatABIII} we see that
\begin{equation}
J_{\rm asy} = \omega |\AI|^2 \Omega_{d-2} R_H^{d-2} .
\end{equation}
\noindent
Therefore, comparing with \eqref{totalJhor}, we get that
\begin{equation}\label{flatfluxcons}
J_{\rm hor} = J_{\rm asy} = J_{\rm in} - J_{\rm out} .
\end{equation}
\noindent
This expresses the fact that the flux is preserved from the horizon to the asymptotic region.

\subsubsection*{Greybody Factor and Absorption Cross--Section}

The greybody factor $\gamma (\omega)$ is given by $J_{\rm hor} / J_{\rm in}$. Using \eqref{flatfluxes} and \eqref{flatfluxcons} along with \eqref{flatABIII}, we see that
\begin{equation} \label{gammaflatbh}
\gamma (\omega) = \frac{J_{\rm hor}}{J_{\rm in}} = 1 - \frac{|C_2|^2}{|C_1|^2} \simeq 4 \frac{C_1-C_2}{C_1+C_2} = \frac{4\pi \omega^{d-2} R_H^{d-2}}{2^{d-2} [ \Gamma (\frac{d-1}{2} ) ]^2}
\end{equation}
\noindent
This is the greybody factor in the low frequency limit \eqref{lowfreq} for asymptotically flat black holes.

To find the absorption cross--section, we need to project a plane--wave wave--function $e^{i\omega z}$ onto an ingoing spherical s--wave $e^{i\omega r} ( r^{d-2} \Omega_{d-2} )^{-1/2} \Psi$. This gives \cite{dgm96}
\begin{equation}
| \Psi |^2 = \frac{(2\pi)^{d-2}}{\omega^{d-2} \Omega_{d-2}}  .
\end{equation}
\noindent
With this, we can write the absorption cross--section as
\begin{equation}
\sigma (\omega) = \gamma (\omega) | \Psi |^2 .
\end{equation}
\noindent
Therefore, we get the absorption cross--section
\begin{equation} \label{acsflat}
\sigma (\omega) = A_H,
\end{equation}
\noindent
where $A_H$ is the area of the horizon given by \eqref{defarea}. Thus, we see from \eqref{acsflat} that the absorption cross--section, for asymptotically flat black holes and in the low frequency limit \eqref{lowfreq}, is universal, since it is precisely equal to the area of the black hole event horizon.


\subsection{Greybody Factors at Asymptotic Frequency}



\subsubsection{The Schwarzschild Solution}


For the Schwarzschild geometry, the asymptotic greybody factors in arbitrary dimension $d$ were computed in \cite{n03}, using the monodromy matching technique first introduced in \cite{mn03}. While this monodromy technique was originally developed to compute asymptotic quasinormal modes, in different spacetime geometries, it is not a difficult exercise to extend it in order to compute asymptotic greybody factors. In fact, the most significant change between computing asymptotic quasinormal modes and asymptotic greybody factors, using monodromy matching, is the change in boundary conditions. The monodromy technique of \cite{mn03} was later applied in the calculation of Schwarzschild asymptotic quasinormal modes for all types of gravitational perturbations, as classified by the IK master equations \cite{ik03b}, in \cite{ns04}. In the present paper, we shall obtain the Schwarzschild gravitational greybody factors by taking the limit $R_C \to + \infty$ in the Schwarzschild dS solution, where $R_C$ is the dS cosmological horizon. As such, we postpone the details of the monodromy calculation for a couple of sections. Let us also point out that it is not always true that valid spacetime limits translate to valid asymptotic quasinormal mode or asymptotic greybody factor limits (\textit{e.g.}, one cannot use a similar approach to compute asymptotic quasinormal modes or asymptotic greybody factors for the extremal RN black holes---see a full discussion on this issue in \cite{ns04}). The Schwarzschild greybody factors at asymptotic frequency, as originally computed in \cite{n03} and as reproduced later by our own calculations, are given by
\begin{eqnarray}
R &=& \frac{\pm 2i}{e^{\frac{2\pi \omega}{k_H}} + 3}, \nonumber \\
T &=& T' = \frac{e^{\frac{2\pi \omega}{k_H}} - 1}{e^{\frac{2\pi \omega}{k_H}} + 3}, \nonumber \\
R' &=& \frac{\pm 2i \left(e^{\frac{2\pi \omega}{k_H}} - 1\right)}{e^{\frac{2\pi \omega}{k_H}} + 3},
\end{eqnarray}
\noindent
where the plus (minus) sign corresponds to tensor and scalar (vector) type perturbations (we are here assuming  $\re(\omega) > 0$). In here, $k_H$ denotes the surface gravity at the event horizon. One may also re--write the expressions above, making use of the equality
\begin{equation}
\pm 2i = 2i \cos \left( \frac{\pi j}{2} \right),
\end{equation}
\noindent
where $j=0$ corresponds to tensor and scalar type perturbations, and where $j=2$ corresponds to vector type perturbations. As we have said, we will later obtain these transmission/reflection coefficients by taking the limit $R_C \to + \infty$ in the Schwarzschild dS solution. One may further compute $\widetilde{T}=1$ and
\begin{equation}
\widetilde{R} = -2i \cos \left( \frac{\pi j}{2} \right).
\end{equation}
\noindent
These very same coefficients will appear again throughout our calculations. The greybody factor finally follows as
\begin{equation}
\gamma(\omega) = T(\omega) \widetilde{T}(\omega) = \frac{e^{\frac{\omega}{T_H}} - 1}{e^{\frac{\omega}{T_H}} + 3},
\end{equation}
\noindent
whose poles precisely correspond to the asymptotic quasinormal frequencies for this geometry \cite{ns04}.


\subsubsection{The Reissner--Nordstr\"om Solution}


In the case of the RN geometry, asymptotic greybody factors have been computed in dimension $d=4$ in \cite{n03}, again using the monodromy technique introduced in \cite{mn03}. As in the situation with the Schwarzschild geometry, we shall here obtain the fully $d$--dimensional RN asymptotic gravitational greybody factors by taking the limit $R_C \to + \infty$ in the RN dS solution. One could instead start by generalizing the original $d=4$ calculation of \cite{n03}, following a similar procedure to the one in \cite{ns04} for the calculation of asymptotic quasinormal modes. The RN greybody factors at asymptotic frequency are given by
\begin{eqnarray}
R &=& \frac{2i \cos\left(\frac{\pi j}2\right) \left(1 + e^{-\frac{2\pi \omega}{k^-}}\right)}{e^{\frac{2\pi \omega}{k^+}} + (1+2\cos(\pi j)) + (2+2\cos(\pi j)) e^{-\frac{2\pi \omega}{k^-}}}, \nonumber \\
T &=& T' = \frac{e^{\frac{2\pi \omega}{k^+}} - 1}{e^{\frac{2\pi \omega}{k^+}} + (1+2\cos(\pi j)) + (2+2\cos(\pi j)) e^{-\frac{2\pi \omega}{k^-}}}, \nonumber \\
R' &=& \frac{2i \cos\left(\frac{\pi j}2\right) \left(e^{\frac{2\pi \omega}{k^+}} - 1 \right)}{e^{\frac{2\pi \omega}{k^+}} + (1+2\cos(\pi j)) + (2+2\cos(\pi j)) e^{-\frac{2\pi \omega}{k^-}}},
\end{eqnarray}
\noindent
where $j$ satisfies $j=\frac{d-3}{2d-5}$ for tensor and scalar type perturbations, and $j=\frac{3d-7}{2d-5}$ for vector type perturbations (we are assuming $\re(\omega) > 0$). In here, $k^\pm$ are the surface gravities at inner and outer horizon (see, \textit{e.g.}, \cite{ns04} for a complete description of the RN geometry). As we have said, we will later obtain these transmission/reflection coefficients by taking the limit $R_C \to + \infty$ in the RN dS solution. One may further compute $\widetilde{T}=1$ and
\begin{equation}
\widetilde{R} = -2i \cos \left( \frac{\pi j}{2} \right).
\end{equation}
\noindent
These two coefficients are exactly the same as in the Schwarzschild case, only the definition of $j$ changes. We shall find them again in later calculations. The greybody factor finally follows as
\begin{equation}
\gamma(\omega) = T(\omega) \widetilde{T}(\omega) = \frac{e^{\frac{\omega}{T_H^+}} - 1}{e^{\frac{\omega}{T_H^+}} + (1+2\cos(\pi j)) + (2+2\cos(\pi j)) e^{-\frac{\omega}{T_H^-}}},
\end{equation}
\noindent
whose poles precisely correspond to the asymptotic quasinormal frequencies for this geometry \cite{ns04}.

\section{Asymptotically de Sitter Spacetimes} \label{sec:ds}


We shall now proceed with the study of asymptotically dS spacetimes, considering both the Schwarzschild dS and the RN dS solutions for $d$--dimensional black holes (we refer the reader to the appendices of \cite{ns04} for a complete description of these geometries). The quantization of a scalar field in dS space was first addressed in \cite{gh77}. While these authors found that the cosmological event horizon is stable, they also found that there is an isotropic background of thermal radiation. The emitted particles are, however, observer dependent, as is the ``cosmological sphere'' of dS. The boundary conditions for the scattering process which computes greybody factors in asymptotically dS spacetimes are simple to understand and are schematically depicted in Figure~\ref{dS}. Blackbody radiation is produced at the black hole horizon, with part of this radiation traveling all the way to the cosmological horizon, and the rest being reflected back to the black hole due to the interaction with the non--trivial spacetime geometry outside of the black hole. At the same time, blackbody radiation is produced at the cosmological event horizon, with part of this radiation traveling all the way to the black hole horizon, and the rest being reflected back to the cosmological horizon due to the interaction with the non--trivial spacetime geometry. In the following, $T'$ and $R'$ are the scattering coefficients associated to black hole emission, while $T$ and $R$ are the scattering coefficients associated to cosmological horizon emission. Interestingly enough, the greybody factor is the same for the emission from both horizons. The spacetime non--trivial geometry translates to the potential in the Schr\"odinger--like equation, and these potentials have been described in \cite{ik03b} (again, we refer to reader to the appendices of \cite{ns04} for a complete listing of all these potentials). Observe that, due to the linearity of the Schr\"odinger equation describing the scattering process, one may study each of these ``types'' of emission, scattering and absorption, from either black hole horizon or cosmological horizon, in separate. We also plot the potential for both scalar field and tensor type gravitational perturbations in the six--dimensional Schwarzschild dS geometry in Figure~\ref{SdS6Pot}.

An important point to have in mind concerns the stability of black holes in asymptotically dS spacetimes to tensor, vector and scalar perturbations, as discussed in \cite{ik03b}. For black holes without charge, tensor and vector perturbations are stable in any dimension. Scalar perturbations are stable up to dimension six but there is no proof of stability in dimension $d \ge 7$. For charged black holes, tensor and vector perturbations are stable in any dimension. Scalar perturbations are stable in four and five dimensions but there is no proof of stability in dimension $d \ge 6$. As we work in generic dimension $d$ we are not guaranteed to always have a stable solution\footnote{See however \cite{kz07}.}. Our results will apply if and only if the spacetime in consideration is stable.

\FIGURE[ht]{\label{dS}
    \centering
    \psfrag{H+}{$\mathcal{H}^+$}
    \psfrag{H-}{$\mathcal{H}^-$}
    \psfrag{I+}{$\mathcal{H}_C^+$}
    \psfrag{I-}{$\mathcal{H}_C^-$}
    \epsfxsize=0.9\textwidth
    \leavevmode
    \epsfbox{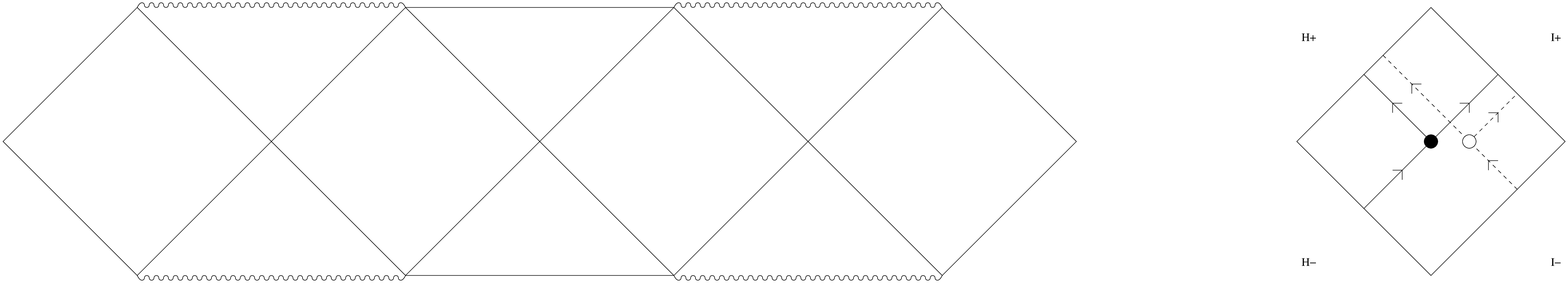}
\caption{Penrose diagram for the Schwarzschild de Sitter spacetime, along with the schematics of the emission problem in the region covered by the tortoise coordinate. The solid line represents emission from the black hole event horizon, while the dashed lined represents emission from the cosmological event horizon. The dots represent the scattering of the emitted waves in the spacetime geometry. The symbol $\mathcal{H}^-$ represents the past black hole horizon, $\mathcal{H}^+$ represents the future black hole horizon, and the symbol $\mathcal{H}_C$ refers to the cosmological horizon.}
}

\FIGURE[ht]{\label{SdS6Pot}
    \centering
    \epsfxsize=0.3\textwidth
    \leavevmode
    \epsfbox{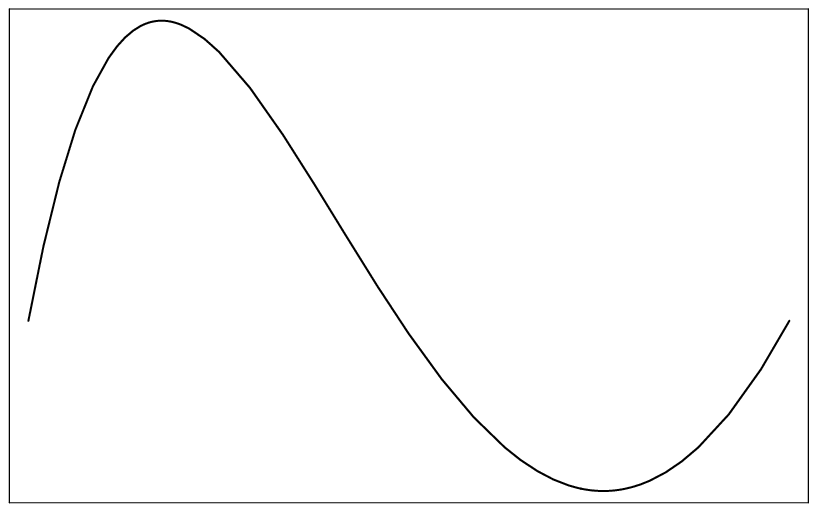}
    $\quad$
    \epsfxsize=0.3\textwidth
    \leavevmode
    \epsfbox{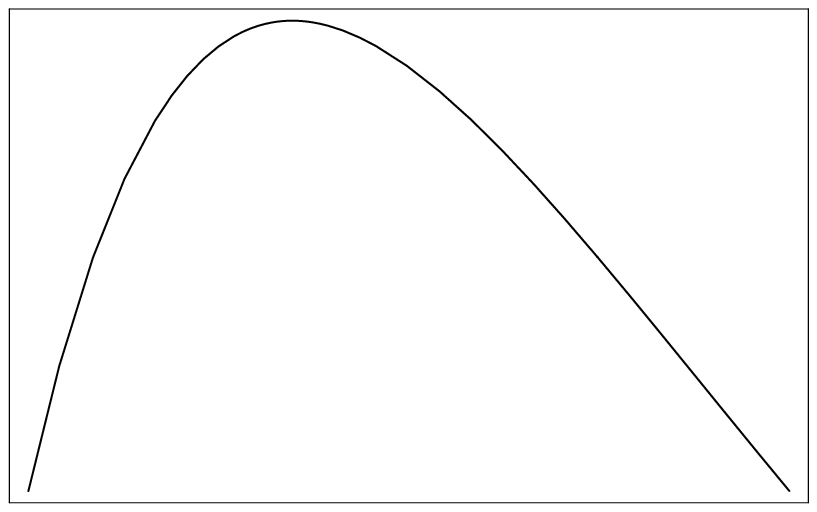}
    $\quad$
    \epsfxsize=0.3\textwidth
    \leavevmode
    \epsfbox{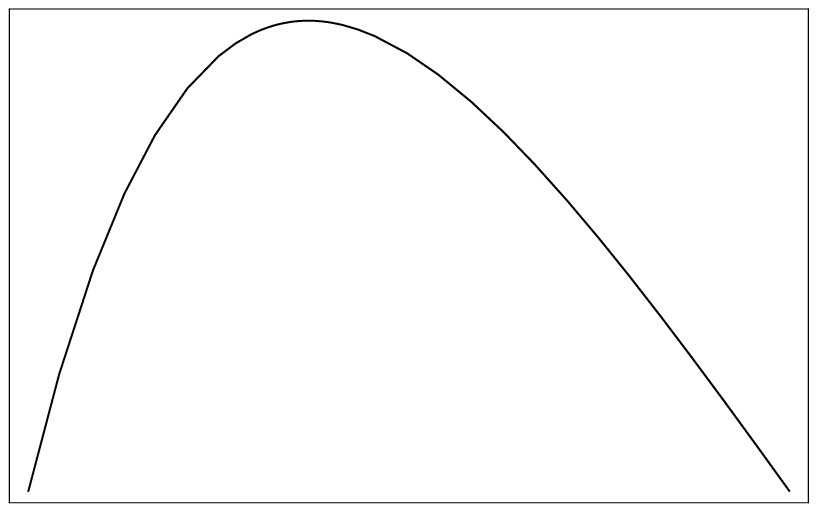}    
\caption{Potential for Schwarzschild de Sitter scalar field and tensor--type perturbations in dimension $d=6$. Plot is in the radial coordinate from the black hole horizon to the cosmological horizon, with $\ell=0,2,4$, respectively.}
}


\subsection{Greybody Factors at Low Frequency} \label{sec:dslow}


In this section we shall find the greybody factor, at low frequencies, for black holes in asymptotically dS spacetimes. We do this in the approximation where we take the cosmological horizon to be far away from the event horizon of the black hole, in order to decouple the behavior of the wave--function in the near horizon region from its behavior in the asymptotic region. This approximation also corresponds to considering small dS black holes, \textit{i.e.}, considering black holes whose size is much smaller than the distance--scale set by the cosmological constant. Greybody factors for Schwarzschild dS black holes have previously been considered in \cite{kgb05}. As we shall further comment below, the methods that we employ are different, but the results that we find match the ones presented in \cite{kgb05} in the strict zero frequency limit.

The class of black hole solutions that we consider have a metric of the form \eqref{bhmet} with the function $f(r)$ of the form \eqref{fdiv}, \textit{i.e.}, we have $f(r) = f_h(r) + f_a(r)$ where, in here, $f_a(r)$ is now given by
\begin{equation} \label{fads}
f_a(r) = 1 - \kappa^2 r^2,
\end{equation}
\noindent
such that setting $f(r)=f_a(r)$ in the metric \eqref{bhmet} corresponds to a dS geometry with the cosmological horizon located at $r=1/\kappa$. As mentioned above, we need to assume that we can separate the near horizon region from the asymptotic region, so that we need to assume
\begin{equation} \label{dsregime}
\kappa R_H \ll 1,
\end{equation}
\noindent
where $r=R_H$ is the location of the event horizon (\textit{i.e.}, we have $f(R_H)=0$). From this, it is clear that we will have that $f_h(r) \ll f_a (r)$ for $r \gg R_H$. This also means that we have the cosmological horizon located at $r=1/\kappa$, since the shift due to the $f_h(r)$ contribution to $f(r)$, for $r \gg R_H$, is negligible.

To compute the leading order greybody factor at low frequencies \eqref{lowfreq}, we consider in the following an $\ell = 0$ scalar wave propagating in the background of an asymptotically dS black hole spacetime, and satisfying \eqref{dsregime}. The wave equation is given by \eqref{sceq2} with $\ell = 0$ and with the potential $V(r)$ given in terms of $f(r)$ by \eqref{potV}. Notice that the tortoise coordinate $x$ is still defined in terms of $f(r)$ by \eqref{deftort}.

The assumption \eqref{dsregime} means that we are able to define an intermediate region, $R_H \ll r \ll 1/\kappa$, in between the near horizon region and the asymptotic region. This region overlaps with region II, earlier defined in section \ref{sec:greysch} as the region where $V(r) \ll \omega^2$. By combining \eqref{genphiII} with \eqref{fads} we learn that, for $r \gg R_H$, $\kappa r \ll 1$ and $r\omega \ll 1$, the wave--function behaves as
\begin{equation} \label{dsmatch}
\Phi_\omega (r) = \AI \left( 1 - i \frac{\omega
R_H^{d-2}}{(d-3)r^{d-3}} \right).
\end{equation}
\noindent
In the following we shall match the wave--function solved in the asymptotic region of the dS geometry, \textit{i.e.}, region III as originally defined in section \ref{sec:greysch}, to the behavior \eqref{dsmatch} of the wave--function in region II. Just like for asymptotically flat spacetimes, this will allow for a direct evaluation of the greybody factors.

\subsubsection*{Scalar Waves in de Sitter Spacetime}

For $r \gg R_H$, \textit{i.e.}, in region III, the tortoise coordinate $x$ as defined by \eqref{deftort} becomes
\begin{equation}
\kappa x = \mbox{arctanh} (\kappa r),
\end{equation}
\noindent
so that $\kappa r = \tanh (\kappa x)$. The potential $V(r)$, defined in \eqref{potV}, is then
\begin{equation}
V(r) = \frac{(d-2)(1-\kappa^2 r^2) (d-4-d\kappa^2r^2)}{4r^2}.
\end{equation}
\noindent
If we now define the coordinate
\begin{equation}
z \equiv \kappa^2 r^2,
\end{equation}
\noindent
we can write the wave equation \eqref{sceq2} as
\begin{equation} \label{dswaveeq}
4z(1-z)^2 \frac{d^2 g}{dz^2} + 2(1-4z+3z^2) \frac{dg}{dz} + \left( \hat{\omega}^2 - \frac{(d-2)(1-z)(d-4-dz)}{4z} \right) g = 0,
\end{equation}
\noindent
where we have defined
\begin{equation}
\hat{\omega} \equiv \frac{\omega}{\kappa}, \qquad g \equiv r^{\frac{d-2}{2}} \Phi_{\omega}.
\end{equation}
\noindent
The general solution to \eqref{dswaveeq} is
\begin{eqnarray} \label{dsgsol}
g &=& C_1\, z^{\frac{d-2}{4}} (1-z)^{-i \frac{1}{2} \hat{\omega}} \hypgeo \left. \left[ -i \frac{\hat{\omega}}{2} , \frac{d-1}{2} -i \frac{\hat{\omega}}{2} ; \frac{d-1}{2} \right| z \right] + \nonumber \\
&+& C_2\, z^{\frac{4-d}{4}} (1-z)^{-i \frac{1}{2} \hat{\omega}} \hypgeo \left. \left[ 1 -i \frac{\hat{\omega}}{2} , - \frac{d-3}{2} -i \frac{\hat{\omega}}{2} ; \frac{5-d}{2} \right| z \right],
\end{eqnarray}
\noindent
where $\hypgeo \left. [ a, b ; c\, \right| z ]$ is the standard hypergeometric function. Alternatively, we may also write the general solution as
\begin{eqnarray} \label{dsgsol2}
g &=& \widetilde{C}_1\, z^{\frac{d-2}{4}} (1-z)^{-i\frac{1}{2} \hat{\omega}} \hypgeo \left. \left[ - i \frac{\hat{\omega}}{2} , \frac{d-1}{2} - i \frac{\hat{\omega}}{2} ; 1- i \hat{\omega} \right| 1-z \right] + \nonumber \\
&+& \widetilde{C}_2\, z^{\frac{d-2}{4}} (1-z)^{i\frac{1}{2} \hat{\omega}} \hypgeo \left. \left[ i \frac{\hat{\omega}}{2} , \frac{d-1}{2} + i \frac{\hat{\omega}}{2} ; 1+ i \hat{\omega} \right| 1-z \right].
\end{eqnarray}
\noindent
Using standard relations for the hypergeometric functions, one can easily see that the relations between the coefficients above are
\begin{equation} \label{Ctildes}
\vecto{\widetilde{C}_1}{\widetilde{C}_2} = \matrto{b_{11}}{b_{12}}{b_{21}}{b_{22}} \vecto{C_1}{C_2},
\end{equation}
\noindent
with
\begin{equation} \label{thebs1}
b_{11} = \frac{\Gamma \left( \frac{d-1}{2} \right) \Gamma \left( i \hat{\omega} \right)}{\Gamma \left( \frac{d-1+i\hat{\omega}}{2} \right)\Gamma \left( \frac{i\hat{\omega}}{2} \right)}, \qquad b_{12} = \frac{\Gamma \left( \frac{5-d}{2} \right) \Gamma \left( i \hat{\omega} \right)}{\Gamma \left( \frac{3-d+i\hat{\omega}}{2} \right)\Gamma \left( 1 + \frac{i\hat{\omega}}{2} \right)},
\end{equation}
\begin{equation} \label{thebs2}
b_{21} = \frac{\Gamma \left( \frac{d-1}{2} \right) \Gamma \left( -i \hat{\omega} \right)}{\Gamma \left( \frac{d-1-i\hat{\omega}}{2} \right)\Gamma \left( -\frac{i\hat{\omega}}{2} \right)}, \qquad b_{22} = \frac{\Gamma \left( \frac{5-d}{2} \right) \Gamma \left( - i \hat{\omega} \right)}{\Gamma \left( \frac{3-d-i\hat{\omega}}{2} \right)\Gamma \left( 1 - \frac{i\hat{\omega}}{2} \right)}.
\end{equation}
\noindent
For use below we further note the important identities
\begin{equation} \label{bs}
b_{21} = b_{11}^*, \qquad b_{22} = b_{12}^*, \qquad b_{11}b_{22} - b_{12}b_{21} = - i \frac{d-3}{2\hat{\omega}}.
\end{equation}

Using \eqref{dsgsol} we see that for $z \rightarrow 0$, or $\kappa r \ll 1$, we have
\begin{equation} \label{dszto0}
\Phi_{\omega} = C_1\, \kappa^{\frac{d-2}{2}} + \frac{C_2\, \kappa^{\frac{4-d}{2}}}{r^{d-3}}.
\end{equation}
\noindent
Using instead \eqref{dsgsol2}, we see that for $z \rightarrow 1$, or $\kappa x \gg 1$, we have
\begin{equation}
\Phi_{\omega} = \widetilde{C}_1\, \kappa^{\frac{d-2}{2}} 2^{-i\hat{\omega}} e^{i\omega x} + \widetilde{C}_2\, \kappa^{\frac{d-2}{2}} 2^{i\hat{\omega}} e^{-i\omega x}.
\end{equation}
\noindent
From these expressions it is simple to obtain that total the flux, for $x \rightarrow \infty$, is given by
\begin{equation} \label{totaldSflux}
J_{\rm asy} = A_C \frac{1}{2i} \left( \Phi_\omega^* \frac{d\Phi_\omega}{dx} - \Phi_\omega \frac{d\Phi_\omega^*}{dx} \right) = J_{\rm in} - J_{\rm out},
\end{equation}
\noindent
where
\begin{equation}\label{dSfluxes}
J_{\rm in} = A_C \omega \kappa^{d-2} |\widetilde{C}_1|^2 , \qquad J_{\rm out} = A_C \omega \kappa^{d-2} |\widetilde{C}_2|^2,
\end{equation}
\noindent
are the incoming and outgoing fluxes, respectively. In order to obtain the total flux we have multiplied with the area of the cosmological horizon $A_C$ above, which is given by
\begin{equation}
A_C = \frac{ \Omega_{d-2}}{\kappa^{d-2}}.
\end{equation}
\noindent
Making further use of \eqref{bs}, it is easily seen that one may write
\begin{equation}
|\widetilde{C}_1|^2 - |\widetilde{C}_2|^2 = (b_{11}b_{22}-b_{12}b_{21} ) (C_1 C_2^* - C_1^* C_2 ) = - i \frac{d-3}{2\hat{\omega}} (C_1 C_2^* - C_1^* C_2 ),
\end{equation}
\noindent
and so, in terms of $C_1$ and $C_2$, the total flux \eqref{totaldSflux} of the wave--function for $x \rightarrow \infty$ is
\begin{equation}\label{dsjasy}
J_{\rm asy} = \frac{d-3}{2i} A_C  \kappa^{d-1} (C_1 C_2^* - C_1^* C_2 ).
\end{equation}

We now proceed to find the coefficients $C_1$ and $C_2$. This we can do by matching the behavior \eqref{dszto0} of the wave--function for $\kappa r \ll 1$ in region III, to the behavior \eqref{dsmatch} for $r\gg R_H$ in region II. A simple calculation yields the result
\begin{equation} \label{dStheC12}
C_1 = \kappa^{\frac{2-d}{2}}\, \AI, \qquad C_2 = - i \kappa^{\frac{d-4}{2}}\, \frac{\omega R_H^{d-2}}{d-3}\, \AI.
\end{equation}
\noindent
Inserting this result in \eqref{dsjasy}, we obtain the following total flux in the asymptotic region,
\begin{equation}
J_{\rm asy} = (\kappa R_H)^{d-2} \omega |\AI|^2 A_C.
\end{equation}
\noindent
Comparing this with \eqref{totalJhor}, we get
\begin{equation}\label{dSfluxcons}
J_{\rm hor} = J_{\rm asy} = J_{\rm in} - J_{\rm out} .
\end{equation}
\noindent
This expresses the fact that the total flux is conserved from the horizon to the asymptotic region $x \rightarrow \infty$ near the cosmological horizon.

\subsubsection*{The Greybody Factor}

The greybody factor $\gamma( \omega )$ is given by $J_{\rm hor}/J_{\rm in}$. Using \eqref{dSfluxes} and \eqref{dSfluxcons} along with \eqref{Ctildes} we get that
\begin{equation}
\gamma (\omega) = \frac{J_{\rm hor}}{J_{\rm in}} = 1 - \frac{|\widetilde{C}_2|^2}{|\widetilde{C}_1|^2} = \left| \frac{b_{21}}{b_{11}} \right|^2  \left| 1 + \frac{b_{11}b_{22}-b_{12}b_{21}}{b_{11}b_{21}} \frac{C_2}{C_1} \right|^2 .
\end{equation}
\noindent
Using now \eqref{bs} and \eqref{dStheC12} we obtain that the greybody factor is given by
\begin{equation} \label{dSgreyfactor}
\gamma (\omega) = 4 h ( \hat{\omega} ) (\kappa R_H )^{d-2} = 4 h ( \hat{\omega} )\, \frac{A_H}{A_C},
\end{equation}
\noindent
where we defined the function $h ( \hat{\omega} )$ by
\begin{equation}
h ( \hat{\omega} ) \equiv \frac{1}{4 |b_{11}|^2}.
\end{equation}
\noindent
For even $d \geq 4$ we have
\begin{equation} \label{hevend}
h ( \hat{\omega} ) = \prod_{n=1}^{\frac{d-2}{2}} \left( 1 + \frac{\hat{\omega}^2}{(2n-1)^2} \right),
\end{equation}
\noindent
while for odd $d \geq 5$ we have instead
\begin{equation} \label{hoddd}
h ( \hat{\omega} ) = \frac{\pi \hat{\omega}}{2} \coth \big( \frac{\pi \hat{\omega}}{2} \big) \prod_{n=1}^{\frac{d-3}{2}} \left( 1 + \frac{\hat{\omega}^2}{(2n)^2} \right).
\end{equation}
\noindent
Equation \eqref{dSgreyfactor}, along with \eqref{hevend} and \eqref{hoddd}, gives the leading contribution to the greybody factor, in the low frequency limit \eqref{lowfreq}, for small asymptotically dS black holes \eqref{dsregime}. Note that $h ( \hat{\omega} ) \rightarrow 1$ as $\hat{\omega} \rightarrow 0$, for both even and odd spacetime dimension. We also see from \eqref{dSgreyfactor} that the greybody factor retains a high degree of universality in that it only depends on $\hat{\omega}$ and $\frac{A_H}{A_C}$, and not on details of the black hole, such as whether it is charged or not. Still, it displays new features as compared to asymptotically flat geometries.

Comparing our result to the results of \cite{kgb05} we find that our final expression for the greybody factor matches the one in \cite{kgb05} in the strict $\omega \to 0$ limit, and in the case of small black holes. This is a nice consistency check on both calculations. However, it is important to notice that the authors of \cite{kgb05} claimed that the absorption cross--section for Schwarzschild dS black holes diverged as $\omega \rightarrow 0$. As we have alluded to before, there is no good notion of absorption cross--section in non--asymptotically flat spacetimes, as one cannot define an S--matrix. As such, there are no divergences of any physical quantities. The divergence found in \cite{kgb05} stems from the fact that the authors defined the dS cross--section using the \textit{flat} space optical theorem, a relation which no longer holds in non--asymptotically flat spacetimes.


\subsection{Greybody Factors at Asymptotic Frequency}



\subsubsection{The Schwarzschild de Sitter Solution}


On what concerns the Schwarzschild dS geometry, asymptotic greybody factors have not been considered in the past literature, and we fill such a gap in the present paper. We shall compute $d$--dimensional asymptotic gravitational greybody--factors for the Schwarzschild dS geometry, using the monodromy--matching technique first developed in \cite{cns04, ns04}. Indeed, and as we have alluded to before, it is not a difficult exercise to extend such monodromy--matching technique from its original quasinormal mode application to the present calculation of asymptotic greybody factors---in some sense, all that is required is an appropriate change in the boundary conditions. This is what we do in the present section, as we shall now explain how to compute the greybody factors at large imaginary frequencies for the Schwarzschild dS black hole. The following calculation heavily relies on \cite{ns04}, where any missing details may be found.

We consider solutions of the Schr\"odinger--like equation in the complex $r$--plane. Near the singularity $r=0$, these solutions behave as
\begin{equation}
\Phi (x) \sim B_+ \sqrt{2\pi\omega x}\ J_{\frac{j}{2}} \left( \omega x \right) + B_- \sqrt{2\pi\omega x}\ J_{-\frac{j}{2}} \left( \omega x \right),
\end{equation}
\noindent
where $x$ is the tortoise coordinate, $J_\nu$ represents a Bessel function of the first kind and $B_\pm$ are (complex) integration constants. The parameter $j$ is left generic for the time being, but will ultimately be set equal to $j=0$ for tensor and scalar type perturbations and equal to $j=2$ for vector type perturbations.

\FIGURE[ht]{\label{StokesSdS}
    \centering
    \psfrag{A}{$A$}
    \psfrag{B}{$B$}
        \psfrag{C}{$C$}
        \psfrag{D}{$D$}
    \psfrag{RH}{$R_H$}
    \psfrag{RC}{$R_C$}
    \psfrag{Re}{$\re$}
    \psfrag{Im}{$\im$}
    \psfrag{Stokes line}{Stokes line}
    \epsfxsize=.6\textwidth
    \leavevmode
    \epsfbox{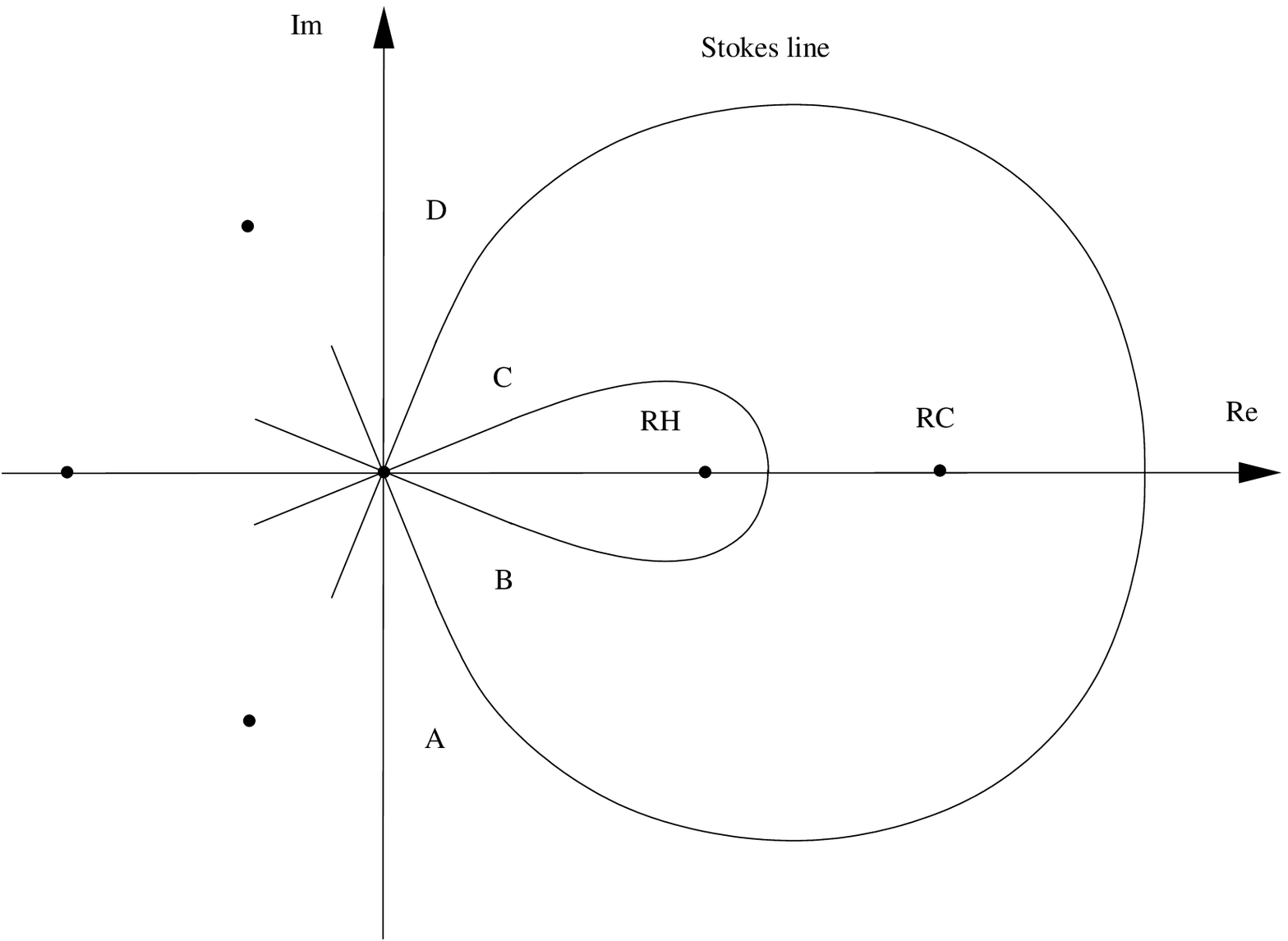}
\caption{Stokes line for the Schwarzschild de Sitter black hole in the case of dimension $d=6$.}
}

Our monodromy calculation must be carried out along the Stokes line $\re(x)=0$, which is sketched in Figure~\ref{StokesSdS}. Starting at point $A$, our solution can be approximated in the limit $\im(\omega) \gg \re(\omega)$ by
\begin{equation}
\Phi (x) \sim \left( B_+ e^{-i\alpha_+} + B_- e^{-i\alpha_-}\right) e^{i \omega x} + \left( B_+ e^{i\alpha_+} + B_- e^{i\alpha_-}\right) e^{-i \omega x},
\end{equation}
\noindent
where $\alpha_\pm = \frac{\pi}4 (1 \pm j)$. The main difference between this calculation and the calculation of the asymptotic quasinormal frequencies in \cite{ns04} is that, unlike the quasinormal modes, our solutions in here will have well defined monodromy only around \textit{one} of the horizons. For this reason we will have to consider the \textit{two} scattering problems corresponding to both incoming or outgoing waves in order to perform the full computation.

Consider the problem of an incoming wave first. In this problem, $\Phi$ has well defined clockwise monodromy $e^{\frac{\pi\omega}{k_H}}$ around the black hole horizon $R_H$, where $k_H$ is the surface gravity of the black hole horizon. As one rotates from point $A$ to point $B$ near the origin, the approximate expression for $\Phi$ changes to
\begin{equation}
\Phi (x) \sim \left( B_+ e^{3i\alpha_+} + B_- e^{3i\alpha_-}\right) e^{i \omega x} + \left( B_+ e^{i\alpha_+} + B_- e^{i\alpha_-}\right) e^{-i \omega x},
\end{equation}
\noindent
and further rotating to point $C$ yields
\begin{equation}
\Phi (x) \sim \left( B_+ e^{3i\alpha_+} + B_- e^{3i\alpha_-}\right) e^{i \omega x} + \left( B_+ e^{5i\alpha_+} + B_- e^{5i\alpha_-}\right) e^{-i \omega x}.
\end{equation}
\noindent
Consider the contour obtained by starting at point $B$, rotating to point $C$ near the origin and returning to point $B$ along the Stokes line. The coefficient of $e^{i\omega x}$ does not change along this contour, and hence this term already has the appropriate monodromy. On the other hand, the monodromy of the term in $e^{-i\omega x}$ will have to match the monodromy of $\Phi$ around $R_H$:
\begin{equation}\label{SdS_one}
\frac{B_+ e^{5i\alpha_+} + B_- e^{5i\alpha_-}}{B_+ e^{i\alpha_+} + B_- e^{i\alpha_-}} e^{-\frac{\pi \omega}{k_H}} = e^{\frac{\pi \omega}{k_H}}.
\end{equation}
\noindent
Since $\re(x)<0$ near $R_H$, we see that for $\im(\omega) \gg \re(\omega)$ the term $e^{i\omega x}$ is exponentially bigger than the term $e^{-i\omega x}$. Since $\Phi(x) \sim T e^{i \omega x}$ near $R_H$, we must have
\begin{equation}
B_+ e^{3i\alpha_+} + B_- e^{3i\alpha_-} = T.
\end{equation}
\noindent
On the other hand, $\re(x)>0$ near the cosmological event horizon $R_C$, and therefore $e^{-i\omega x}$ exponentially dominates $e^{i\omega x}$ in this region. Since $\Phi(x) \sim e^{i \omega x} + Re^{i \omega x}$ near $R_C$, we must also have
\begin{equation}
B_+ e^{i\alpha_+} + B_- e^{i\alpha_-} = R.
\end{equation}
\noindent
Consequently,
\begin{equation}\label{SdS_two}
\frac{B_+ e^{3i\alpha_+} + B_- e^{3i\alpha_-}}{B_+ e^{i\alpha_+} + B_- e^{i\alpha_-}} = \frac{T}{R}.
\end{equation}
\noindent
Seen as a linear system for $(B_+,B_-)$, equations (\ref{SdS_one}) and (\ref{SdS_two}) can only have non--trivial solutions if
\begin{equation}
\left|
\begin{matrix}
e^{5i\alpha_+} - e^{\frac{2\pi \omega}{k_H}} e^{i\alpha_+} & & e^{5i\alpha_-} - e^{\frac{2\pi \omega}{k_H}} e^{i\alpha_-} \\
& & \\
e^{3i\alpha_+} - \frac{T}{R} e^{i\alpha_+} & & e^{3i\alpha_-} - \frac{T}{R} e^{i\alpha_-}
\end{matrix}
\right| = 0,
\end{equation}
\noindent
which yields
\begin{equation}
\frac{T}{R} = \mp \frac{i}2 \left(e^{\frac{2\pi \omega}{k_H}}-1\right),
\end{equation}
\noindent
where the minus (plus) sign corresponds to $j=0$ ($j=2$) and tensor or scalar (vector) type perturbations.

Let us now consider the problem of an outgoing wave. In this problem, $\Phi$ has well defined clockwise monodromy $e^{-\frac{\pi\omega}{k_C}}$ around the cosmological horizon $R_C$, where $k_C$ is the (negative) surface gravity of the cosmological horizon. Again as one starts out at point $A$ the solution $\Phi$ has the approximate expression
\begin{equation}
\Phi (x) \sim \left( B'_+ e^{-i\alpha_+} + B'_- e^{-i\alpha_-}\right) e^{i \omega x} + \left( B'_+ e^{i\alpha_+} + B'_- e^{i\alpha_-}\right) e^{-i \omega x}.
\end{equation}
\noindent
As one rotates from point $A$ to point $B$ near the origin, this changes to
\begin{equation}
\Phi (x) \sim \left( B'_+ e^{3i\alpha_+} + B'_- e^{3i\alpha_-}\right) e^{i \omega x} + \left( B'_+ e^{i\alpha_+} + B'_- e^{i\alpha_-}\right) e^{-i \omega x}.
\end{equation}
\noindent
To compute the monodromy of $\Phi$ around $R_C$ we must follow a contour which encloses only this singularity. Therefore we proceed to point $C$ along the branch of the Stokes line which goes around $R_H$. As we do this, $x$ increases by $\frac{i \pi}{k_H}$, and consequently at point $C$ one has
\begin{eqnarray}
\Phi (x) &\sim& \left( C'_+ e^{-i\alpha_+} + C'_- e^{-i\alpha_-}\right) e^{i \omega \left(x-\frac{i \pi}{k_H} \right)} + \left( C'_+ e^{i\alpha_+} + C'_- e^{i\alpha_-}\right) e^{-i \omega \left(x-\frac{i \pi}{k_H} \right)} \nonumber \\
&=& \left( C'_+ e^{-i\alpha_+} + C'_- e^{-i\alpha_-}\right) e^{\frac{\pi \omega}{k_H}} e^{i \omega x} + \left( C'_+ e^{i\alpha_+} + C'_- e^{i\alpha_-}\right) e^{-\frac{\pi\omega}{k_H}} e^{-i \omega x}.
\end{eqnarray}
\noindent
Further rotating to point $D$ yields
\begin{equation}
\Phi (x) \sim \left( C'_+ e^{3i\alpha_+} + C'_- e^{3i\alpha_-}\right) e^{\frac{\pi \omega}{k_H}} e^{i \omega x} + \left( C'_+ e^{i\alpha_+} + C'_- e^{i\alpha_-}\right) e^{-\frac{\pi\omega}{k_H}} e^{-i \omega x}.
\end{equation}
\noindent
Closing the contour by returning to point $A$ along the Stokes line, we see that the coefficient of $e^{-i\omega x}$ does not change along this contour, and hence this term already has the appropriate monodromy. On the other hand, the monodromy of the term in $e^{i\omega x}$ will have to match the monodromy of $\Phi$ around $R_C$:
\begin{equation}\label{SdS_three}
\frac{C'_+ e^{3i\alpha_+} + C'_- e^{3i\alpha_-}}{B'_+ e^{-i\alpha_+} + B'_- e^{-i\alpha_-}} e^{\frac{\pi \omega}{k_H}} e^{\frac{\pi \omega}{k_C}}= e^{-\frac{\pi \omega}{k_C}}.
\end{equation}
\noindent
Since $\Phi(x) \sim e^{-i \omega x} + R' e^{i \omega x}$ near $R_H$, we must have
\begin{equation}
B'_+ e^{3i\alpha_+} + B'_- e^{3i\alpha_-} = R'.
\end{equation}
\noindent
Since $\Phi(x) \sim T'e^{-i \omega x}$ near $R_C$, we must also have
\begin{equation}
B'_+ e^{i\alpha_+} + B'_- e^{i\alpha_-} = T'.
\end{equation}
\noindent
Consequently,
\begin{equation}\label{SdS_four}
\frac{B'_+ e^{3i\alpha_+} + B'_- e^{3i\alpha_-}}{B'_+ e^{i\alpha_+} + B'_- e^{i\alpha_-}} = \frac{R'}{T'}.
\end{equation}
\noindent
Finally, the approximate expressions for $\Phi$ at points $B$ and $C$ must be matched, yielding
\begin{eqnarray}
B'_+ e^{3i\alpha_+} + B'_- e^{3i\alpha_-} = C'_+ e^{-i\alpha_+} e^{\frac{\pi \omega}{k_H}} + C'_- e^{-i\alpha_-} e^{\frac{\pi \omega}{k_H}}, \\
\label{SdS_five}
B'_+ e^{i\alpha_+} + B'_- e^{i\alpha_-} = C'_+ e^{i\alpha_+} e^{-\frac{\pi\omega}{k_H}} + C'_- e^{i\alpha_-} e^{-\frac{\pi\omega}{k_H}}. \label{SdS_six}
\end{eqnarray}
\noindent
Seen as a linear system for $(B'_+,B'_-,C'_+,C'_-)$, equations (\ref{SdS_three}), (\ref{SdS_four}), (\ref{SdS_five}) and (\ref{SdS_six}) can only have non--trivial solutions if
\begin{equation}
\left|
\begin{matrix}
e^{-i\alpha_+} & & e^{-i\alpha_-} & & e^{3i\alpha_+} e^{\frac{\pi \omega}{k_H}+\frac{2\pi \omega}{k_C}} & & e^{3i\alpha_-} e^{\frac{\pi \omega}{k_H}+\frac{2\pi \omega}{k_C}} \\
e^{3i\alpha_+} - \frac{R'}{T'} e^{i\alpha_+} & & e^{3i\alpha_-} - \frac{R'}{T'} e^{i\alpha_-} & & 0 & & 0 \\
e^{3i\alpha_+} & &  e^{3i\alpha_-} & &  e^{-i\alpha_+} e^{\frac{\pi \omega}{k_H}} & & e^{-i\alpha_-} e^{\frac{\pi \omega}{k_H}} \\
e^{i\alpha_+} & &  e^{i\alpha_-} & &  e^{i\alpha_+} e^{-\frac{\pi \omega}{k_H}} & & e^{i\alpha_-} e^{-\frac{\pi \omega}{k_H}}
\end{matrix}
\right| = 0,
\end{equation}
\noindent
which yields
\begin{equation}
\frac{R'}{T'} = \mp 2i \frac{e^{\frac{2\pi \omega}{k_C}+\frac{2\pi \omega}{k_H}}-1}{e^{\frac{2\pi \omega}{k_C}}-1},
\end{equation}
\noindent
where the minus (plus) sign corresponds to $j=0$ ($j=2$) and tensor or scalar (vector) type perturbations.

To close the system, and end the calculation, we must now consider an incoming wave in the limit $-\im(\omega) \gg \re(\omega)$. In this limit, the solution of the Schr\"odinger--like equation near the origin is approximated by
\begin{equation}
\Phi (x) \sim \left( \widetilde{B}_+ e^{i\alpha_+} + \widetilde{B}_- e^{i\alpha_-}\right) e^{i \omega x} + \left( \widetilde{B}_+ e^{-i\alpha_+} + \widetilde{B}_- e^{-i\alpha_-}\right) e^{-i \omega x}
\end{equation}
\noindent
in the branch of the Stokes line containing point $A$. As one rotates to point $B$ near the origin, the approximate expression for $\Phi$ changes to
\begin{equation}
\Phi (x) \sim \left( \widetilde{B}_+ e^{i\alpha_+} + \widetilde{B}_- e^{i\alpha_-}\right) e^{i \omega x} + \left( \widetilde{B}_+ e^{3i\alpha_+} + \widetilde{B}_- e^{3i\alpha_-}\right) e^{-i \omega x}.
\end{equation}
\noindent
Since $\re(x)<0$ near $R_H$, we see that for $-\im(\omega) \gg \re(\omega)$ the term $e^{-i\omega x}$ is exponentially bigger than the term $e^{i\omega x}$. However, since $\Phi(x) \sim \widetilde{T} e^{i \omega x}$ near $R_H$, we must have
\begin{equation}
\widetilde{B}_+ e^{3i\alpha_+} + \widetilde{B}_- e^{3i\alpha_-} = 0,
\end{equation}
\noindent
and consequently the term in $e^{-i\omega x}$ is not present. We can therefore match the coefficient of the term in $e^{i\omega x}$, yielding
\begin{equation}
B_+ e^{i\alpha_+} + B_- e^{i\alpha_-} = \widetilde{T}.
\end{equation}
\noindent
On the other hand, $\re(x)>0$ near $R_C$, and therefore $e^{i\omega x}$ exponentially dominates $e^{-i\omega x}$ in this region. Since $\Phi(x) \sim e^{i \omega x} + \widetilde{R}e^{i \omega x}$ near $R_C$, we must also have
\begin{equation}
B_+ e^{i\alpha_+} + B_- e^{i\alpha_-} = 1.
\end{equation}
\noindent
Consequently, $\widetilde{T} = 1$. This equation, together with
\begin{eqnarray}
R \widetilde{R} + T \widetilde{T} &=& 1, \nonumber \\
\frac{R'}{T'} &=& - \frac{\widetilde{R}}{\widetilde{T}}, \nonumber \\
T' &=& T,
\end{eqnarray}
\noindent
closes the system, as we now have $6$ equations for the $6$ unknowns $R,T,R',T',\widetilde{R},\widetilde{T}$. These are readily solved to yield
\begin{eqnarray}
R &=& - \frac{\pm 2i e^{-\frac{\pi \omega}{k_H}}\sinh\left( \frac{\pi \omega}{k_C} \right)}{3\cosh\left( \frac{\pi \omega}{k_H} + \frac{\pi \omega}{k_C} \right) + \cosh\left( \frac{\pi \omega}{k_H} - \frac{\pi \omega}{k_C} \right)}, \nonumber \\
T &=& T' = \frac{-2\sinh\left( \frac{\pi \omega}{k_H} \right)\sinh\left( \frac{\pi \omega}{k_C} \right)}{3\cosh\left( \frac{\pi \omega}{k_H} + \frac{\pi \omega}{k_C} \right) + \cosh\left( \frac{\pi \omega}{k_H} - \frac{\pi \omega}{k_C} \right)}, \nonumber \\
R' &=& \frac{\pm 2i \left( e^{\frac{2\pi \omega}{k_H}} - 1 \right)\cosh\left( \frac{\pi \omega}{k_H} + \frac{\pi \omega}{k_C} \right)}{3\cosh\left( \frac{\pi \omega}{k_H} + \frac{\pi \omega}{k_C} \right) + \cosh\left( \frac{\pi \omega}{k_H} - \frac{\pi \omega}{k_C} \right)},
\end{eqnarray}
\noindent
where the plus (minus) sign corresponds to $j=0$ ($j=2$) and tensor or scalar (vector) type perturbations.

Notice that the poles of these coefficients are the frequencies of the asymptotic quasinormal modes (see \cite{ns04} for further details), as it should be. On the other hand, the limit $R_C \to + \infty$, which is to say $k_C \to 0^-$, assuming $\re(\omega) > 0$, yields the Schwarzschild coefficients
\begin{eqnarray}
R &=& \frac{\pm 2i}{e^{\frac{2\pi \omega}{k_H}} + 3}, \nonumber \\
T &=& T' = \frac{e^{\frac{2\pi \omega}{k_H}} - 1}{e^{\frac{2\pi \omega}{k_H}} + 3}, \nonumber \\
R' &=& \frac{\pm 2i \left(e^{\frac{2\pi \omega}{k_H}} - 1\right)}{e^{\frac{2\pi \omega}{k_H}} + 3},
\end{eqnarray}
\noindent
where the plus (minus) sign corresponds to $j=0$ ($j=2$) and tensor or scalar (vector) type perturbations, and as we have advertised for before (these were also obtained in \cite{n03}).

The calculation above changes slightly in the case $d=5$, as explained in \cite{ns04}. The end result is
\begin{eqnarray}
R &=& - \frac{\pm 2i e^{-\frac{\pi \omega}{k_H}}\cosh\left( \frac{\pi \omega}{k_C} \right)}{3\sinh\left( \frac{\pi \omega}{k_H} + \frac{\pi \omega}{k_C} \right) - \sinh\left( \frac{\pi \omega}{k_H} - \frac{\pi \omega}{k_C} \right)}, \nonumber \\
T &=& T' = \frac{-2\sinh\left( \frac{\pi \omega}{k_H} \right)\cosh\left( \frac{\pi \omega}{k_C} \right)}{3\sinh\left( \frac{\pi \omega}{k_H} + \frac{\pi \omega}{k_C} \right) - \sinh\left( \frac{\pi \omega}{k_H} - \frac{\pi \omega}{k_C} \right)}, \nonumber \\
R' &=& \frac{\pm 2i \left( e^{\frac{2\pi \omega}{k_H}} - 1 \right)\sinh\left( \frac{\pi \omega}{k_H} + \frac{\pi \omega}{k_C} \right)}{3\sinh\left( \frac{\pi \omega}{k_H} + \frac{\pi \omega}{k_C} \right) - \sinh\left( \frac{\pi \omega}{k_H} - \frac{\pi \omega}{k_C} \right)},
\end{eqnarray}
\noindent
where the plus (minus) sign corresponds to $j=0$ ($j=2$) and tensor or scalar (vector) type perturbations. Again, notice that the poles of these coefficients are the frequencies of the asymptotic quasinormal modes \cite{ns04}, as it should be. Again the limit $R_C \to + \infty$ yields the Schwarzschild coefficients in $d=5$.


\subsubsection{The Reissner--Nordstr\"om de Sitter Solution}


As in the previous case of the Schwarzschild dS geometry, the RN dS black hole asymptotic greybody factors have not been considered in the past literature, and we fill such a gap in the present paper. We shall compute $d$--dimensional asymptotic gravitational greybody--factors for the RN dS geometry, using the monodromy--matching technique first developed in \cite{ns04}. Again, the main difference with respect to the calculation in \cite{ns04} is an appropriate change in the boundary conditions, from quasinormal to greybody boundary conditions. This is what we do in the present section, as we shall now explain how to compute the greybody factors at large imaginary frequencies for the RN dS black hole. The following calculation relies heavily on \cite{ns04}, where any missing details may be found.

We consider solutions of the Schr\"odinger--like equation in the complex $r$--plane. Near the singularity $r=0$, these solutions behave as
\begin{equation}
\Phi (x) \sim B_+ \sqrt{2\pi\omega x}\ J_{\frac{j}{2}} \left( \omega x \right) + B_- \sqrt{2\pi\omega x}\ J_{-\frac{j}{2}} \left( \omega x \right),
\end{equation}
\noindent
where $x$ is the tortoise coordinate, $J_\nu$ represents a Bessel function of the first kind and $B_\pm$ are (complex) integration constants. The parameter $j$ satisfies $j=\frac{d-3}{2d-5}$ for tensor and scalar type perturbations and $j=\frac{3d-7}{2d-5}$ for vector type perturbations.

\FIGURE[ht]{\label{StokesRNdS}
    \centering
    \psfrag{A}{$A$}
    \psfrag{B}{$B$}
        \psfrag{C}{$C$}
        \psfrag{D}{$D$}
        \psfrag{E}{$E$}
        \psfrag{F}{$F$}
    \psfrag{R+}{$R_+$}
    \psfrag{RC}{$R_C$}
    \psfrag{Re}{$\re$}
    \psfrag{Im}{$\im$}
    \psfrag{Stokes line}{Stokes line}
    \epsfxsize=.6\textwidth
    \leavevmode
    \epsfbox{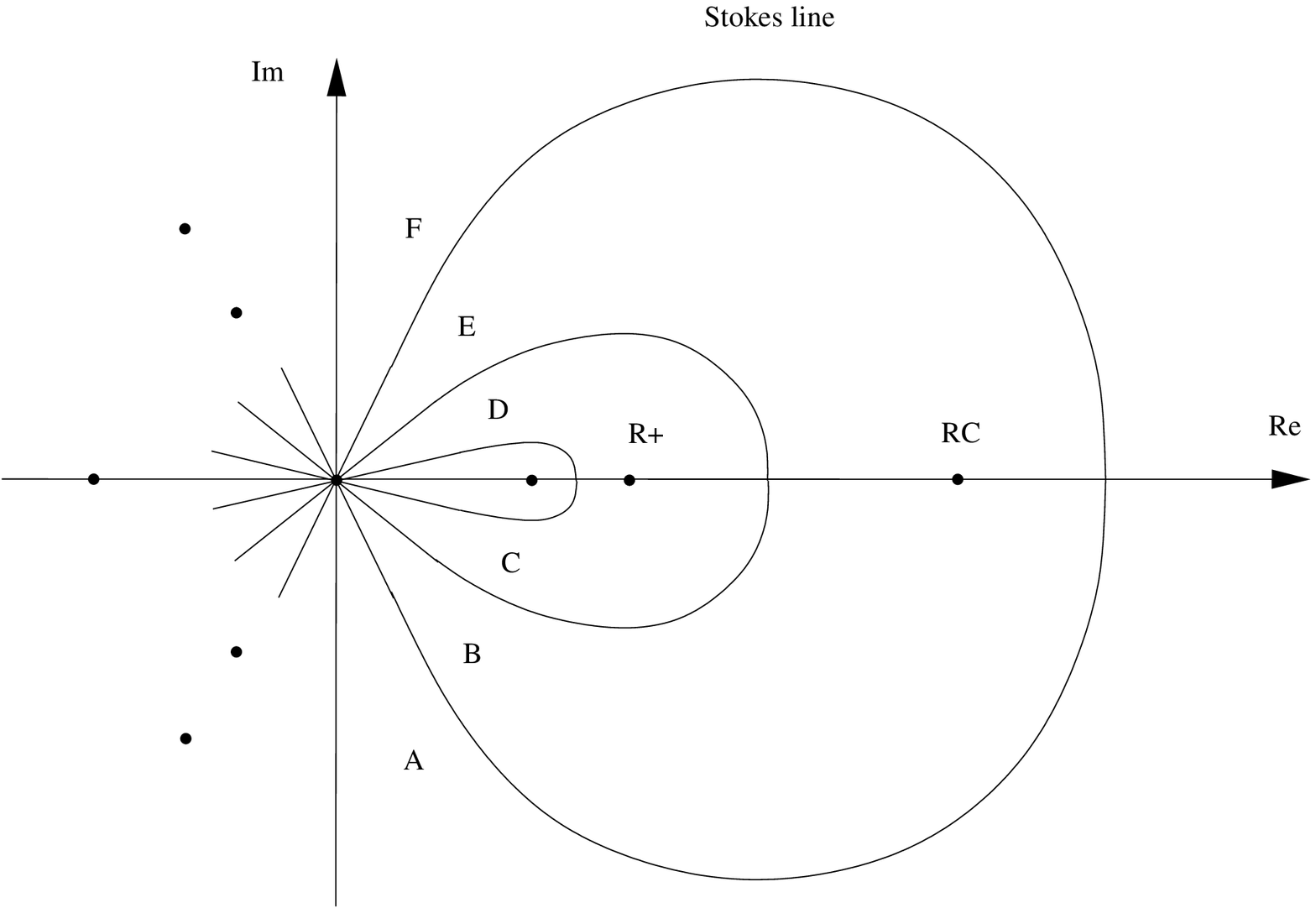}
\caption{Stokes line for the Reissner--Nordstr\"om de Sitter black hole in the case of dimension $d=6$.}
}

Our monodromy calculation must be carried out along the Stokes line $\re(x)=0$, which is sketched in Figure~\ref{StokesRNdS}. Starting at point $A$, our solution can be approximated in the limit $\im(\omega) \gg \re(\omega)$ by
\begin{equation}
\Phi (x) \sim \left( B_+ e^{-i\alpha_+} + B_- e^{-i\alpha_-}\right) e^{i \omega x} + \left( B_+ e^{i\alpha_+} + B_- e^{i\alpha_-}\right) e^{-i \omega x},
\end{equation}
\noindent
where $\alpha_\pm = \frac{\pi}4 (1 \pm j)$. The main difference between this calculation and the calculation of the asymptotic quasinormal frequencies in \cite{ns04} is that, unlike the quasinormal modes, our solutions will here have well defined monodromy only around one of the horizons. For this reason we will have to consider the two scattering problems corresponding to incoming or outgoing waves in order to perform the full computation.

Consider the problem of an incoming wave first. In this problem, $\Phi$ has well defined clockwise monodromy $e^{\frac{\pi\omega}{k^+}}$ around the black hole outer horizon $R_+$, where $k^+$ is the surface gravity of the black hole outer horizon. As one rotates from point $A$ to point $B$ near the origin, the approximate expression for $\Phi$ changes to
\begin{equation}
\Phi (x) \sim \left( B_+ e^{3i\alpha_+} + B_- e^{3i\alpha_-}\right) e^{i \omega x} + \left( B_+ e^{i\alpha_+} + B_- e^{i\alpha_-}\right) e^{-i \omega x},
\end{equation}
\noindent
and further rotating to point $C$ yields
\begin{equation}
\Phi (x) \sim \left( B_+ e^{3i\alpha_+} + B_- e^{3i\alpha_-}\right) e^{i \omega x} + \left( B_+ e^{5i\alpha_+} + B_- e^{5i\alpha_-}\right) e^{-i \omega x}.
\end{equation}
\noindent
To compute the monodromy of $\Phi$ around the black hole outer horizon $R_+$ we must follow a contour which encloses only this singularity. Therefore we start at point $B$, rotate to point $C$ near the origin and proceed to point $D$ along the branch of the Stokes line which goes around the inner horizon $R_-$. As we do this, $x$ increases by $\frac{i \pi}{k^-}$, where $k^-$ is the (negative) surface gravity of the black hole inner horizon, and consequently at point $C$ one has
\begin{eqnarray}
\Phi (x) &\sim& \left( C_+ e^{i\alpha_+} + C_- e^{i\alpha_-}\right) e^{i \omega \left(x-\frac{i \pi}{k^-} \right)} + \left( C_+ e^{-i\alpha_+} + C_- e^{-i\alpha_-}\right) e^{-i \omega \left(x-\frac{i \pi}{k^-} \right)} \nonumber \\
&=& \left( C_+ e^{i\alpha_+} + C_- e^{i\alpha_-}\right) e^{\frac{\pi \omega}{k^-}} e^{i \omega x} + \left( C_+ e^{-i\alpha_+} + C_- e^{-i\alpha_-}\right) e^{-\frac{\pi\omega}{k^-}} e^{-i \omega x}.
\end{eqnarray}
\noindent
Further rotating to point $E$ yields
\begin{equation}
\Phi (x) \sim \left( C_+ e^{i\alpha_+} + C_- e^{i\alpha_-}\right) e^{\frac{\pi \omega}{k^-}} e^{i \omega x} + \left( C_+ e^{-i\alpha_+} + C_- e^{-i\alpha_-}\right) e^{-\frac{\pi\omega}{k^-}} e^{-i \omega x}.
\end{equation}
\noindent
Closing the contour by returning to point $B$ along the Stokes line, we see that the coefficient of $e^{i\omega x}$ does not change along this contour, and hence this term already has the appropriate monodromy. On the other hand, the monodromy of the term in $e^{-i\omega x}$ will have to match the monodromy of $\Phi$ around $R_+$:
\begin{equation}\label{RNdS_three}
\frac{C_+ e^{3i\alpha_+} + C_- e^{3i\alpha_-}}{B_+ e^{i\alpha_+} + B_- e^{i\alpha_-}} e^{-\frac{\pi \omega}{k^-}} e^{-\frac{\pi \omega}{k^+}}= e^{\frac{\pi \omega}{k^+}}.
\end{equation}
\noindent
Since $\Phi(x) \sim e^{i \omega x} + R e^{-i \omega x}$ near $R_C$, we must have
\begin{equation}
B_+ e^{i\alpha_+} + B_- e^{i\alpha_-} = R.
\end{equation}
\noindent
Since $\Phi(x) \sim T e^{i \omega x}$ near $R_+$, we must also have
\begin{equation}
B_+ e^{3i\alpha_+} + B_- e^{3i\alpha_-} = T.
\end{equation}
\noindent
Consequently,
\begin{equation}\label{RNdS_four}
\frac{B_+ e^{i\alpha_+} + B_- e^{i\alpha_-}}{B_+ e^{3i\alpha_+} + B_- e^{3i\alpha_-}} = \frac{R}{T}.
\end{equation}
\noindent
Finally, the approximate expressions for $\Phi$ at points $B$ and $C$ must be matched, yielding
\begin{eqnarray}
B_+ e^{3i\alpha_+} + B_- e^{3i\alpha_-} &=& C_+ e^{i\alpha_+} e^{\frac{\pi \omega}{k^-}} + C_- e^{i\alpha_-} e^{\frac{\pi \omega}{k^-}}, \label{RNdS_five} \\
B_+ e^{5i\alpha_+} + B_- e^{5i\alpha_-} &=& C_+ e^{-i\alpha_+} e^{-\frac{\pi\omega}{k^-}} + C_- e^{-i\alpha_-} e^{-\frac{\pi\omega}{k^-}}. \label{RNdS_six}
\end{eqnarray}
\noindent
Seen as a linear system for $(B_+,B_-,C_+,C_-)$, equations (\ref{RNdS_three}), (\ref{RNdS_four}), (\ref{RNdS_five}) and (\ref{RNdS_six}) can only have non--trivial solutions if
\begin{equation}
\left|
\begin{matrix}
e^{i\alpha_+} & & e^{i\alpha_-} & & e^{3i\alpha_+} e^{-\frac{\pi \omega}{k^-}-\frac{2\pi \omega}{k^+}} & & e^{3i\alpha_-} e^{-\frac{\pi \omega}{k^-}-\frac{2\pi \omega}{k^+}} \\
e^{i\alpha_+} - \frac{R}{T} e^{3i\alpha_+} & & e^{i\alpha_-} - \frac{R}{T} e^{3i\alpha_-} & & 0 & & 0 \\
e^{3i\alpha_+} & &  e^{3i\alpha_-} & &  e^{i\alpha_+} e^{\frac{\pi \omega}{k^-}} & & e^{i\alpha_-} e^{\frac{\pi \omega}{k^-}} \\
e^{5i\alpha_+} & &  e^{5i\alpha_-} & &  e^{-i\alpha_+} e^{-\frac{\pi \omega}{k^-}} & & e^{-i\alpha_-} e^{-\frac{\pi \omega}{k^-}}
\end{matrix}
\right| = 0,
\end{equation}
\noindent
which yields
\begin{equation}
\frac{R}{T} = \frac{ 2i \cos\left(\frac{\pi j}2\right)\left(1 + e^{\frac{-2\pi \omega}{k^-}}\right)}{e^{\frac{2\pi \omega}{k^+}}-1}.
\end{equation}

Let us now consider the problem of an outgoing wave. In this problem, $\Phi$ has well defined clockwise monodromy $e^{-\frac{\pi\omega}{k_C}}$ around the cosmological horizon $R_C$, where $k_C$ is the (negative) surface gravity of the cosmological horizon. Again as one starts out at point $A$ the solution $\Phi$ has the approximate expression
\begin{equation}
\Phi (x) \sim \left( B'_+ e^{-i\alpha_+} + B'_- e^{-i\alpha_-}\right) e^{i \omega x} + \left( B'_+ e^{i\alpha_+} + B'_- e^{i\alpha_-}\right) e^{-i \omega x}.
\end{equation}
\noindent
As one rotates from point $A$ to point $B$ near the origin, this changes to
\begin{equation}
\Phi (x) \sim \left( B'_+ e^{3i\alpha_+} + B'_- e^{3i\alpha_-}\right) e^{i \omega x} + \left( B'_+ e^{i\alpha_+} + B'_- e^{i\alpha_-}\right) e^{-i \omega x}.
\end{equation}
\noindent
To compute the monodromy of $\Phi$ around $R_C$ we must follow a contour which encloses only this singularity. Therefore we proceed to point $E$ along the branch of the Stokes line which goes around $R_+$. As we do this, $x$ increases by $\frac{i \pi}{k^+}+\frac{i \pi}{k^-}$, and consequently at point $E$ one has
\begin{eqnarray}
\Phi (x) &\sim& \left( C'_+ e^{-i\alpha_+} + C'_- e^{-i\alpha_-}\right) e^{i \omega \left(x-\frac{i \pi}{k^+} - \frac{i \pi}{k^-} \right)} + \left( C'_+ e^{i\alpha_+} + C'_- e^{i\alpha_-}\right) e^{-i \omega \left(x-\frac{i \pi}{k^+} - \frac{i \pi}{k^-}\right)} \nonumber \\
&=& \left( C'_+ e^{-i\alpha_+} + C'_- e^{-i\alpha_-}\right) e^{\frac{\pi \omega}{k^+}} e^{\frac{\pi \omega}{k^-}} e^{i \omega x} + \left( C'_+ e^{i\alpha_+} + C'_- e^{i\alpha_-}\right) e^{-\frac{\pi\omega}{k^+}} e^{-\frac{\pi \omega}{k^-}} e^{-i \omega x}.
\end{eqnarray}
\noindent
Further rotating to point $F$ yields
\begin{equation}
\Phi (x) \sim \left( C'_+ e^{3i\alpha_+} + C'_- e^{3i\alpha_-}\right) e^{\frac{\pi \omega}{k^+}} e^{\frac{\pi \omega}{k^-}} e^{i \omega x} + \left( C'_+ e^{i\alpha_+} + C'_- e^{i\alpha_-}\right) e^{-\frac{\pi\omega}{k^+}} e^{-\frac{\pi\omega}{k^-}} e^{-i \omega x}.
\end{equation}
\noindent
Closing the contour by returning to point $A$ along the Stokes line, we see that the coefficient of $e^{-i\omega x}$ does not change along this contour, and hence this term already has the appropriate monodromy. On the other hand, the monodromy of the term in $e^{i\omega x}$ will have to match the monodromy of $\Phi$ around $R_C$:
\begin{equation}\label{RNdS_three_again}
\frac{C'_+ e^{3i\alpha_+} + C'_- e^{3i\alpha_-}}{B'_+ e^{-i\alpha_+} + B'_- e^{-i\alpha_-}} e^{\frac{\pi \omega}{k^+}} e^{\frac{\pi \omega}{k^-}} e^{\frac{\pi \omega}{k_C}}= e^{-\frac{\pi \omega}{k_C}}.
\end{equation}
\noindent
Since $\Phi(x) \sim e^{-i \omega x} + R' e^{i \omega x}$ near $R_H$, we must have
\begin{equation}
B'_+ e^{3i\alpha_+} + B'_- e^{3i\alpha_-} = R'.
\end{equation}
\noindent
Since $\Phi(x) \sim T'e^{-i \omega x}$ near $R_C$, we must also have
\begin{equation}
B'_+ e^{i\alpha_+} + B'_- e^{i\alpha_-} = T'.
\end{equation}
\noindent
Consequently,
\begin{equation}
\frac{B'_+ e^{3i\alpha_+} + B'_- e^{3i\alpha_-}}{B'_+ e^{i\alpha_+} + B'_- e^{i\alpha_-}} = \frac{R'}{T'}. \label{RNdS_four_again}
\end{equation}
\noindent
Finally, the approximate expressions for $\Phi$ at points $B$ and $C$ must be matched, yielding
\begin{eqnarray}
B'_+ e^{3i\alpha_+} + B'_- e^{3i\alpha_-} &=& C'_+ e^{-i\alpha_+} e^{\frac{\pi \omega}{k^+}} e^{\frac{\pi \omega}{k^-}} + C'_- e^{-i\alpha_-} e^{\frac{\pi \omega}{k^+}} e^{\frac{\pi \omega}{k^-}}, \label{RNdS_five_again} \\
B'_+ e^{i\alpha_+} + B'_- e^{i\alpha_-} &=& C'_+ e^{i\alpha_+} e^{-\frac{\pi\omega}{k^+}} e^{-\frac{\pi\omega}{k^-}} + C'_- e^{i\alpha_-} e^{-\frac{\pi\omega}{k^+}} e^{-\frac{\pi\omega}{k^-}}. \label{RNdS_six_again}
\end{eqnarray}
\noindent
Seen as a linear system for $(B'_+,B'_-,C'_+,C'_-)$, equations (\ref{RNdS_three_again}), (\ref{RNdS_four_again}), (\ref{RNdS_five_again}) and (\ref{RNdS_six_again}) can only have non--trivial solutions if
\begin{equation}
\left|
\begin{matrix}
e^{-i\alpha_+} & & e^{-i\alpha_-} & & e^{3i\alpha_+} e^{\frac{\pi \omega}{k^+}+\frac{\pi \omega}{k^-}+\frac{2\pi \omega}{k_C}} & & e^{3i\alpha_-} e^{\frac{\pi \omega}{k^+}+\frac{\pi \omega}{k^-}+\frac{2\pi \omega}{k_C}} \\
e^{3i\alpha_+} - \frac{R'}{T'} e^{i\alpha_+} & & e^{3i\alpha_-} - \frac{R'}{T'} e^{i\alpha_-} & & 0 & & 0 \\
e^{3i\alpha_+} & &  e^{3i\alpha_-} & &  e^{-i\alpha_+} e^{\frac{\pi \omega}{k^+}+\frac{\pi \omega}{k^-}} & & e^{-i\alpha_-} e^{\frac{\pi \omega}{k^+}+\frac{\pi \omega}{k^-}} \\
e^{i\alpha_+} & &  e^{i\alpha_-} & &  e^{i\alpha_+} e^{-\frac{\pi \omega}{k^+}-\frac{\pi \omega}{k^-}} & & e^{i\alpha_-} e^{-\frac{\pi \omega}{k^+}-\frac{\pi \omega}{k^-}}
\end{matrix}
\right| = 0,
\end{equation}
\noindent
which yields
\begin{equation}
\frac{R'}{T'} = \frac{ -2i \cos\left(\frac{\pi j}2\right)\left(e^{\frac{2\pi \omega}{k^-} + \frac{2\pi \omega}{k^+} + \frac{2\pi \omega}{k_C}} + 1 \right)}{e^{\frac{2\pi \omega}{k_C}}-1}.
\end{equation}

To close the system, and end the calculation, we must now consider an incoming wave in the limit $-\im(\omega) \gg \re(\omega)$. In this limit, the solution of the Schr\"odinger--like equation near the origin is approximated by
\begin{equation}
\Phi (x) \sim \left( \widetilde{B}_+ e^{i\alpha_+} + \widetilde{B}_- e^{i\alpha_-}\right) e^{i \omega x} + \left( \widetilde{B}_+ e^{-i\alpha_+} + \widetilde{B}_- e^{-i\alpha_-}\right) e^{-i \omega x}
\end{equation}
\noindent
in the branch of the Stokes line containing point $A$. As one rotates to point $B$ near the origin, the approximate expression for $\Phi$ changes to
\begin{equation}
\Phi (x) \sim \left( \widetilde{B}_+ e^{i\alpha_+} + \widetilde{B}_- e^{i\alpha_-}\right) e^{i \omega x} + \left( \widetilde{B}_+ e^{3i\alpha_+} + \widetilde{B}_- e^{3i\alpha_-}\right) e^{-i \omega x}.
\end{equation}
\noindent
Since $\re(x)<0$ near $R_+$, we see that for $-\im(\omega) \gg \re(\omega)$ the term $e^{-i\omega x}$ is exponentially bigger than the term $e^{i\omega x}$. However, since $\Phi(x) \sim \widetilde{T} e^{i \omega x}$ near $R_H$, we must have
\begin{equation}
\widetilde{B}_+ e^{3i\alpha_+} + \widetilde{B}_- e^{3i\alpha_-} = 0,
\end{equation}
\noindent
and consequently the term in $e^{-i\omega x}$ is not present. We can therefore match the coefficient of the term in $e^{i\omega x}$, yielding
\begin{equation}
B_+ e^{i\alpha_+} + B_- e^{i\alpha_-} = \widetilde{T}.
\end{equation}
\noindent
On the other hand, $\re(x)>0$ near $R_C$, and therefore $e^{i\omega x}$ exponentially dominates $e^{-i\omega x}$ in this region. Since $\Phi(x) \sim e^{i \omega x} + \widetilde{R}e^{i \omega x}$ near $R_C$, we must also have
\begin{equation}
B_+ e^{i\alpha_+} + B_- e^{i\alpha_-} = 1.
\end{equation}
\noindent
Consequently, $\widetilde{T} = 1$. This equation, together with
\begin{eqnarray}
R \widetilde{R} + T \widetilde{T} &=& 1, \nonumber \\
\frac{R'}{T'} &=& - \frac{\widetilde{R}}{\widetilde{T}}, \nonumber \\
T' &=& T,
\end{eqnarray}
\noindent
closes the system, as we now have $6$ equations for the $6$ unknowns $R,T,R',T',\widetilde{R},\widetilde{T}$. These are readily solved to yield
\begin{eqnarray}
R &=& \frac{-2i \cos\left(\frac{\pi j}2\right) \left(1 + e^{-\frac{2\pi \omega}{k^-}}\right) e^{-\frac{\pi \omega}{k^+}}\sinh\left( \frac{\pi \omega}{k_C} \right)}{\cosh\left( \frac{\pi\omega}{k^+}-\frac{\pi\omega}{k_C}\right) + (1+2\cos(\pi j)) \cosh\left( \frac{\pi\omega}{k^+}+\frac{\pi\omega}{k_C}\right) + (2+2\cos(\pi j)) \cosh\left( \frac{2\pi\omega}{k^-}+\frac{\pi\omega}{k^+}+\frac{\pi\omega}{k_C}\right)}, \nonumber \\
T &=& T' = \frac{-2\sinh\left( \frac{\pi \omega}{k^+} \right)\sinh\left( \frac{\pi \omega}{k_C} \right)}{\cosh\left( \frac{\pi\omega}{k^+}-\frac{\pi\omega}{k_C}\right) + (1+2\cos(\pi j)) \cosh\left( \frac{\pi\omega}{k^+}+\frac{\pi\omega}{k_C}\right) + (2+2\cos(\pi j)) \cosh\left( \frac{2\pi\omega}{k^-}+\frac{\pi\omega}{k^+}+\frac{\pi\omega}{k_C}\right)}, \nonumber \\
R' &=& \frac{2i \cos\left(\frac{\pi j}2\right) e^{\frac{\pi \omega}{k^-}} \left( e^{\frac{2\pi \omega}{k^+}} - 1 \right)\cosh\left( \frac{\pi \omega}{k^-} + \frac{\pi \omega}{k^+} + \frac{\pi \omega}{k_C} \right)}{\cosh\left( \frac{\pi\omega}{k^+}-\frac{\pi\omega}{k_C}\right) + (1+2\cos(\pi j)) \cosh\left( \frac{\pi\omega}{k^+}+\frac{\pi\omega}{k_C}\right) + (2+2\cos(\pi j)) \cosh\left( \frac{2\pi\omega}{k^-}+\frac{\pi\omega}{k^+}+\frac{\pi\omega}{k_C}\right)}.
\end{eqnarray}

Notice that the poles of these coefficients are the frequencies of the asymptotic quasinormal modes (see \cite{ns04} for further details), as it should be. On the other hand, the limit $R_C \to + \infty$, which is to say $k_C \to 0^-$, assuming $\re(\omega) > 0$, yields the RN coefficients
\begin{eqnarray}
R &=& \frac{2i \cos\left(\frac{\pi j}2\right) \left(1 + e^{-\frac{2\pi \omega}{k^-}}\right)}{e^{\frac{2\pi \omega}{k^+}} + (1+2\cos(\pi j)) + (2+2\cos(\pi j)) e^{-\frac{2\pi \omega}{k^-}}}, \nonumber \\
T &=& T' = \frac{e^{\frac{2\pi \omega}{k^+}} - 1}{e^{\frac{2\pi \omega}{k^+}} + (1+2\cos(\pi j)) + (2+2\cos(\pi j)) e^{-\frac{2\pi \omega}{k^-}}}, \nonumber \\
R' &=& \frac{2i \cos\left(\frac{\pi j}2\right) \left(e^{\frac{2\pi \omega}{k^+}} - 1 \right)}{e^{\frac{2\pi \omega}{k^+}} + (1+2\cos(\pi j)) + (2+2\cos(\pi j)) e^{-\frac{2\pi \omega}{k^-}}},
\end{eqnarray}
\noindent
which can also be obtained by an easy generalization of the calculation for $d=4$ in \cite{n03}.

The calculation above changes slightly in the case $d=5$, as explained in \cite{ns04}. The end result is
\begin{eqnarray}
R &=& \frac{-2i \cos\left(\frac{\pi j}2\right) \left(1 + e^{-\frac{2\pi \omega}{k^-}}\right) e^{-\frac{\pi \omega}{k^+}}\cosh\left( \frac{\pi \omega}{k_C} \right)}{\sinh\left( \frac{\pi\omega}{k_C}-\frac{\pi\omega}{k^+}\right) + (1+2\cos(\pi j)) \sinh\left( \frac{\pi\omega}{k^+}+\frac{\pi\omega}{k_C}\right) + (2+2\cos(\pi j)) \sinh\left( \frac{2\pi\omega}{k^-}+\frac{\pi\omega}{k^+}+\frac{\pi\omega}{k_C}\right)}, \nonumber \\
T &=& T' = \frac{-2\sinh\left( \frac{\pi \omega}{k^+} \right)\cosh\left( \frac{\pi \omega}{k_C} \right)}{\sinh\left( \frac{\pi\omega}{k_C}-\frac{\pi\omega}{k^+}\right) + (1+2\cos(\pi j)) \sinh\left( \frac{\pi\omega}{k^+}+\frac{\pi\omega}{k_C}\right) + (2+2\cos(\pi j)) \sinh\left( \frac{2\pi\omega}{k^-}+\frac{\pi\omega}{k^+}+\frac{\pi\omega}{k_C}\right)}, \nonumber \\
R' &=& \frac{2i \cos\left(\frac{\pi j}2\right) e^{\frac{\pi \omega}{k^-}} \left( e^{\frac{2\pi \omega}{k^+}} - 1 \right)\cosh\left( \frac{\pi \omega}{k^-} + \frac{\pi \omega}{k^+} + \frac{\pi \omega}{k_C} \right)}{\sinh\left( \frac{\pi\omega}{k_C}-\frac{\pi\omega}{k^+}\right) + (1+2\cos(\pi j)) \sinh\left( \frac{\pi\omega}{k^+}+\frac{\pi\omega}{k_C}\right) + (2+2\cos(\pi j)) \sinh\left( \frac{2\pi\omega}{k^-}+\frac{\pi\omega}{k^+}+\frac{\pi\omega}{k_C}\right)},
\end{eqnarray}
\noindent
with $j=\frac25$ for tensor and scalar type perturbations and $j=2-\frac25$ for vector type perturbations. Notice that the poles of these coefficients are the frequencies of the asymptotic quasinormal modes \cite{ns04}, as it should be. Again the limit $R_C \to + \infty$ yields the RN coefficients in $d=5$.


\section{Asymptotically Anti--de Sitter Spacetimes} \label{sec:ads}


This final section is dedicated to the study of asymptotically AdS spacetimes, considering both the Schwarzschild AdS and the RN AdS solutions for $d$--dimensional black holes (we refer the reader to the appendices of \cite{ns04} for a complete description of these geometries). The quantization of a scalar field in AdS was first addressed in \cite{ais78}, where considerable attention was given to the question of what are the AdS boundary conditions. In fact, in AdS, light rays can reach spatial infinity and return to the origin in finite time, as measured by the observer at the origin (crossing AdS within half the natural period). As it turns out, the only sensible boundary condition  to impose on quasinormal modes is the usual incoming waves at the black hole event horizon and the new requirement of vanishing of the wave--function at infinity. These boundary conditions were explored in \cite{ns04} to compute asymptotic quasinormal modes. The boundary conditions for the scattering process which computes greybody factors in asymptotically AdS spacetimes are a bit more subtle than in the previous cases (asymptotically flat and asymptotically dS) and are schematically depicted in Figure~\ref{AdS}. Black holes in AdS are in thermal equilibrium with their environment; the radiation which is produced at the black hole horizon is all re--absorbed. This is clear from Figure~\ref{AdS}, where blackbody radiation is produced at the black hole horizon, with part of this radiation traveling all the way to spatial infinity, and the rest being reflected back to the black hole due to the interaction with the non--trivial spacetime geometry outside of the black hole. But in AdS, the radiation which reaches spatial infinity is reflected back, with part of this radiation traveling all the way through to the black hole horizon, and the rest being reflected back to spatial infinity due to the interaction with the non--trivial spacetime geometry, and so on \textit{ad infinitum}. This is the physical picture which ensures thermal equilibrium. In the following, $T'$ and $R'$ are the scattering coefficients associated to black hole emission, while $T$ and $R$ are the scattering coefficients associated to ``emission'' of the reflected wave at spatial infinity. Interestingly enough, the greybody factor is the same regardless of which process one considers. The background non--trivial geometry translates to the potential in the Schr\"odinger--like equation, and these potentials have been described in \cite{ik03b} (as usual, we refer the reader to the appendices of \cite{ns04} for a complete listing of all these potentials). Observe that, due to the linearity of the Schr\"odinger equation describing the scattering process, one may study each of the infinite series of reflections/interactions in separate. We shall explore such linear properties in the following. We also plot the potential for both scalar field and tensor type gravitational perturbations in the six--dimensional Schwarzschild AdS geometry in Figure~\ref{SAdS6Pot}.

An important point to have in mind concerns the stability of black holes in asymptotically AdS spacetimes to tensor, vector and scalar perturbations, as discussed in \cite{ik03b}. For black holes without charge, tensor and vector perturbations are stable in any dimension. Scalar perturbations are stable in dimension four but there is no proof of stability in dimension $d \ge 5$. For charged black holes, tensor and vector perturbations are stable in any dimension. Scalar perturbations are stable in four dimensions but there is no proof of stability in dimension $d \ge 5$. As we work in generic dimension $d$ we are thus not guaranteed to always have a stable solution. Our results will apply if and only if the spacetime in consideration is stable.

\FIGURE[ht]{\label{AdS}
    \centering
    \psfrag{H+}{$\mathcal{H}^+$}
    \psfrag{H-}{$\mathcal{H}^-$}
    \psfrag{I}{$\mathcal{I}$}
    \epsfxsize=0.6\textwidth
    \leavevmode
    \epsfbox{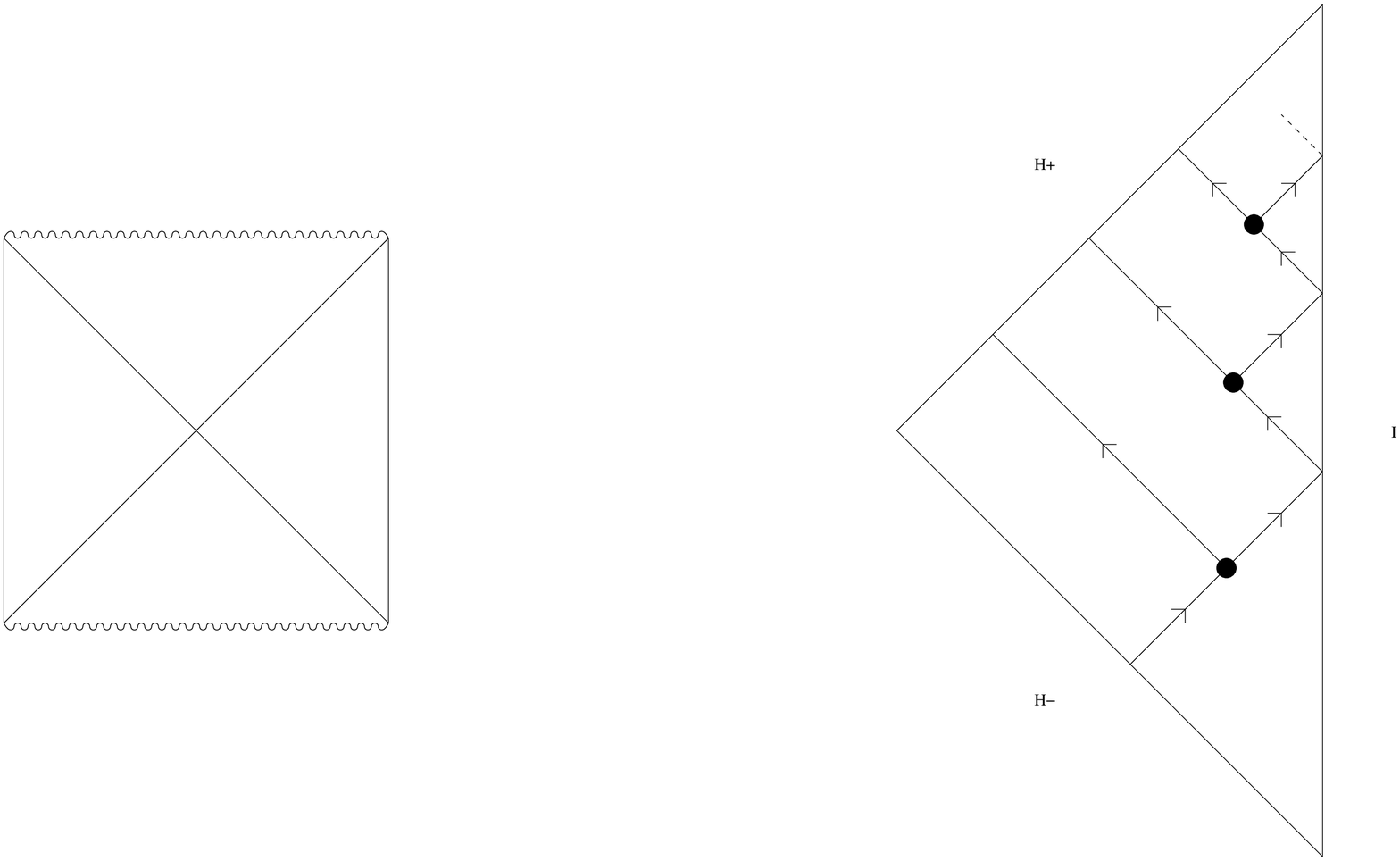}
\caption{Penrose diagram for the Schwarzschild Anti--de Sitter spacetime, along with the schematics of the emission problem in the region covered by the tortoise coordinate. The solid line represents emission from the black hole event horizon, while the dots represent the scattering of waves in the spacetime geometry. After each reflection at the spacetime boundary, there is a new interaction of the emitted wave with the spacetime geometry.}
}

\FIGURE[ht]{\label{SAdS6Pot}
    \centering
    \epsfxsize=0.3\textwidth
    \leavevmode
    \epsfbox{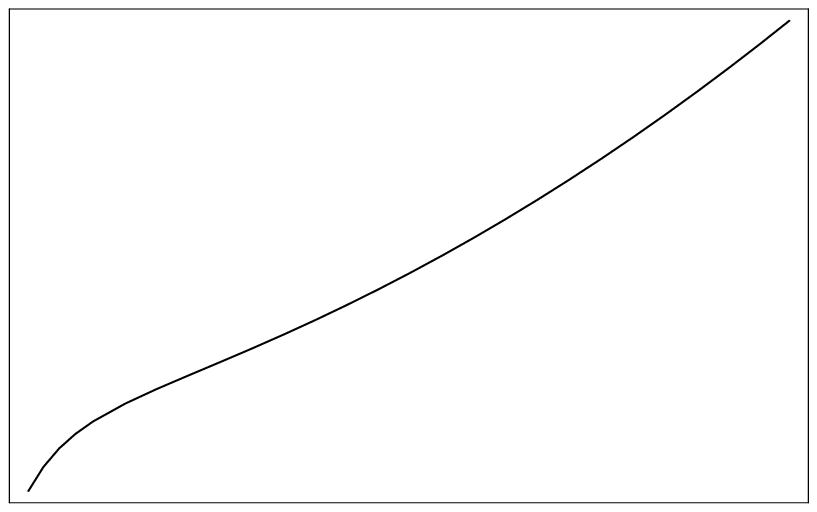}
    $\quad$
    \epsfxsize=0.3\textwidth
    \leavevmode
    \epsfbox{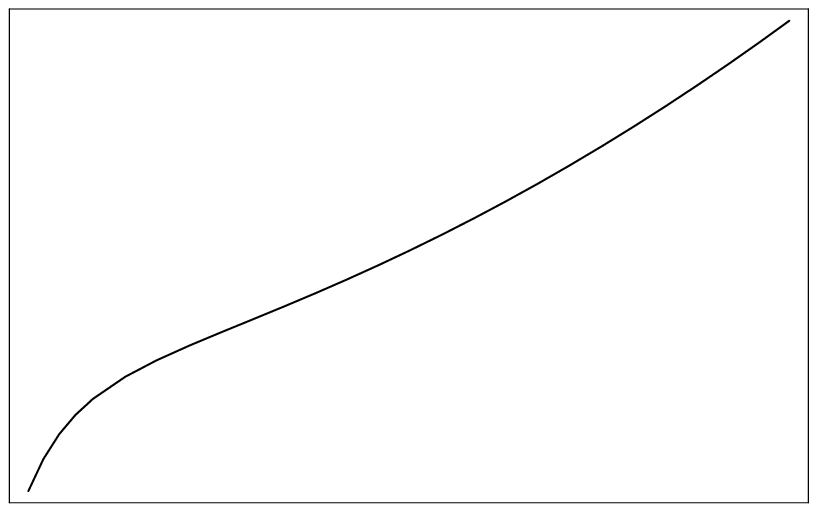}
    $\quad$
    \epsfxsize=0.3\textwidth
    \leavevmode
    \epsfbox{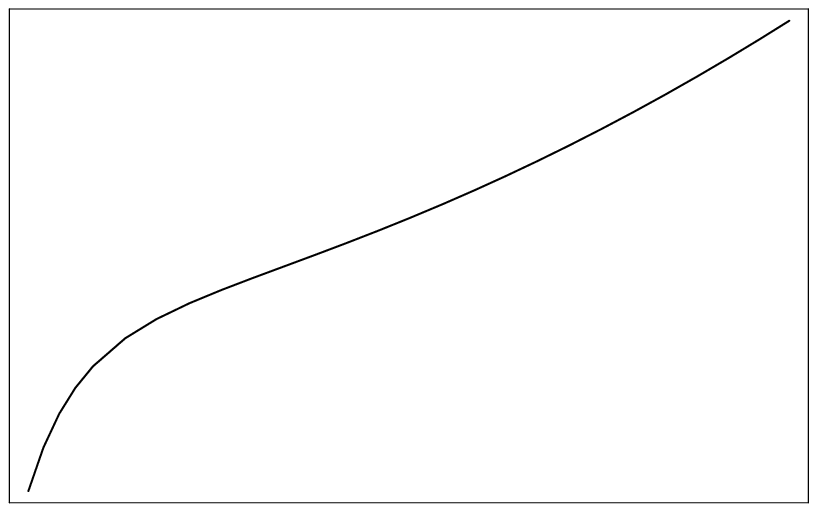}    
\caption{Potential for Schwarzschild Anti--de Sitter scalar field and tensor--type perturbations in dimension $d=6$. Plot is in the radial coordinate from the black hole horizon to asymptotic infinity, with $\ell=0,2,4$, respectively.}
}


\subsection{Greybody Factors at Low Frequency} \label{sec:adslow}


In this section we turn to the greybody factor, at low frequencies, for black holes in asymptotically AdS spacetimes. We do this in two different approximations. In one approximation we consider black holes of arbitrary size, but we focus on the case where the frequency of the emitted radiation is much smaller than the scale set by the cosmological constant. In the other approximation we consider small AdS black holes, \textit{i.e.}, black holes whose size is much smaller than the distance--scale set by the cosmological constant. Reliable computations of greybody factors for black holes in AdS geometries, in the low frequency approximation, do not seem to have been previously performed in the literature. For high frequencies, however, greybody factors have been considered in \cite{hk00}, in a geometrical optics approximation.

The class of black hole solutions that we consider have a metric of the form \eqref{bhmet} with the function $f(r)$ of the form \eqref{fdiv}, \textit{i.e.}, we have $f(r) = f_h(r) + f_a(r)$ where, in here, $f_a(r)$ is given by
\begin{equation}
\label{faads} f_a(r) = 1 + \kappa^2 r^2,
\end{equation}
\noindent
such that setting $f(r)=f_a(r)$ in the metric \eqref{bhmet} corresponds to an AdS geometry.

To compute the leading order greybody factor at low frequencies \eqref{lowfreq}, we shall consider in the following an $\ell = 0$ scalar wave propagating in the background of an asymptotically AdS black hole spacetime. The wave equation is given by \eqref{sceq2} with $\ell = 0$ and with the potential $V(r)$ given in terms of $f(r)$ by \eqref{potV}. Notice that the tortoise coordinate $x$ is still defined in terms of $f(r)$ by \eqref{deftort}.

Consider the general scalar wave equation \eqref{sceq2} for $\ell=0$. When $r \gg R_H$, the tortoise coordinate $x$, defined in \eqref{deftort}, can be written as
\begin{equation}
x = \frac{1}{\kappa} \arctan (\kappa r).
\end{equation}
\noindent
In particular, this expression tells us that $x < \frac{\pi}{2\kappa}$. The potential $V(r)$, defined in \eqref{potV}, is now given by
\begin{equation} \label{adsV}
V(r) = \frac{(d-2)(1+\kappa^2 r^2)(d-4+d \kappa^2 r^2 )}{4r^2}.
\end{equation}
\noindent
In the following, we shall find it useful to work in terms of a dimensionless variable for the frequency. We therefore define
\begin{equation} \label{hatom}
\hat{\omega} \equiv \frac{\omega}{\kappa}.
\end{equation}
In addition, it will also be useful to consider the dimensionless quantities $\frac{T_H}{\kappa}$ and $\kappa R_H$.

\subsubsection*{Measurement of Asymptotic Fluxes}

Before proceeding with the determination of the greybody factor for black holes in AdS, we first consider how to measure the incoming and outgoing fluxes in an asymptotically AdS spacetime. This has also been briefly discussed in the introduction.

The scalar equation \eqref{sceq2}, with $V(r)$ given by \eqref{adsV}, has the following general solution for $\kappa r \gg 1$
\begin{equation} \label{phiasyads}
\Phi_\omega (u) = u^{\frac{d-1}{2}} \left( \widehat{C}_1 H^{(1)}_{\frac{d-1}{2}} (u) + \widehat{C}_2 H^{(2)}_{\frac{d-1}{2}} (u) \right) ,
\end{equation}
\noindent
where
\begin{equation}
u \equiv \frac{\omega}{\kappa^2 r} .
\end{equation}
\noindent
It is apparent from \eqref{phiasyads} that we can identify $H^{(1)}_{\frac{d-1}{2}} (u)$ as the outgoing part and $H^{(2)}_{\frac{d-1}{2}} (u)$ as the incoming part of the wave--function $\Phi_\omega$. Now, we have that $x \simeq \frac{\pi}{2\kappa} - \frac{u}{\omega}$ for $\kappa r \gg 1$. Therefore, the asymptotic region $\kappa x \rightarrow \frac{\pi}{2}$ corresponds to $u \rightarrow 0$. In this region \eqref{phiasyads} reduces to
\begin{equation} \label{thephiads}
\Phi_\omega = \frac{\widehat{C}_1+\widehat{C}_2}{2^{\frac{d-1}{2}} \Gamma \left( \frac{d+1}{2} \right) } \left( \frac{\omega}{\kappa^2 r} \right)^{d-1} + i (\widehat{C}_2-\widehat{C}_1) \frac{2^{\frac{d-1}{2}} \Gamma \left( \frac{d-1}{2} \right) }{\pi} .
\end{equation}
\noindent
Using the fact that $d/dx \simeq - \omega d/du$ for $\kappa r \gg 1$, we see that the total asymptotic flux is then given by
\begin{equation}
J_{\rm asy} = - \Omega_{d-2} r^{d-2} \frac{\omega}{2i} \left( \Phi_\omega^* \frac{d\Phi_\omega}{du} - \Phi_\omega \frac{d\Phi_\omega^*}{du} \right) = J_{\rm in} - J_{\rm out} ,
\end{equation}
\noindent
with the incoming and outgoing fluxes given by the following expressions:
\begin{equation} \label{AdSfluxes}
J_{\rm in} =  \frac{2 \Omega_{d-2} \omega^{d-1}}{\pi \kappa^{2d-4} } | \widehat{C}_2 |^2, \qquad J_{\rm out} = \frac{2 \Omega_{d-2} \omega^{d-1}}{\pi \kappa^{2d-4} } | \widehat{C}_1 |^2 .
\end{equation}
\noindent
One may alternatively write the total asymptotic flux as
\begin{equation} \label{adsJasy}
J_{\rm asy} = \frac{\Omega_{d-2} \omega^{d-1}}{\pi \kappa^{2d-4} } \left[ (\widehat{C}_2-\widehat{C}_1)(\widehat{C}^*_1+\widehat{C}^*_2) + (\widehat{C}_1+\widehat{C}_2) (\widehat{C}^*_2-\widehat{C}^*_1) \right],
\end{equation}
\noindent
an expression which will be of some use in our subsequent analysis.

\subsubsection*{Greybody Factor for $\omega \ll \kappa$}

We shall begin by considering the case of computing greybody factor for AdS black holes such that $\omega \ll \kappa$. Naturally, this requirement must also be supplemented with the usual low frequency requirement, $\omega \ll T_H$, since we have made use of this condition when matching regions I and II earlier. To summarize, we shall consider the greybody factor in the specific regime where
\begin{equation} \label{otherreg}
\hat{\omega} \ll \frac{T_H}{\kappa}, \qquad \hat{\omega} \ll 1.
\end{equation}
\noindent
Here we used the rescaled frequency defined in \eqref{hatom}. An obvious interest in this particular regime is that it also includes large AdS black holes, as these are characterized by having $\kappa R_H \gg 1$. We shall comment more on this point below.

Considering region III, which is the standard asymptotic region where $r \gg R_H$, we have that the potential $V(r)$ is simply given by \eqref{adsV}. From this expression it is immediate to realize that, in the regime \eqref{otherreg}, we have
\begin{equation}
V(r) \geq 2(d-2) \kappa^2 \gg \omega^2,
\end{equation}
\noindent
when $r \gg R_H$. Thus, we see that region III is now included in region II, a region which was defined in section \ref{sec:greysch} as the region where $V(r) \gg \omega^2$. Therefore, it follows from \eqref{genphiII} and \eqref{faads} that the wave--function, for $r \gg R_H$, is given by
\begin{equation}
\Phi_\omega (r) = \AI \left( 1 + i \omega R_H^{d-2} \int^r_\infty \frac{dr'}{(r')^{d-2} (1+\kappa^2 (r')^2)} \right),
\end{equation}
\noindent
and in the particular limit where $\kappa r \gg 1$ this expression becomes
\begin{equation} \label{wfa}
\Phi_\omega (r) = \AI \left( 1 - \frac{i \omega R_H^{d-2}}{(d-1) \kappa^2 r^{d-1}} \right).
\end{equation}
\noindent
From this, we may now easily determine both $\widehat{C}_2 - \widehat{C}_1$ and $\widehat{C}_1 + \widehat{C}_2$ by comparison with equation \eqref{thephiads}, which also is valid for $\kappa r \gg 1$, with the result that
\begin{equation} \label{theChats}
\widehat{C}_2 - \widehat{C}_1 = - i \frac{\pi}{2^{\frac{d-1}{2}} \Gamma ( \frac{d-1}{2} ) } \AI , \qquad \widehat{C}_1 + \widehat{C}_2 = - i\, 2^{\frac{d-3}{2} } \Gamma ( \frac{d-1}{2} ) \left( \frac{\kappa^2 R_H}{\omega} \right)^{d-2} \AI.
\end{equation}
\noindent
Inserting this result in the expression for the total asymptotic flux \eqref{adsJasy}, it is simple to obtain
\begin{equation}
J_{\rm asy} = A_H \omega |\AI|^2 ,
\end{equation}
\noindent
where $A_H$ is the area of the black hole event horizon, defined in \eqref{defarea}. Comparing this expression with \eqref{totalJhor}, we find
\begin{equation}\label{AdSfluxcons}
J_{\rm hor} = J_{\rm asy} = J_{\rm in} - J_{\rm out},
\end{equation}
\noindent
which expresses the fact that the total flux is preserved from the horizon to the asymptotic region.

Let us now define the quantity
\begin{equation} \label{defz}
z( \hat{\omega} ) \equiv \frac{\widehat{C}_2-\widehat{C}_1}{\widehat{C}_1 +\widehat{C}_2} .
\end{equation}
\noindent
In terms of $z(\hat{\omega})$, one can write the greybody factor $\gamma (\hat{\omega})$ as
\begin{equation} \label{greyz}
\gamma ( \hat{\omega} ) = \frac{J_{\rm hor}}{J_{\rm in}} = 1 - \frac{|\widehat{C}_1|^2}{|\widehat{C}_2|^2} = 1 - \left| \frac{1-z(\hat{\omega})}{1+z(\hat{\omega})} \right|^2 ,
\end{equation}
\noindent
where we made use of \eqref{AdSfluxes} and \eqref{AdSfluxcons}. On the other hand, using \eqref{theChats} in \eqref{defz}, we obtain
\begin{equation} \label{explz}
z(\hat{\omega} ) = \frac{\pi}{2^{d-2} [ \Gamma(\frac{d-1}{2}) ]^2}\, \frac{\hat{\omega}^{d-2}}{(\kappa R_H)^{d-2}}.
\end{equation}
\noindent
Inserting this result in \eqref{greyz}, we have fully computed the greybody factor for asymptotically AdS black holes, in the low frequency regime \eqref{otherreg}. Observe that since $z(\hat{\omega}) > 0$ we always have $0 < \gamma (\hat{\omega} ) \leq 1$.

One interesting feature to notice is that we can define a critical frequency, $\hat{\omega}_c$, by the equation
\begin{equation} \label{critom}
z ( \hat{\omega}_c ) = 1.
\end{equation}
\noindent
From \eqref{explz} one simply determines
\begin{equation} \label{omegac}
\hat{\omega}_c = \frac{2 [ \Gamma(\frac{d-1}{2}) ]^\frac{2}{d-2}}{\pi^{\frac{1}{d-2}}}\, \kappa R_H.
\end{equation}
\noindent
What one learns from this result is that, since from \eqref{greyz} $\gamma (\hat{\omega}_c ) = 1$, there is no reflection of radiation at the critical frequency $\hat{\omega}_c$, \textit{i.e.}, the black hole absorbs all of the radiation which is sent towards it. Equivalently, in the reverse process, in which there is emission of radiation from the black hole, it means that all of the emitted radiation will reach the asymptotic region. Moreover, from \eqref{omegac} we see that having $\hat{\omega} = \hat{\omega}_c$ implies $\hat{\omega} \sim \kappa R_H$, and since from \eqref{otherreg} we have $\hat{\omega} \ll 1$, we obtain that $\kappa R_H \ll 1$. This result just means that we are dealing with a small AdS black hole. Therefore, we may conclude that only for small AdS black holes one can achieve the critical frequency $\hat{\omega}_c$, at least when working in the specific regime \eqref{otherreg}.

If we now consider frequencies much lower than the critical frequency, $\hat{\omega} \ll \hat{\omega}_c$, we find that
\begin{equation}
\gamma (\hat{\omega} ) = 4 z(\hat{\omega} ) = \frac{\pi}{2^{d-2} [ \Gamma(\frac{d-1}{2}) ]^2}\, \frac{\hat{\omega}^{d-2}}{(\kappa R_H)^{d-2}}.
\end{equation}
\noindent
In this case the greybody factor is inversely proportional to the area of the black hole, whereas it is proportional to $\omega^{d-2}$. We thus see that $\hat{\omega} \ll \hat{\omega}_c$ is equivalent to $\hat{\omega} \ll \kappa R_H $. In other words, we have that large AdS black holes, with $\kappa R_H \gg 1$, are always in a regime such that $\hat{\omega} \ll \hat{\omega}_c$.

Considering instead frequencies much higher than the critical frequency, $\hat{\omega} \gg \hat{\omega}_c$, we now find that
\begin{equation}
\gamma (\hat{\omega} ) = \frac{4}{z(\hat{\omega} )} = \frac{2^{d-2} [ \Gamma(\frac{d-1}{2}) ]^2}{\pi} \frac{(\kappa R_H)^{d-2}}{\hat{\omega}^{d-2}}.
\end{equation}
\noindent
In this case the greybody factor is proportional to the area of the black hole, whereas it is instead inversely proportional to $\omega^{d-2}$. We thus see that $\hat{\omega} \gg \hat{\omega}_c$ is equivalent to $\hat{\omega} \gg \kappa R_H $. Since from \eqref{otherreg} we have $\hat{\omega} \ll 1$, one may infer that the condition $\hat{\omega} \gg \hat{\omega}_c$ is only possible for small AdS black holes with $\kappa R_H \ll 1$.

What we learn from the above results is that, both for $\hat{\omega} \ll \hat{\omega}_c$ and for $\hat{\omega} \gg \hat{\omega}_c$, the greybody factor behaves in a remarkably different fashion from the case of an asymptotically flat black hole \eqref{gammaflatbh}. Indeed, for an asymptotically flat black hole we found that $\gamma (\omega ) \sim \omega^{d-2} A_H$. This is quite contrary to the behavior of $\gamma(\hat{\omega})$ which we now find for AdS black holes, both for $\hat{\omega} \ll \hat{\omega}_c$ and $\hat{\omega} \gg \hat{\omega}_c$. It is furthermore also quite different from the case of  asymptotically dS black holes, for which $\gamma(\hat{\omega} )\sim A_H /A_C$, see equation \eqref{dSgreyfactor}.

Finally, one may ask whether the condition $\hat{\omega} \ll T_H/\kappa$ in \eqref{otherreg} is consistent with the above considerations concerning the critical frequency $\hat{\omega}_c$. In the particular case of a neutral AdS black hole, the temperature is given by
\begin{equation} \label{Tadsbh}
\frac{T_H}{\kappa} = \frac{d-3 + (d-1) (\kappa R_H)^2}{4\pi \kappa R_H}.
\end{equation}
\noindent
Now, for a small AdS black hole, this implies $T_H/\kappa \sim (\kappa R_H)^{-1}$. Therefore $\hat{\omega} \ll T_H/\kappa$ is equivalent to $\hat{\omega} \ll (\kappa R_H)^{-1}$. Having $\hat{\omega} = \hat{\omega}_c$ means that $\hat{\omega} \sim \kappa R_H$, and thus it is clearly possible to have both $\hat{\omega} \ll T_H/\kappa$ and $\hat{\omega}=\hat{\omega}_c$ for a small AdS black hole. For a large AdS black hole, we have instead $T_H/\kappa \sim \kappa R_H$. Therefore $\hat{\omega} \ll T_H /\kappa$ implies that $\hat{\omega} \ll \kappa R_H$ which is equivalent to $\hat{\omega} \ll \hat{\omega}_c$. We conclude that the bound $\hat{\omega} \ll T_H/\kappa$ is indeed consistent with the above considerations.

\subsubsection*{Greybody Factor for Small AdS Black Holes}

We finally turn to the case of small black holes in AdS, \textit{i.e.}, black holes with $\kappa R_H \ll 1$. Combining this condition with the low frequency requirement, \eqref{lowfreq}, we see that we are considering the regime where
\begin{equation} \label{smallads}
\hat{\omega} \ll \frac{T_H}{\kappa}, \qquad \hat{\omega} \ll \frac{1}{\kappa R_H}, \qquad \kappa R_H \ll 1.
\end{equation}
\noindent
Since $\kappa R_H \ll 1$, we may consider an intermediate region defined via $R_H \ll r \ll 1/\kappa$. This region overlaps with region II, which was previously defined in section \ref{sec:greysch} as the region where $V(r) \ll \omega^2$. By combining \eqref{genphiII} with \eqref{faads} we learn that, for $r \gg R_H$, $r\omega \ll 1$ and $\kappa r \ll 1$, the wave--function behaves as
\begin{equation} \label{adsmatch}
\Phi_\omega (r) = \AI \left( 1 - i \frac{\omega R_H^{d-2}}{(d-3)r^{d-3}} \right).
\end{equation}
\noindent
In the following we shall match the wave--function solved in the asymptotic region of the AdS geometry, \textit{i.e.}, region III as originally defined in section \ref{sec:greysch}, to the behavior \eqref{adsmatch} of the wave--function in region II. Just like in the earlier cases, this will allow for a direct evaluation of the greybody factors. At this stage, it is useful to re--write the scalar wave equation \eqref{sceq2} in terms of some more appropriate variables. To this end, let us define the coordinate
\begin{equation}
z = \sin^2 (\kappa x) = \frac{\kappa^2 r^2}{1+\kappa^2 r^2}.
\end{equation}
\noindent
In this case, the scalar wave equation \eqref{sceq2} becomes
\begin{equation} \label{ggeq}
4 z (1-z) \frac{d^2 g}{dz^2} + 2(1-2z) \frac{dg}{dz} + \frac{4\hat{\omega}^2z(1-z) - (d-2)(d-4+4z)}{4z(1-z)} g = 0,
\end{equation}
\noindent
where we have defined
\begin{equation}
g \equiv r^{\frac{d-2}{2}} \Phi_{\omega}.
\end{equation}
\noindent
The general solution to \eqref{ggeq} is the familiar hypergeometric solution,
\begin{eqnarray}\label{gsol}
g &=& C_1\, z^{\frac{d-2}{4}} (1-z)^{\frac{2-d}{4}} \hypgeo \left. \left[ - \frac{\hat{\omega}}{2}, \frac{\hat{\omega}}{2} ; \frac{d-1}{2} \right| z \right] \nonumber \\ 
&& + C_2\, z^{\frac{4-d}{4}} (1-z)^{\frac{d}{4}} \hypgeo \left. \left[ 1 - \frac{\hat{\omega}}{2},1 + \frac{\hat{\omega}}{2} ; \frac{d+1}{2} \right| 1-z \right].
\end{eqnarray}

Let us start by considering the $z \rightarrow 0$ limit. This limit corresponds to having $\kappa r \ll 1$, which implies that $f_a \simeq 1$. Thus, in this limit,  the wave equation reduces to that of flat spacetime. Moreover, in this same limit, we also have that $z \simeq \kappa^2 r^2$ and we thus see from the general solution \eqref{gsol} that the wave--function $\Phi_{\omega}(r)$ becomes
\begin{equation} \label{zto0}
\Phi_{\omega} = C_1 \kappa^{\frac{d-2}{2}} + C_2 \kappa^{\frac{4-d}{2}} \frac{\Gamma (\frac{d+1}{2} ) \Gamma (\frac{d-3}{2} )}{\Gamma(\frac{d-1+\hat{\omega}}{2} )\Gamma(\frac{d-1-\hat{\omega}}{2} )} \frac{1}{r^{d-3}} ,
\end{equation}
\noindent
for $\kappa r \ll 1$. Next, consider instead the $z \rightarrow 1$ limit. This corresponds to having $\kappa r \gg 1$, implying that $f_a \simeq \kappa^2 r^2$. Thus, in this limit, we have that $1-z \simeq 1/(\kappa^2 r^2)$ and we obtain from the general solution \eqref{gsol} that the wave--function $\Phi_{\omega}(r)$ becomes
\begin{equation} \label{zto1}
\Phi_{\omega} = C_1 \kappa^{\frac{d-2}{2}} \frac{[\Gamma (\frac{d-1}{2} )]^2}{\Gamma(\frac{d-1+\hat{\omega}}{2} )\Gamma(\frac{d-1-\hat{\omega}}{2} )} + \frac{C_2}{ \kappa^{\frac{d}{2}} r^{d-1}} ,
\end{equation}
\noindent
when $\kappa r \gg 1$. We can find the $C_1$, $C_2$ coefficients by matching \eqref{zto0} with \eqref{adsmatch} in region II, since both these expressions are valid in the regime where $R_H \ll r \ll 1/\kappa$ and $r \ll 1/\omega$. This gives
\begin{equation}
C_1 = \kappa^{\frac{2-d}{2}} \AI, \qquad C_2 = - i \kappa^{\frac{d-4}{2}} \frac{\Gamma(\frac{d-1+\hat{\omega}}{2} )\Gamma(\frac{d-1-\hat{\omega}}{2} )}{\Gamma (\frac{d+1}{2} ) \Gamma (\frac{d-3}{2} )} \frac{\omega R_H^{d-2}}{d-3} \AI.
\end{equation}
\noindent
Inserting this in the expression \eqref{zto1} for the wave--function for $\kappa r \gg 1$ we can read off $\widehat{C}_1$ and $\widehat{C}_2$ by \eqref{thephiads}, obtaining
\begin{eqnarray} \label{theChats2}
\widehat{C}_2 - \widehat{C}_1 &=& - i \frac{\pi \Gamma ( \frac{d-1}{2} )}{2^{\frac{d-1}{2}}\Gamma(\frac{d-1+\hat{\omega}}{2} )\Gamma(\frac{d-1-\hat{\omega}}{2} )} \AI , \nonumber \\
\widehat{C}_1 + \widehat{C}_2 &=& - i\, 2^{\frac{d-3}{2} } \frac{\Gamma(\frac{d-1+\hat{\omega}}{2} )\Gamma(\frac{d-1-\hat{\omega}}{2} )}{ \Gamma ( \frac{d-1}{2} ) } \left( \frac{\kappa^2 R_H}{\omega} \right)^{d-2} \AI .
\end{eqnarray}
\noindent
At this point we notice that \eqref{theChats2} reduces to \eqref{theChats} for $\hat{\omega} \ll 1$. This is a good consistency check since both regimes \eqref{otherreg} and \eqref{smallads} are valid for $\omega \ll T_H$, $\omega R_H \ll 1$, $\kappa R_H \ll 1$ and $\hat{\omega} \ll 1$.

Inserting the result \eqref{theChats2} into \eqref{adsJasy} we can now find the total asymptotic flux
\begin{equation}
J_{\rm asy} = A_H \omega | \AI |^2.
\end{equation}
\noindent
with $A_H$ being the area of the event horizon, defined in \eqref{defarea}. Comparing this result with \eqref{totalJhor}, we again find that the flux is conserved
\begin{equation}\label{AdSfluxcons2}
J_{\rm hor} = J_{\rm asy} = J_{\rm in} - J_{\rm out} .
\end{equation}
\noindent
Finally turning to the greybody factor, we obtain from \eqref{theChats2} that
the quantity $z(\hat{\omega})$, defined in \eqref{defz}, is given by
\begin{equation} \label{fullz}
z(\hat{\omega} ) = \frac{\pi}{2^{d-2}}\, \frac{[ \Gamma (\frac{d-1}{2} )]^2} {[ \Gamma(\frac{d-1+\hat{\omega}}{2} )\Gamma(\frac{d-1-\hat{\omega}}{2} ) ]^2}\, \frac{\hat{\omega}^{d-2}}{(\kappa R_H)^{d-2}}.
\end{equation}
\noindent
Thus, the greybody factor $\gamma (\hat{\omega})$ is given in terms of $z(\hat{\omega})$ by \eqref{greyz}. From \eqref{smallads} we moreover have that the greybody factor which we hereby have computed is accurate as long as $\hat{\omega} \ll 1/(\kappa R_H)$.

Analyzing $z(\hat{\omega} )$ in \eqref{fullz} as a function of $\hat{\omega}$, and for fixed $\kappa R_H$, we find that $z(\hat{\omega}) = 0$ for $\hat{\omega} = 2n + d - 1$ with $n \in \{ 0, 1, 2, \cdots \}$. Therefore we see from \eqref{greyz} that one has
\begin{equation}
\gamma (\hat{\omega}) = 0 \quad \mbox{for} \quad \hat{\omega} = 2n + d - 1 \quad \mbox{with} \quad n \in \{ 0, 1, 2, \cdots \}.
\end{equation}
\noindent
Thus, at these critical frequencies, we find that the greybody factor vanishes. This implies, for the specific absorption process that we are considering, that the radiation that we are sending towards the black hole is completely reflected. In the reverse process, where one considers emission of radiation from the black hole, it instead means that at these critical frequencies the radiation cannot overcome the potential barrier. Notice that the values of the critical frequencies precisely match the values of the normal frequencies of scalar wave perturbations in \textit{pure} AdS spacetime, as computed in \cite{ns04}. It would be interesting to understand this match both in light of black hole physics, as what we study in the present paper, as well as in light of using our techniques in order to compute dual correlation functions via AdS/CFT.

In addition to the critical values of $\hat{\omega}$ for which $\gamma( \hat{\omega} )=0$ we also find critical frequencies for which $\gamma( \hat{\omega} )=1$. These critical frequencies are, as above, solutions to the equation $z(\hat{\omega})=1$. We can write this equation as
\begin{equation} \label{hatomeq}
\frac{\hat{\omega}^{d-2}}{[ \Gamma(\frac{d-1+\hat{\omega}}{2} )\Gamma(\frac{d-1-\hat{\omega}}{2} ) ]^2} = \frac{2^{d-2}}{\pi [ \Gamma (\frac{d-1}{2} )]^2} (\kappa R_H)^{d-2}.
\end{equation}
\noindent
For small $\hat{\omega}$, we find that $\hat{\omega}_c$ given by \eqref{omegac} is a solution, in accordance with the results above for the regime \eqref{otherreg}. However, there are also other solutions to \eqref{hatomeq}. By \eqref{smallads}, we see that the right--hand side is required to be small. On the other hand, the left--hand side precisely vanishes for $\hat{\omega} = 2n + d - 1$, with $n \in \{ 0, 1, 2, \cdots \}$. This means that, for sufficiently small $\kappa R_H$, there can be several solutions to equation \eqref{hatomeq}. In detail, these solutions occur for $(2n+d-1-\hat{\omega})^2 \sim (\kappa R_H)^{d-2}$. Therefore, there are several possible critical frequencies, allowed at small $\kappa R_H$, for which we have that the greybody factor is $\gamma = 1$. This is consistent with the fact that the smaller values of $\kappa R_H$ we have, the larger values of $\hat{\omega}$ we can consider, as we see from \eqref{smallads}. In order to illustrate the behavior of $\gamma(\hat{\omega})$ we have depicted it in Figure \ref{fig:adsgam} for $d=4$ and $\kappa R_H = 0.05$. As one can see in the figure, for every critical value of $\hat{\omega}$ with $\gamma(\hat{\omega})=0$ we have two critical values for which $\gamma(\hat{\omega})=1$.

\FIGURE[ht]{\label{fig:adsgam}
    \centering
    \epsfxsize=.6\textwidth
    \put(-10,205){\Large{$\gamma$}}
    \put(315,5){\Large{$\hat{\omega}$}}
    \leavevmode
    \epsfbox{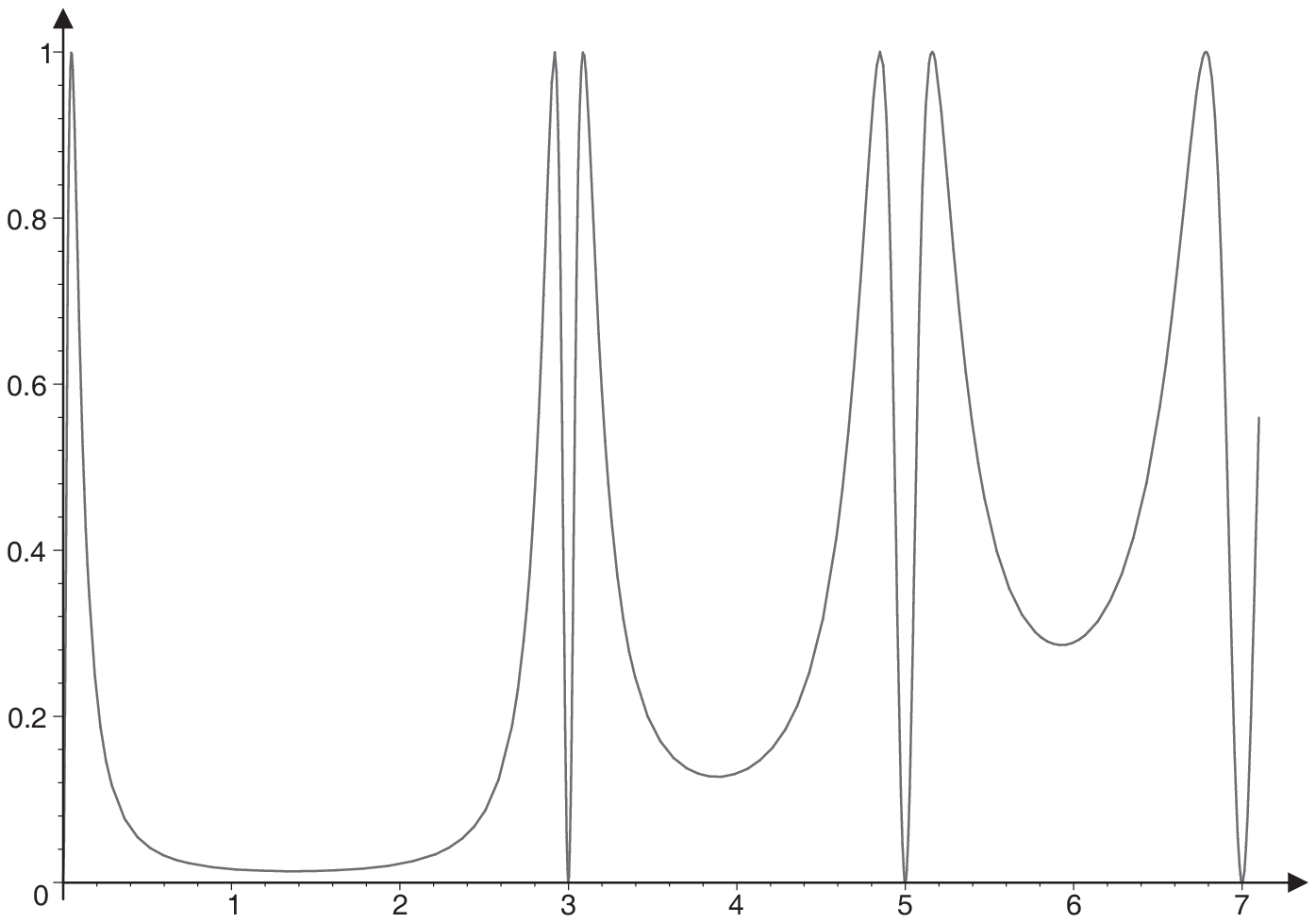}
\caption{The greybody factor $\gamma(\hat{\omega})$ for $d=4$ and $\kappa R_H = 0.05$.}
}

Let us end this section with one last comment concerning AdS/CFT. As we have discussed earlier, the greybody factor is a useful quantity also to compute physical observables, such as emission rates of particles off a black hole. In the particular case of an asymptotically flat spacetime, these observables are associated to the concept of an S--matrix, and as such the primary quantity to extract out of the greybody factor is the absorption cross--section. As has become clear in recent years (see, \textit{e.g.}, \cite{agmoo99}), the good physical observables for perturbative quantum gravity in an asymptotically AdS spacetime are the boundary correlation functions of the dual gauge theory. As such, the AdS greybody factor we have just computed is a first step in order to evaluate these thermal correlators (one still needs to adapt the calculation in order to allow boundary insertions of arbitrary gauge theory operators), and such a calculation should be considered in the future, both in the present case of low frequency as well as in the case of asymptotic frequencies which we shall consider in the following.


\subsection{Greybody Factors at Asymptotic Frequency}



\subsubsection{The Schwarzschild Anti--de Sitter Solution}


For the Schwarzschild AdS geometry, asymptotic greybody factors have not been considered in the past literature, and we fill such a gap in the present paper. We shall compute $d$--dimensional asymptotic gravitational greybody--factors for the Schwarzschild AdS geometry, using the monodromy--matching technique first developed in \cite{cns04, ns04}. We already know from the previous sections that it is not a difficult exercise to extend the monodromy--matching technique from its original quasinormal mode application to the present calculation of asymptotic greybody factors, by paying special attention to the appropriate change in the boundary conditions. This is, however, a subtle issue in asymptotically AdS geometries, but one which we shall resolve in the following. In the present section we shall explain how to compute the greybody factors at large imaginary frequencies for the Schwarzschild AdS black hole. The following calculation relies heavily on \cite{ns04}, where any missing details may be found.

We consider solutions of the Schr\"odinger--like equation in the complex $r$--plane. Near the singularity $r=0$, these solutions behave as
\begin{equation}
\Phi (x) \sim B_+ \sqrt{2\pi\omega x}\ J_{\frac{j}{2}} \left( \omega x \right) + B_- \sqrt{2\pi\omega x}\ J_{-\frac{j}{2}} \left( \omega x \right),
\end{equation}
\noindent
where $x$ is the tortoise coordinate, $J_\nu$ represents a Bessel function of the first kind and $B_\pm$ are (complex) integration constants. The parameter $j$ is left generic for the time being, but will ultimately be set equal to $j=0$ for tensor and scalar type perturbations and equal to $j=2$ for vector type perturbations.

\FIGURE[ht]{\label{StokesSAdS}
    \centering
    \psfrag{A}{$A$}
    \psfrag{B}{$B$}
    \psfrag{RH}{$R_H$}
    \psfrag{g1}{}
    \psfrag{og1}{}
    \psfrag{g2}{}
    \psfrag{og2}{}
    \psfrag{Re}{$\re$}
    \psfrag{Im}{$\im$}
    \psfrag{contour}{contour}
    \psfrag{Stokes line}{Stokes line}
    \epsfxsize=.6\textwidth
    \leavevmode
    \epsfbox{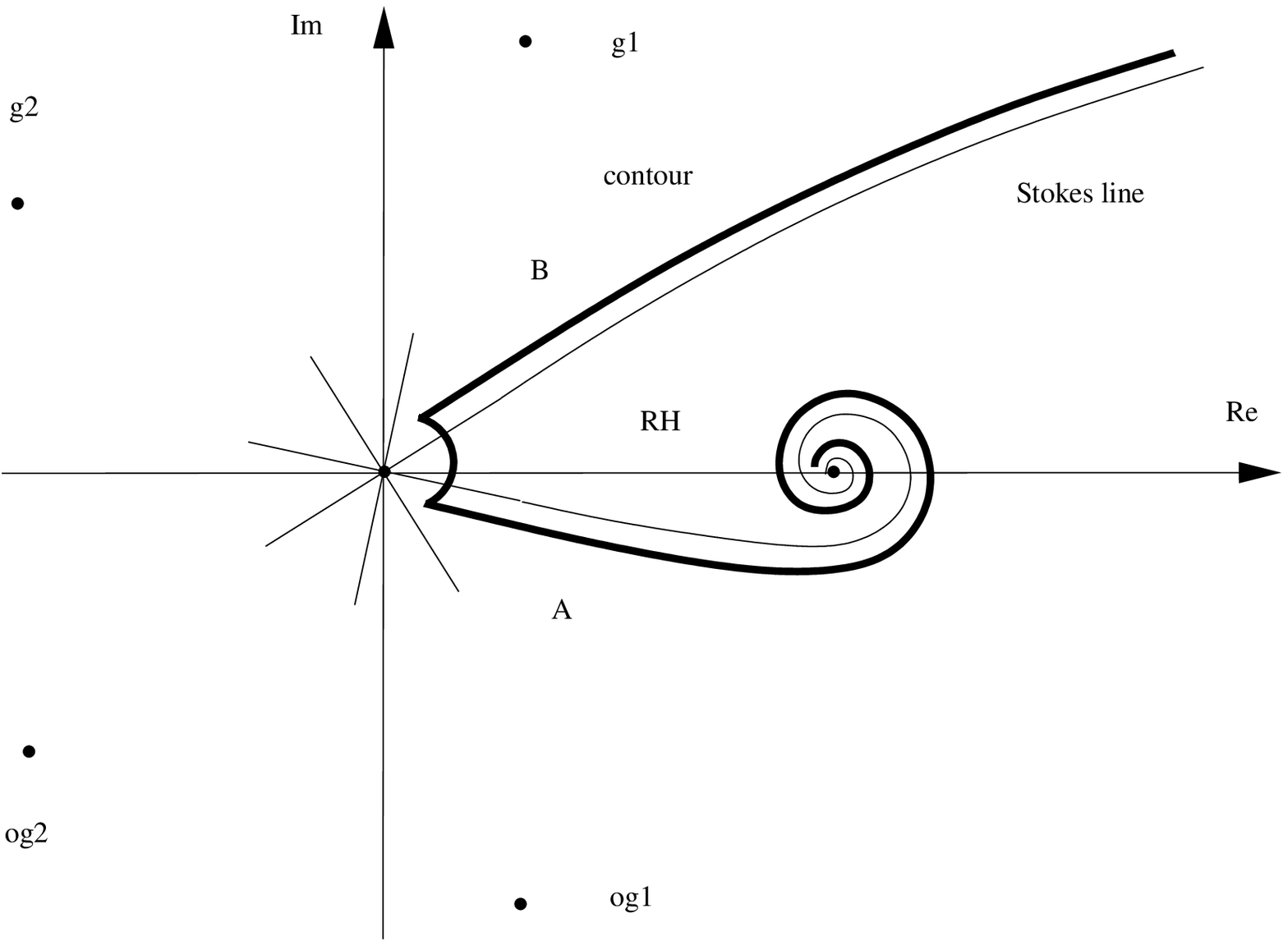}
\caption{Stokes line for the Schwarzschild Anti--de Sitter black hole, along with the chosen contour for monodromy matching, in the case of dimension $d=6$.}
}

Our monodromy calculation must be carried out along the standard contour in Figure~\ref{StokesSAdS}. Starting at point $B$, our solution can be approximated in the limit $\im(\omega) \gg \re(\omega)$ by
\begin{equation}
\Phi (x) \sim \left( B_+ e^{-i\alpha_+} + B_- e^{-i\alpha_-}\right) e^{i \omega x} + \left( B_+ e^{i\alpha_+} + B_- e^{i\alpha_-}\right) e^{-i \omega x},
\end{equation}
\noindent
where $\alpha_\pm = \frac{\pi}4 (1 \pm j)$. This is to be matched to
\begin{equation}
\Phi (x) \sim e^{i \omega x} + R e^{-i \omega x}
\end{equation}
\noindent
for an incoming wave at infinity. As one rotates from point $B$ to point $A$ near the origin, the approximate expression for $\Phi$ changes to
\begin{equation}
\Phi (x) \sim \left( B_+ e^{-i\alpha_+} + B_- e^{-i\alpha_-}\right) e^{i \omega x} + \left( B_+ e^{-3i\alpha_+} + B_- e^{-3i\alpha_-}\right) e^{-i \omega x},
\end{equation}
\noindent
which is to be matched to the expression for $\Phi$ near the horizon,
\begin{equation}
\Phi (x) \sim T e^{i \omega x}.
\end{equation}
\noindent
Therefore we have the system
\begin{eqnarray}
B_+ e^{-i\alpha_+} + B_- e^{-i\alpha_-} &=& 1, \nonumber \\
B_+ e^{i\alpha_+} + B_- e^{i\alpha_-} &=& R, \nonumber \\
B_+ e^{-i\alpha_+} + B_- e^{-i\alpha_-} &=& T, \nonumber \\
B_+ e^{-3i\alpha_+} + B_- e^{-3i\alpha_-} &=& 0,
\end{eqnarray}
\noindent
from which $T=1$ and
\begin{equation}
R = 2i \cos\left(\frac{\pi j}2\right) = \pm 2i,
\end{equation}
\noindent
where the plus (minus) sign corresponds to $j=0$ ($j=2$) and tensor or scalar (vector) type perturbations.

In this case it is important to point out that the quasinormal modes are \textit{not} the poles of the coefficients, as the boundary condition at infinity is different from the asymptotically flat or asymptotically dS case. They can, however, be obtained by matching the approximate expression at $r \sim \infty$
\begin{equation}
\Phi (x) \sim \left( C_+ e^{i\beta_+} + C_- e^{i\beta_-}\right) e^{i \omega (x-x_0)} + \left( C_+ e^{-i\beta_+} + C_- e^{-i\beta_-}\right) e^{-i \omega (x-x_0)}
\end{equation}
\noindent
(see \cite{ns04} for further details) to
\begin{equation}
\Phi (x) \sim e^{i \omega x} + R e^{-i \omega x},
\end{equation}
\noindent
and requiring $C_-=0$. The resulting condition,
\begin{equation}
e^{2i \omega x_0 -2i\beta_+} = R,
\end{equation}
\noindent
is easily seen to yield the asymptotic quasinormal frequencies in \cite{ns04}.

In the limit $\im(\omega) \gg \re(\omega)$, our solution can be approximated at point $B$ by
\begin{equation}
\Phi (x) \sim \left( B_+ e^{i\alpha_+} + B_- e^{i\alpha_-}\right) e^{i \omega x} + \left( B_+ e^{-i\alpha_+} + B_- e^{-i\alpha_-}\right) e^{-i \omega x},
\end{equation}
\noindent
which is to be matched to
\begin{equation}
\Phi (x) \sim e^{i \omega x} + \widetilde{R} e^{-i \omega x}.
\end{equation}
\noindent
As one rotates from point $B$ to point $A$ near the origin, the approximate expression for $\Phi$ changes to
\begin{equation}
\Phi (x) \sim \left( B_+ e^{-3\alpha_+} + B_- e^{-3\alpha_-}\right) e^{i \omega x} + \left( B_+ e^{-i\alpha_+} + B_- e^{-i\alpha_-}\right) e^{-i \omega x},
\end{equation}
\noindent
which is to be matched to
\begin{equation}
\Phi (x) \sim \widetilde{T} e^{i \omega x}.
\end{equation}
\noindent
Therefore we have the system
\begin{eqnarray}
B_+ e^{i\alpha_+} + B_- e^{i\alpha_-} &=& 1, \nonumber \\
B_+ e^{-i\alpha_+} + B_- e^{-i\alpha_-} &=& \widetilde{R}, \nonumber \\
B_+ e^{-3i\alpha_+} + B_- e^{-3i\alpha_-} &=& \widetilde{T}, \nonumber \\
B_+ e^{-i\alpha_+} + B_- e^{-i\alpha_-} &=& 0,
\end{eqnarray}
\noindent
from which $\widetilde{T} = 1$ and $\widetilde{R} = 0$. Notice that we have the consistency check
\begin{equation}
R \widetilde{R} + T \widetilde{T} = 1.
\end{equation}
\noindent
These very same coefficients will appear again in the RN AdS calculation. The greybody factor finally follows as
\begin{equation}
\gamma(\omega) = T(\omega) \widetilde{T}(\omega) = 1.
\end{equation}


\subsubsection{The Reissner--Nordstr\"om Anti--de Sitter Solution}


As in all non--asymptotically flat spacetime geometries, the RN AdS black hole asymptotic greybody factors have not been considered in the past literature, and we fill such a gap in the present paper. We shall compute $d$--dimensional asymptotic gravitational greybody--factors for the RN AdS geometry, using the monodromy--matching technique first developed in \cite{ns04}. As usual, the main difference with respect to the calculation in \cite{ns04} is an appropriate change in the boundary conditions, from quasinormal to greybody boundary conditions, a subtle issue in asymptotically AdS geometries, but one which we shall resolve in the following. This is what we do in the present section, as we shall now explain how to compute the greybody factors at large imaginary frequencies for the RN AdS black hole. The following calculation heavily relies on \cite{ns04}, where any missing details may be found.

We consider solutions of the Schr\"odinger--like equation in the complex $r$--plane. Near the singularity $r=0$, these solutions behave as
\begin{equation}
\Phi (x) \sim B_+ \sqrt{2\pi\omega x}\ J_{\frac{j}{2}} \left( \omega x \right) + B_- \sqrt{2\pi\omega x}\ J_{-\frac{j}{2}} \left( \omega x \right),
\end{equation}
\noindent
where $x$ is the tortoise coordinate, $J_\nu$ represents a Bessel function of the first kind and $B_\pm$ are (complex) integration constants. The parameter $j$ satisfies $j=\frac{d-3}{2d-5}$ for tensor and scalar type perturbations and $j=\frac{3d-7}{2d-5}$ for vector type perturbations.

\FIGURE[ht]{\label{StokesRNAdS}
    \centering
    \psfrag{A}{$A$}
    \psfrag{B}{$B$}
    \psfrag{R+}{$R^+$}
    \psfrag{R-}{$R^-$}
    \psfrag{g1}{}
    \psfrag{og1}{}
    \psfrag{g2}{}
    \psfrag{og2}{}
    \psfrag{g3}{}
    \psfrag{og3}{}
    \psfrag{Re}{$\re$}
    \psfrag{Im}{$\im$}
    \psfrag{contour}{contour}
    \psfrag{Stokes line}{Stokes line}
    \epsfxsize=.6\textwidth
    \leavevmode
    \epsfbox{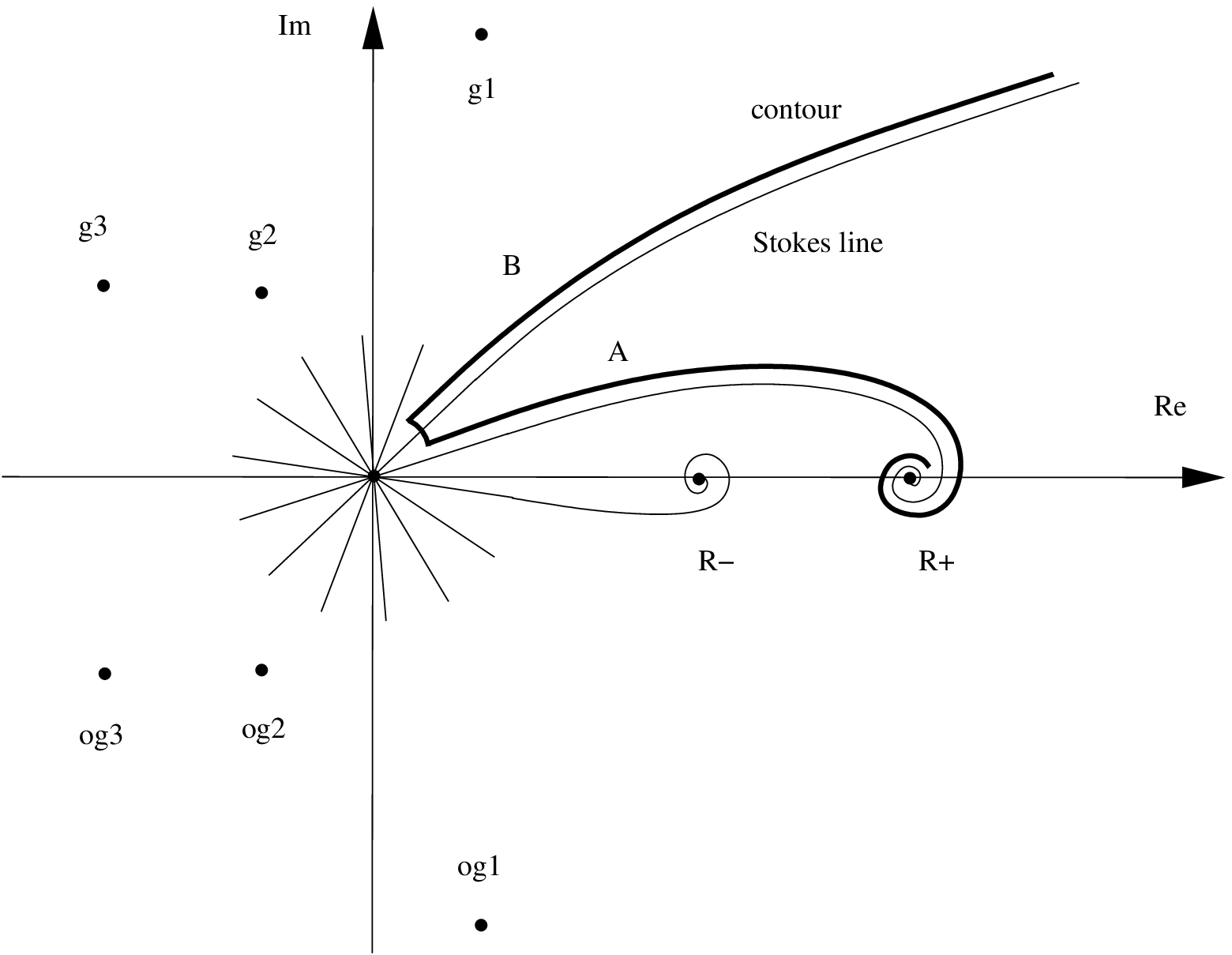}
\caption{Stokes line for the Reissner--Nordstr\"om Anti--de Sitter black hole, along with the chosen contour for monodromy matching, in the case of dimension $d=6$.}
}

Our monodromy calculation must be carried out along the standard contour in Figure~\ref{StokesRNAdS}. Starting at point $B$, our solution can be approximated in the limit $\im(\omega) \gg \re(\omega)$ by
\begin{equation}
\Phi (x) \sim \left( B_+ e^{-i\alpha_+} + B_- e^{-i\alpha_-}\right) e^{i \omega x} + \left( B_+ e^{i\alpha_+} + B_- e^{i\alpha_-}\right) e^{-i \omega x},
\end{equation}
\noindent
where $\alpha_\pm = \frac{\pi}4 (1 \pm j)$. This is to be matched to
\begin{equation}
\Phi (x) \sim e^{i \omega x} + R e^{-i \omega x}
\end{equation}
\noindent
for an incoming wave at infinity. As one rotates from point $B$ to point $A$ near the origin, the approximate expression for $\Phi$ changes to
\begin{equation}
\Phi (x) \sim \left( B_+ e^{-i\alpha_+} + B_- e^{-i\alpha_-}\right) e^{i \omega x} + \left( B_+ e^{-3i\alpha_+} + B_- e^{-3i\alpha_-}\right) e^{-i \omega x},
\end{equation}
\noindent
which is to be matched to the expression for $\Phi$ near the horizon,
\begin{equation}
\Phi (x) \sim T e^{i \omega x}.
\end{equation}
\noindent
Therefore we have the system
\begin{eqnarray}
B_+ e^{-i\alpha_+} + B_- e^{-i\alpha_-} &=& 1, \nonumber \\
B_+ e^{i\alpha_+} + B_- e^{i\alpha_-} &=& R, \nonumber \\
B_+ e^{-i\alpha_+} + B_- e^{-i\alpha_-} &=& T, \nonumber \\
B_+ e^{-3i\alpha_+} + B_- e^{-3i\alpha_-} &=& 0,
\end{eqnarray}
\noindent
from which $T = 1$ and
\begin{equation}
R = 2i \cos\left(\frac{\pi j}2\right).
\end{equation}

In this asymptotically AdS case it is important to point out that the quasinormal modes are not the poles of the scattering coefficients, as the boundary condition at infinity is different from the asymptotically flat or asymptotically de Sitter case. They can, however, be obtained by matching the approximate expression at $r \sim \infty$
\begin{equation}
\Phi (x) \sim \left( C_+ e^{i\beta_+} + C_- e^{i\beta_-}\right) e^{i \omega (x-x_0)} + \left( C_+ e^{-i\beta_+} + C_- e^{-i\beta_-}\right) e^{-i \omega (x-x_0)}
\end{equation}
\noindent
(see \cite{ns04} for further details) to
\begin{equation}
\Phi (x) \sim e^{i \omega x} + R e^{-i \omega x},
\end{equation}
\noindent
and requiring $C_-=0$. The resulting condition,
\begin{equation}
e^{2i \omega x_0 -2i\beta_+} = R,
\end{equation}
\noindent
is easily seen to yield the asymptotic quasinormal frequencies in \cite{ns04}.

In the limit $\im(\omega) \gg \re(\omega)$, our solution can be approximated at point $B$ by
\begin{equation}
\Phi (x) \sim \left( B_+ e^{i\alpha_+} + B_- e^{i\alpha_-}\right) e^{i \omega x} + \left( B_+ e^{-i\alpha_+} + B_- e^{-i\alpha_-}\right) e^{-i \omega x},
\end{equation}
\noindent
which is to be matched to
\begin{equation}
\Phi (x) \sim e^{i \omega x} + \widetilde{R} e^{-i \omega x}.
\end{equation}
\noindent
As one rotates from point $B$ to point $A$ near the origin, the approximate expression for $\Phi$ changes to
\begin{equation}
\Phi (x) \sim \left( B_+ e^{-3\alpha_+} + B_- e^{-3\alpha_-}\right) e^{i \omega x} + \left( B_+ e^{-i\alpha_+} + B_- e^{-i\alpha_-}\right) e^{-i \omega x},
\end{equation}
\noindent
which is to be matched to
\begin{equation}
\Phi (x) \sim \widetilde{T} e^{i \omega x}.
\end{equation}
\noindent
Therefore we have the system
\begin{eqnarray}
B_+ e^{i\alpha_+} + B_- e^{i\alpha_-} &=& 1, \nonumber \\
B_+ e^{-i\alpha_+} + B_- e^{-i\alpha_-} &=& \widetilde{R}, \nonumber \\
B_+ e^{-3i\alpha_+} + B_- e^{-3i\alpha_-} &=& \widetilde{T}, \nonumber \\
B_+ e^{-i\alpha_+} + B_- e^{-i\alpha_-} &=& 0,
\end{eqnarray}
\noindent
from which $\widetilde{T} = 1$ and $\widetilde{R} = 0$. Notice that we have the consistency check
\begin{equation}
R \widetilde{R} + T \widetilde{T} = 1.
\end{equation}
\noindent
Incidentally, this is the exact same result as for the Schwarzschild AdS solution. This result shows that, for asymptotically AdS spacetimes, the scattering coefficients are universal. The same thing happens for the greybody factor, which finally follows as
\begin{equation}
\gamma(\omega) = T(\omega) \widetilde{T}(\omega) = 1.
\end{equation}

\section*{Acknowledgments}
The work of JN and RS was partially supported by the Funda\c{c}\~ao para a Ci\^encia e a Tecnologia (Portugal) through the Programme POCI/2010/FEDER and the Project POCI/MAT/58549/2004. The work of TH is partially supported by the European Community's Human Potential Programme under contract MRTN-CT-2004-005104 ``Constituents, Fundamental Forces and Symmetries of the Universe''. TH would like to thank the Carlsberg Foundation for support.

\vfill

\eject


\vfill

\eject

\bibliographystyle{plain}

\end{document}